\documentclass[12pt]{article}
\usepackage{url}
\usepackage{setspace}
\usepackage{natbib}

\usepackage[colorlinks,citecolor=blue,urlcolor=blue]{hyperref}
\usepackage[utf8]{inputenc}
\usepackage[margin=2.5cm]{geometry}
\usepackage{setspace}
\usepackage{rotating}
\usepackage{color}
\usepackage{amsmath}
\usepackage{pdflscape}
\usepackage{float}

\usepackage{natbib}
\usepackage{subfig}
\usepackage{amsfonts}

\usepackage{url}
\usepackage{amsmath}
\usepackage{amssymb}
\usepackage{graphicx,psfrag,epsf}
\usepackage{xcolor}
\usepackage{subfig}

\usepackage[algo2e,ruled]{algorithm2e}

\usepackage{natbib}
\usepackage[margin=2.5cm]{geometry}

\newtheorem{theorem}{Theorem}

\begin{document}
	
	\title{Mastering Rare Event Analysis: Optimal Subsample Size in Logistic and Cox Regressions}
	
	\author{TAL AGASSI, NIR KERET, MALKA GORFINE$^\ast$\\[4pt]
		\textit{Department of Statistics and Operations Research} 	\\[2pt]			
			\textit{Tel Aviv University, Tel Aviv 69978, Israel.}	\\[2pt]
		\\[2pt]
		{gorfinem@tauex.tau.ac.il}}
	
	\markboth%
	{T. Agassi  and others}
	{Optimal Subsample Size}
	
	\maketitle
	
	\footnotetext{To whom correspondence should be addressed.}
	
	\begin{abstract}
{In the realm of contemporary data analysis, the use of massive  datasets has taken on heightened significance, albeit often entailing considerable demands on computational time and memory. While a multitude of existing works offer optimal subsampling methods for conducting analyses on subsamples with minimized efficiency loss, they notably lack tools for judiciously selecting the optimal subsample size. To bridge this gap, our work introduces tools designed for  choosing the optimal subsample size. We focus on three settings: the Cox regression model for survival data with rare events and logistic regression for both balanced and imbalanced datasets. Additionally, we present a novel optimal subsampling procedure tailored for logistic regression with imbalanced data. The efficacy of these tools and procedures is  demonstrated through an extensive simulation study and meticulous analyses of two sizable datasets.}{Hypothesis testing; Imbalanced data; Time to event analysis; Relative efficiency; }
	\end{abstract}

\clearpage

\doublespacing

\section{Introduction}
\label{sec:intro}
The escalating demand to analyze massive datasets with millions of observations often leads to considerable computational time and memory requirements, presenting significant challenges in implementing statistical analyses. In response, subsampling has become a widely adopted and effective method for expediting computation, for various regression models. These models encompass least-squares regression models \citep{dhillon2013new,ma2015statistical}, logistic regression \citep{wang2018optimal,wang2021nonuniform}, generalized linear models \citep{ai2021optimal}, quantile regression \citep{wang2021optimal}, quasi-likelihood estimators \citep{yu2020optimal}, time-to-event regression under the additive-hazards model \citep{zuo2021sampling}, semi-competing risks \citep{gorfine2021marginalized}, the Cox proportional-hazards (PH) model \citep{keret2023analyzing} and accelerated failure time model \citep{yang2024subsampling}.

In this work, we mainly concentrate on two prominent scenarios associated with rare events: (1) addressing the challenge of highly imbalanced data in logistic regression, where one of the classes is rare, and (2) employing Cox proportional-hazards (PH) regression \citep{cox1972regression} for survival data characterized by a notably high right-censoring rate. 


\cite{wang2018optimal} introduced an innovative subsampling method optimized for logistic regression, demonstrating high effectiveness for balanced data but acknowledging its limited efficacy for highly imbalanced data. In the context of rare-event data, a natural approach involves subsampling exclusively within the majority group (the common class or censored observations) to prevent the loss of crucial information.
Addressing this concern, \cite{wang2021nonuniform} focused on logistic regression and proposed an optimal subsampling procedure  targeting the rare-event setting, ensuring retention of all events in binary outcome scenarios. However, the underlying assumption is that the proportion of rare events decreases as the sample size increases, a condition that is often considered undesirable.

For survival analysis involving rare events,  \cite{gorfine2021marginalized} advocated subsampling solely the observations that have not yet experienced the event (i.e., the censored observations) and implemented a uniform subsampling approach. In a similar vein, \cite{keret2023analyzing} presented an optimal subsampling strategy for the Cox PH model within the rare-events framework. In this approach, optimal subsampling exclusively targets censored observations, combining all observed events with the subsample set of censored observations. These optimal subsampling techniques have been convincingly demonstrated to significantly reduce computational burden compared to analyzing the entire dataset, with minimal loss of efficiency.

However, a notable aspect left unaddressed in the aforementioned works  is the lack of practical guidelines for determining the subsample size. While our primary goal is to reduce computation time, we are equally committed to maintaining the  statistical power or efficiency for answering the research questions and avoiding a substantial increase in standard errors. Hence,  it is valuable to offer researchers a tool for choosing the subsample size that aligns with their research objectives. 

This work offers notable contributions in two key aspects: 
\begin{enumerate}
	\item We introduce tools designed to optimize the process of selecting subsample sizes in the realm of optimal subsampling. These tools are versatile, and applied here to Cox regression models dealing with rare events and logistic regression models, regardless of the presence of rare events. 
	\item We present optimal subsampling methods specifically tailored for logistic regression models handling rare events. Notably, our approach assumes that the proportion of rare events converges to a positive constant with increasing sample size, a substantial departure from the assumption made by \cite{wang2021nonuniform}.	
\end{enumerate}

This paper is structured as follows: Section \ref{Cox} begins by summarizing the key findings on optimal subsampling from \cite{keret2023analyzing} to ensure the current paper is self-contained. It then introduces new methodologies for determining optimal subsample size. Section \ref{logistic_rare_events} proposes a two-step subsampling algorithm specifically designed for logistic regression in scenarios involving rare events, including techniques for selecting the subsample size. Section \ref{non_rare_logistic} focuses on  Wang et al.'s (2018) two-step algorithm to nearly balanced datasets and offers strategies for identifying the optimal subsample size. Section \ref{sims} summarizes a comprehensive simulation study to evaluate the effectiveness of the proposed approaches. Sections \ref{data_survival} and \ref{data_logistic} focus on analyzing two large-scale datasets, a survival regression model with around 350 million records and a logistic regression model with approximately 28 million observations. The paper concludes with a short discussion in Section \ref{discussion}.

\section{Optimal Subsample Size for Cox Regression with Optimal Subsampling}\label{Cox}

\subsection{Notation, Formulation and Reservoir-Sampling \citep{keret2023analyzing}}\label{sec_cox_notation}
For the sake of clarity, this section presents the model formulation and pertinent findings from \cite{keret2023analyzing}. Consider a set of $n$ independent and identically distributed observations. Let $V_i$ represent the failure time for the $i$th observation, $C_i$ denote the right-censoring time, and $T_i$ signify the observed time,  $T_i=\min(V_i,C_i)$. Define $\Delta_i=I(V_i \leq C_i)$, and let $\mathbf{X}_i$ be a vector of potentially time-dependent covariates of size $r$. The observed dataset is denoted by $\mathcal{D}_n=\{T_i,\Delta_i,\mathbf{X}_i \, ; \, i=1,\dots,n\}$. Among the $n$ observations, there are $n_e$ instances where the failure times are observed, termed ``events". It is assumed that as $n \rightarrow \infty$, the ratio $n_e/n$ converges to a small positive constant. The count of censored observations is represented by $n_c=n-n_e$, and $\tau$ denotes the maximum follow-up time.

In time-to-event data, the predominant source of information comes from events rather than censored observations. This rationale underlies the two-step algorithm introduced by \cite{keret2023analyzing}, which utilizes all observed events while sampling a subset of censored observations. Let $q_n$ be the number of censored observations sampled from the full data, where $q_n$ is typically much smaller than $n$, and it is assumed that $q_n/n$ converges to a small positive constant as $q_n, n \rightarrow \infty$. Define $\mathcal{C}$ as the index set containing all censored observations in the full data, and $\mathcal{Q}$ as the index set encompassing all observations with observed failure times and all censored observations included in the subsample. Due to computational and theoretical considerations, censored observations are sampled with replacement, implying that a censored observation in the original sample may appear more than once in the subsample. 

 Let $\boldsymbol{\beta}$ be a vector of size $r$ of unknown coefficients. Then, under the Cox PH regression, the instantaneous hazard rate of observation $i$ at time $t$ is given by
\begin{align*}
	\lambda(t|\mathbf{X}_i)=\lambda_0(t)e^{\boldsymbol{\beta}^T\mathbf{X}_i} \quad \quad i=1,\dots,n
\end{align*}
where $\lambda_0(\cdot)$ is an unspecified non-negative function and $\Lambda_0(t)=\int_0^t \lambda_0(u)du$ is the cumulative baseline hazard function. The goal is estimating the unknown parameters  $\boldsymbol{\beta}$ and $\Lambda_0$. Define $\mathbf{S}^{(k)}(\boldsymbol{\beta},t)=\sum_{i=1}^n e^{\boldsymbol{\beta}^T\mathbf{X}_i}Y_i(t)\mathbf{X}_i^{\otimes k}, k=0,1,2$, where $\mathbf{X}^{\otimes 0}=1,\mathbf{X}^{\otimes 1}=\mathbf{X}, \mathbf{X}^{\otimes 2}=\mathbf{X}\mathbf{X}^T$ and $Y_i(t)=I(T_i\geq t)$ is the at-risk process of observation $i$  at time $t$. Denote $\widehat{\boldsymbol{\beta}}_{PL}$ as the full-sample partial-likelihood (PL) estimator of $\boldsymbol{\beta}$ that solves
\begin{equation*}
	\frac{\partial l(\boldsymbol{\beta})}{\partial\boldsymbol{\beta}^T}=\sum_{i=1}^n \Delta_i \bigg\{\mathbf{X}_i-\frac{\mathbf{S}^{(1)}(\boldsymbol{\beta},T_i)}{S^{(0)}(\boldsymbol{\beta},T_i)}\bigg\}=\mathbf{0} \, .
\end{equation*}

Suppose that the data are organized such that the censored observations precede the failure times, namely $\mathcal{C}=\{1,\dots,n_c\}$, and $\mathcal{E}=\{n_c+1,\dots,n\}$. Let $\mathbf{p}=(p_1,\dots,p_{n_c})^T$ be a vector of the sampling probabilities for the censored observations, where $\sum_{i=1}^{n_c} p_i=1$, and set
\begin{equation*}
	w_i= \begin{cases}
		(p_i q_n)^{-1} & \text{if } \Delta_i=0, p_i>0\\
		1 & \text{if } \Delta_i=1
	\end{cases} \quad \quad i=1,\dots,n \, .
\end{equation*}
The subsample-based counterpart of $\mathbf{S}^{(k)}(\boldsymbol{\beta},t)$ is $\mathbf{S}_w^{(k)}(\boldsymbol{\beta},t)=\sum_{i\in\mathcal{Q}}w_i e^{\boldsymbol{\beta}^T\mathbf{X}_i}Y_i(t)\mathbf{X}_i^{\otimes k}, k=0,1,2$. Then, $\widetilde{\boldsymbol{\beta}}$ is defined as the estimator derived from the subsample $\mathcal{Q}$,  by solving
\begin{equation}\label{cox_weighted_likelihood}
	\frac{\partial l^*(\boldsymbol{\beta})}{\partial\boldsymbol{\beta}^T}\equiv\sum_{i\in\mathcal{Q}} \Delta_i \bigg\{\mathbf{X}_i-\frac{\mathbf{S}_w^{(1)}(\boldsymbol{\beta},T_i)}{S_w^{(0)}(\boldsymbol{\beta},T_i)}\bigg\}=\mathbf{0}
\end{equation}
where $l^*$ is the log PL based on the subsample $\mathcal{Q}$. Finally, for a given vector  $\boldsymbol{\beta}$, define
$
	\widehat{\Lambda}_0(t,\boldsymbol{\beta})=\sum_{i=1}^n {\Delta_i I(T_i\leq t)}/{S^{(0)}(\boldsymbol{\beta},T_i)},
$
and the Breslow estimator \citep{breslow1972contribution} of $\Lambda_0$ function is produced by $\widehat{\Lambda}_0(t,\widehat{\boldsymbol{\beta}}_{PL})$.

Consistency and asymptotic normality of $\widetilde{\boldsymbol{\beta}}$ and $\widehat{\Lambda}_0$ were established by \cite{keret2023analyzing} under some regularity assumptions. Specifically,  given the true  $\boldsymbol{\beta}^o$, 
\begin{equation*}
	\sqrt{n}\mathbb{V}(\textbf{p},\boldsymbol{\beta}^o)^{-1/2}(\widetilde{\boldsymbol{\beta}}-\boldsymbol{\beta}^o)\xrightarrow[]{D} N(0,\mathbf{I})
\end{equation*}
as $n,q_n \rightarrow\infty$, where $\mathbf{I}$ is the identity matrix and
\begin{equation*}
	\mathbb{V}(\textbf{p},\boldsymbol{\beta})=\boldsymbol{\mathcal{I}}^{-1}(\boldsymbol{\beta})+\frac{n}{q_n}\boldsymbol{\mathcal{I}}^{-1}(\boldsymbol{\beta})\boldsymbol{\varphi}(\mathbf{p},\boldsymbol{\beta})\boldsymbol{\mathcal{I}}^{-1}(\boldsymbol{\beta}) \, ,
\end{equation*}
\begin{equation*}
	\boldsymbol{\varphi}(\mathbf{p},\boldsymbol{\beta})=\frac{1}{n^2}\Bigg\{\sum_{i\in\mathcal{C}}\frac{\mathbf{a}_i(\boldsymbol{\beta})\mathbf{a}_i(\boldsymbol{\beta})^T}{p_i}-\sum_{i,j\in\mathcal{C}}\mathbf{a}_i(\boldsymbol{\beta})\mathbf{a}_j(\boldsymbol{\beta})^T\Bigg\} \, ,
\end{equation*}
\begin{equation*}
	\mathbf{a}_i(\boldsymbol{\beta})=\int_{0}^{\tau}\bigg\{\mathbf{X}_i-\frac{\mathbf{S}^{(1)}(\boldsymbol{\beta},t)}{S^{(0)}(\boldsymbol{\beta},t)}\bigg\}\frac{Y_i(t)e^{\boldsymbol{\beta}^T\mathbf{X}_i}}{S^{(0)}(\boldsymbol{\beta},t)}dN.(t) \, ,
\end{equation*}
where $N_{.}(t) =  \sum_{i=1}^n \Delta_i I(T_i \leq t)$ and
\begin{equation*}
	\boldsymbol{\boldsymbol{\mathcal{I}}}(\boldsymbol{\beta})=\frac{1}{n}\frac{\partial^2l(\boldsymbol{\beta})}{\partial\boldsymbol{\beta}^T\partial\boldsymbol{\beta}}=-\frac{1}{n}\int_0^\tau \Bigg\{\frac{\mathbf{S}^{(2)}(\boldsymbol{\beta},t)}{S^{(0)}(\boldsymbol{\beta},t)}-\Big(\frac{\mathbf{S}^{(1)}(\boldsymbol{\beta},t)}{S^{(0)}(\boldsymbol{\beta},t)}\Big)\Big(\frac{\mathbf{S}^{(1)}(\boldsymbol{\beta},t)}{S^{(0)}(\boldsymbol{\beta},t)}\Big)^T\Bigg\}dN.(t) \, .
\end{equation*}
As $\boldsymbol{\mathcal{I}}$ and $\boldsymbol{\varphi}$ involve the entire dataset, their subsampling-based counterparts, $\widetilde{\boldsymbol{\mathcal{I}}}$ and $\widetilde{\boldsymbol{\varphi}}$, will be utilized in the variance estimator $\widetilde{\mathbb{V}}(\textbf{p},\widetilde{\boldsymbol{\beta}})$. The Exact expressions of $\widetilde{\boldsymbol{\mathcal{I}}}$ and $\widetilde{\boldsymbol{\varphi}}$ can be found in the Supplementary Material (SM) file Section S1.

The above results laid the foundation for establishing the subsequent optimal subsampling probabilities. The A-optimal sampling probabilities vector, denoted as $\mathbf{p}^A$ and derived from minimizing the trace of $\mathbb{V}(\mathbf{p},\boldsymbol{\beta}^o)$, is expressed as follows 
\begin{equation}\label{A_optimal}
	p_m^A=\frac{\|\boldsymbol{\mathcal{I}}^{-1}(\boldsymbol{\beta}^o)\mathbf{a}_m(\boldsymbol{\beta}^o)\|_2}{\sum_{i\in \mathcal{C}} \|\boldsymbol{\mathcal{I}}^{-1}(\boldsymbol{\beta}^o)\mathbf{a}_i(\boldsymbol{\beta}^o)\|_2} \quad \textit{ for all m} \in \mathcal{C}
\end{equation}
where $\|\cdot\|_2$ is the $l_2$ euclidean norm. The L-optimal sampling probabilities vector, denoted as $\mathbf{p}^L$ and obtained from minimizing the trace of \(\boldsymbol{\varphi}(\textbf{p},\boldsymbol{\beta}^o)\), is given by
\begin{equation} \label{L_optimal}
	p_m^L=\frac{\|\mathbf{a}_m(\boldsymbol{\beta}^o)\|_2}{\sum_{i\in \mathcal{C}} \|\mathbf{a}_i(\boldsymbol{\beta}^o)\|_2}  \quad  \textit{ for all m} \in \mathcal{C} \, .
\end{equation}
Evidently, $\mathbf{p}^A$ incorporate $\boldsymbol{\mathcal{I}}^{-1}$, enabling a more efficient estimation of $\boldsymbol{\beta}$ in contrast to the estimator using probabilities from the L-optimal criterion. Nevertheless, for the same reason, $\mathbf{p}^A$ demands a greater computational time.

As both $\mathbf{p}^A$ and $\mathbf{p}^L$ rely on the true unknown regression vector, $\boldsymbol{\beta}^o$, the following two-step procedure has been proposed. It commences with a quick and straightforward consistent estimator of the regression vector to estimate the optimal sampling probabilities. The complete implementation of the two-step procedure is outlined below:

\begin{algorithm2e}
	\caption{Cox Regression - Two Step Optimal Subsampling}\label{alg:algorithm1}
	\-\hspace{5mm}
		{\bf Step 1:}
		Select $q_0$ observations uniformly from $\mathcal{C}$ and combine them with the observed events to get $\mathcal{Q}{pilot}$. Conduct a weighted Cox regression on $\mathcal{Q}{pilot}$ and obtain  $\widetilde{\boldsymbol{\beta}}_U$ based on Eq. \eqref{cox_weighted_likelihood}.  Compute approximated optimal sampling probabilities by substituting $\boldsymbol{\beta}^o$ with $\widetilde{\boldsymbol{\beta}}_U$ in Eq. \eqref{A_optimal} or \eqref{L_optimal}.
		
			\-\hspace{5mm}
		{\bf Step 2:}
		Select $q_n$ observations from $\mathcal{C}$ based on the sampling probabilities of  Step 1. Combine these selected observations with the observed events and get $\mathcal{Q}$. Perform a weighted Cox regression on $\mathcal{Q}$, based on Eq. \eqref{cox_weighted_likelihood}, and get  the two-step estimator $\widetilde{\boldsymbol{\beta}}_{TS}.$
\end{algorithm2e}

The original algorithm employs  $q_0=q_n$. However, here we  suggest using a small value of $q_0$ for the initial uniform sampling of Step 1. Subsequently,  the methods outlined in the following subsections focus on determining the value of  $q_n$ under various criteria. To maintain computational efficiency, we recommend setting $q_0$ as $c_0  n_e$, where $c_0$ is a small scalar (e.g., $c_0 < 5$). Our simulation study and real-data analysis indicate that this recommendation is generally sufficient. Furthermore, our findings suggest that when none of the covariates exhibit long-tailed distributions, setting $c_0 = 1$ is often adequate. 

The asymptotic properties of $\widetilde{\boldsymbol{\beta}}_{TS}$ and $\widehat{\Lambda}_0(t,\widetilde{\boldsymbol{\beta}}_{TS})$ were established   \citep{keret2023analyzing}. Specifically, it was shown that under standard assumptions,
\begin{equation*}
	\sqrt{n}\mathbb{V}(\mathbf{p}^{opt},\boldsymbol{\beta}^o)^{-1/2}(\widetilde{\boldsymbol{\beta}}_{TS}-\boldsymbol{\beta}^o)\xrightarrow[]{D}N(\mathbf{0},\mathbf{I})
\end{equation*}
as $q_n, n \rightarrow\infty$, where $\mathbf{p}^{opt}$ is either $\mathbf{p}^A$ or $\mathbf{p}^L$. Moreover, the asymptotic theory  accommodates left truncation, stratified analysis, time-dependent covariates and time-dependent coefficients. However, a practical methodology for selecting the size of $q_n$ was not studied, despite its considerable importance. This motivates us to propose the following frameworks for determining the necessary size of $q_n$ based on specific objectives.

Often, datasets are too voluminous to fit within the RAM limitations of standard computers, a challenge highlighted in the real-data analyses of Section 6. \cite{keret2023analyzing}  proposed a speedy and memory-efficient approach for batch-based reservoir sampling, designed to operate on conventional computer systems. Summarizing,  the observed-events dataset, $\mathcal{E}$, is consistently maintained in the RAM. Conversely, the censored observations are split into $B$ batches, labeled as $\mathcal{B}_1,\ldots,\mathcal{B}_B$. At any given moment, only one batch is in the RAM. In every batch $b$, $b=1,\ldots,B$, we approximate $\mathbf{p}^A$ or $\mathbf{p}^L$ by considering the dataset comprised of $\mathcal{E} \cup \mathcal{B}_b$. The approximation employs distinct weights, 1 for an event and $n_c/|\mathcal{B}_b|$ for each censored observation. The reservoir-sampling algorithm selects $q_n$ observations with replacement in a single iteration. The key idea involves keeping a sample of $q_n$ observations (referred to as the ``reservoir"), where replacements can occur as new batches are loaded. For an in-depth description and proof of the algorithm validity, refer to Section 2.6 in \cite{keret2023analyzing}. This reservoir-sampling algorithm can be applied to any scenario of sampling with replacement, independently of the original regression problem. In this work, we employ it for  survival regression with approximately 350 million records.

\subsection{Subsample Size Based on  Relative Efficiency}\label{RE_Cox}
What is the optimal size of $q_n$ that maintains a small efficiency loss? Here, we introduce a tool that enables us to evaluate the efficiency loss resulting from the subsampling approach. We begin by defining an estimator of the relative efficiency (RE) of the two-step estimator compared to the full PL estimator by
\begin{equation}\label{RR}
	RE(q_n)=\frac{\|n^{-1}\boldsymbol{\mathcal{I}}^{-1}(\widetilde{\boldsymbol{\beta}}_{TS})+q_n^{-1}\boldsymbol{\mathcal{I}}^{-1}
		(\widetilde{\boldsymbol{\beta}}_{TS})\boldsymbol{\varphi}(\mathbf{p}^{opt},\widetilde{\boldsymbol{\beta}}_{TS})\boldsymbol{\mathcal{I}}^{-1}(\widetilde{\boldsymbol{\beta}}_{TS})\|_F}{\|n^{-1}\boldsymbol{\mathcal{I}}^{-1}(\widehat{\boldsymbol{\beta}}_{PL})\|_F}
\end{equation}
where $\|A\|_F=\sqrt{\sum_{i,j} |a_{i,j}|^2}$. The lower limit of Eq. \eqref{RR} is close to $1$. 

If interest lies in the effect of a particular covariate, e.g., the $p$-th covariate, then we may utilize
\begin{equation}\label{RR_single}
	RE_p(q_n)=\frac{\Big[n^{-1}\boldsymbol{\mathcal{I}}^{-1}(\widetilde{\boldsymbol{\beta}}_{TS})+q_n^{-1}\boldsymbol{\mathcal{I}}^{-1}(\widetilde{\boldsymbol{\beta}}_{TS})\boldsymbol{\varphi}(\mathbf{p}^{opt},\widetilde{\boldsymbol{\beta}}_{TS})\boldsymbol{\mathcal{I}}^{-1}(\widetilde{\boldsymbol{\beta}}_{TS})\Big]_{pp}}{\Big[n^{-1}\boldsymbol{\mathcal{I}}^{-1}(\widehat{\boldsymbol{\beta}}_{PL})\Big]_{pp}}
\end{equation}
where $\big[A\big]_{pp}$ is the $pp$ element of the matrix $A$. Adjusted optimal sampling probabilities to target a subset of covariates, while retaining the rest in the model to control for confounders, are available in \cite{keret2023analyzing} (see, Equations 7 and 8).

The equations above pose practical challenges: firstly, they include $\widehat{\boldsymbol{\beta}}_{PL}$—whose calculations we aim to avoid; secondly, they involve $\widetilde{\boldsymbol{\beta}}_{TS}$, which can be computed \emph{after} determining the subsample size, $q_n$. However, leveraging  the consistent estimator $\widetilde{\boldsymbol{\beta}}_U$ from Step 1 resolves it by substituting $\widetilde{\boldsymbol{\beta}}_{TS}$ and $\widehat{\boldsymbol{\beta}}_{PL}$ with $\widetilde{\boldsymbol{\beta}}_U$. An additional challenge arises as ${\boldsymbol{\mathcal{I}}}^{-1}(\widetilde{\boldsymbol{\beta}}_U)$ and ${\boldsymbol{\varphi}}(\mathbf{p}^{opt},\widetilde{\boldsymbol{\beta}}_U)$ involve the full data. Alternatively, their subsampling-based counterparts, $\widetilde{\boldsymbol{\mathcal{I}}}^{-1}(\widetilde{\boldsymbol{\beta}}_U)$ and $\widetilde{\boldsymbol{\varphi}}(\mathbf{p}^{opt},\widetilde{\boldsymbol{\beta}}_U)$,
can be used. However, it is advisable to refrain from using $\widetilde{\boldsymbol{\mathcal{I}}}^{-1}(\widetilde{\boldsymbol{\beta}}_U)$ and $\widetilde{\boldsymbol{\varphi}}(\mathbf{p}^{opt},\widetilde{\boldsymbol{\beta}}_U)$ based on the uniform subsample of Step 1, since uniform sampling allows the selection of observations with extremely small optimal probabilities $\mathbf{p}^{opt}$. Consequently, dividing by these probabilities often renders Eq. \eqref{RR} or \eqref{RR_single} numerically unstable. Our proposed approach involves sampling an additional small subsample of size $q_0$, but this time using the approximated optimal probabilities obtained by substituting $\boldsymbol{\beta}^o$, in Eq.s \eqref{A_optimal} and \eqref{L_optimal}, with $\widetilde{\boldsymbol{\beta}}_U$. Let $\mathcal{Q}_{1.5}$ be the index set containing all observations whose failure time was observed, along with the censored observations included in this new subsample of size $q_0$. We denote the counterparts of $\boldsymbol{\mathcal{I}}^{-1}$ and $\boldsymbol{\varphi}$ for this subsample as $\widetilde{\boldsymbol{\mathcal{I}}}^{-1}_{Q_{1.5}}(\widetilde{\boldsymbol{\beta}}_U)$ and $\widetilde{\boldsymbol{\varphi}}_{Q_{1.5}}(\mathbf{p}^{opt},\widetilde{\boldsymbol{\beta}}_U)$ (see the SM, Section S1, for details).

Hence, the proposed RE estimator is given by
\begin{equation}\label{RR_est}
	\widehat{RE}(q_n)=\frac{\|n^{-1}\widetilde{\boldsymbol{\mathcal{I}}}^{-1}_{Q_{1.5}}(\widetilde{\boldsymbol{\beta}}_U)
		+q_n^{-1}\widetilde{\boldsymbol{\mathcal{I}}}^{-1}_{Q_{1.5}}(\widetilde{\boldsymbol{\beta}}_U)\widetilde{\boldsymbol{\varphi}}_{Q_{1.5}}(\widetilde{\mathbf{p}}^{opt},\widetilde{\boldsymbol{\beta}}_U)\widetilde{\boldsymbol{\mathcal{I}}}^{-1}_{Q_{1.5}}(\widetilde{\boldsymbol{\beta}}_U)\|_F}{\|n^{-1}\widetilde{\boldsymbol{\mathcal{I}}}^{-1}_{Q_{1.5}}(\widetilde{\boldsymbol{\beta}}_U)\|_F}
\end{equation}
where $\widetilde{\mathbf{p}}^{opt}$ is the estimated optimal-probabilities vector calculated in Step 1. To save computational time, we utilize $\widetilde{\mathbf{p}}^{opt}$ and $\widetilde{\boldsymbol{\beta}}_U$ from Step 1 instead of re-estimating the optimal probabilities and the coefficient vector based on $\mathcal{Q}_{1.5}$. The computational time for this additional step is very short, since $q_0$ is very small compared to $n$. Lastly,  by calculating $\widetilde{\boldsymbol{\mathcal{I}}}^{-1}_{Q_{1.5}}(\widetilde{\boldsymbol{\beta}}_U)$ and $\widetilde{\boldsymbol{\varphi}}_{Q_{1.5}}(\widetilde{\mathbf{p}}^{opt},\widetilde{\boldsymbol{\beta}}_U)$ only once, a plot of  $\widehat{RE}(q_n)$ as a function of $q_n$ can be generated quickly and effortlessly. This additional step can be added easily in the above two-step Algorithm \ref{alg:algorithm1}, between Steps 1 and 2, as follows:

\textbf{Step 1.5}: Sample $q_0$ observation from $\mathcal{C}$ using the optimal sampling probabilities computed at Step 1. Combine these observations with the observed failure times to form $\mathcal{Q}_{1.5}$ and compute $\widetilde{\boldsymbol{\mathcal{I}}}^{-1}_{Q_{1.5}}(\widetilde{\boldsymbol{\beta}}_U)$ and $\widetilde{\boldsymbol{\varphi}}_{Q_{1.5}}(\mathbf{p}^{opt},\widetilde{\boldsymbol{\beta}}_U)$.  Plot $\widehat{RE}(q_n)$ as a function of $q_n$. Choose the minimal $q_n$ that provides the required RE. 

In practice, with a large $n$, the curve of $\widehat{RE}(q_n)$ is anticipated to show a rapid decrease followed by a gradual decline, resembling an `elbow' shape. A sensible selection for $q_n$ would be in the region where the decline becomes moderate, as the incremental efficiency gain from further increasing $q_n$ is likely to be minimal. Comprehensive examples from simulations and real data analysis are presented in Sections 5--7 for further insights.

\subsection{Subsample Size Based on Hypothesis Testing}\label{hypothesis_testing_cox}

Let \(\beta^o_p\) be the \(p^{th}\) element of  \(\boldsymbol{\beta}^o\). Suppose we aim to test the hypothesis $H_0:\beta^o_p=0$ against $H_1:\beta_p^o \neq 0$ at a significance level of $\alpha$ with a power of $\gamma$. Our current objective is to determine the necessary subsample size, given that $\beta^o_p=\beta_p^*$, where $\beta_p^*$ is specified by the researcher. Since the minimal $n$ should satisfy
$$
n = \left\lceil{(Z_{1-\alpha/2}+Z_{\gamma})^2 \left\{\boldsymbol{\mathcal{I}}^{-1}({\boldsymbol{\beta}}^o)+
\frac{n}{q_n}{\boldsymbol{\mathcal{I}}}^{-1}({\boldsymbol{\beta}}^o){\boldsymbol{\varphi}}({\mathbf{p}}^{opt},{\boldsymbol{\beta}}^o){\boldsymbol{\mathcal{I}}}^{-1}(\boldsymbol{\beta}^o)\right\}_{pp}  \beta_p^{*-2}}\right\rceil \, ,
$$
where $\left\lceil{.}\right\rceil$ is the ceiling function, the required $q_n$ is obtained by solving for $q_n$ and using estimated quantities, namely by 

\begin{equation} \label{q_size}
	\widetilde{q}_n = \left \lceil \frac{\left\{ \widetilde{\boldsymbol{\mathcal{I}}}^{-1}_{Q_{1.5}}(\widetilde{\boldsymbol{\beta}}_U) \widetilde{\boldsymbol{\varphi}}_{Q_{1.5}}(\widetilde{\mathbf{p}}^{opt}, \widetilde{\boldsymbol{\beta}}_U) \widetilde{\boldsymbol{\mathcal{I}}}^{-1}_{Q_{1.5}}(\widetilde{\boldsymbol{\beta}}_U) \right\}_{pp} (Z_{1-\alpha/2} + Z_{\gamma})^2}{\left({{\beta_p^*}}^2 - n^{-1} \widetilde{\boldsymbol{\mathcal{I}}}^{-1}_{Q_{1.5}}(\widetilde{\boldsymbol{\beta}}_U)_{pp}\right) (Z_{1-\alpha/2} + Z_{\gamma})^2} \right \rceil.
\end{equation}

This formula is convenient and practically valuable because, upon completing Steps 1 and 1.5, we can straightforwardly plot $\widetilde{q}_n$ as a function of $\gamma$. A negative value of $\widetilde{q}_n$ indicates that the required power cannot be achieved even with the entire sample $n$. Our simulation study demonstrates that in scenarios where the required power is attainable, typically only a small fraction of the censored observations is necessary.

We summarize the additional step for a single-covariate hypothesis testing by adding the following mid-step to the original two-step Algorithm \ref{alg:algorithm1}:

\textbf{Step 1.5*}: Sample $q_0$ observations from $\mathcal{C}$ using the optimal sampling probabilities computed in Step 1. Combine these observations with the observed events to create $\mathcal{Q}_{1.5}$. Compute $\widetilde{\boldsymbol{\mathcal{I}}}^{-1}_{Q_{1.5}}(\widetilde{\boldsymbol{\beta}}_U)$, $\widetilde{\boldsymbol{\varphi}}_{Q_{1.5}}(\mathbf{p}^{opt},\widetilde{\boldsymbol{\beta}}_U)$, and $\widetilde{q}_n$, with the desired values of $\alpha$ and $\gamma$. If $\widetilde{q}_n<0$, achieving the required power is not feasible even with the entire sample. Otherwise, set $q_n=\widetilde{q}_n$.

\section{Logistic Regression with Rare Events}\label{logistic_rare_events}

\subsection{Two-Step Algorithm}

While \cite{wang2018optimal} introduced a two-stage optimal subsampling algorithm for logistic regression, it was observed that their asymptotic variance may not perform well in cases of highly imbalanced data (i.e., when the rate of cases is below 15\%). Section 4 presents a method for choosing a subsample size for their optimal subsampling algorithm, and our simulation results indeed indicate its effectiveness primarily when the event is not rare.

In the realm of subsampling for imbalanced binary data, various methods have been explored and developed \citep{wang2020logistic, wang2021nonuniform}. However, their results were derived under the assumption that the intercept approaches zero as the sample size goes to infinity, and the other coefficients are fixed, leading to the probability of experiencing an event decreasing to zero as the sample size goes to infinity.
Our current work, akin to  \cite{wang2021nonuniform}, is based on subsampling only among non-cases observations while retaining all cases. Notably, our approach does not necessitate the undesired assumption that the event probability approaches zero as the sample size increases. Furthermore, our method yields a simpler formula for the asymptotic variance, enabling evaluation of the required subsample size in a practically efficient manner.

Let $D_i\in\{0,1\}$ be the response of individual $i$,  $i=1,\dots,n$. In order to include an intercept term, we extend the vector of covariates of each individual from $r$ to $r+1$  to include the value of 1 in its first element, and for simplicity of presentation we continue using the notation $\mathbf{X}_i$.  Let $\mathcal{N}=\{i \, ; \, D_i=0\}$ and $n_0=|\mathcal{N}|$.  The logistic regression model is of the form
\begin{equation*}
	\Pr(D_i=1|\mathbf{X}_i)=\mu_i(\boldsymbol{\beta})={e^{\mathbf{X}_i^T\boldsymbol{\beta}}}\left(1+e^{\mathbf{X}_i^T\boldsymbol{\beta}}\right)^{-1} 
	\quad \quad i=1,\ldots,n \, .
\end{equation*} 
and the maximum likelihood estimator (MLE) is given by
\begin{equation*}
	\widehat{\boldsymbol{\beta}}_{MLE}=\arg\max_{\boldsymbol{\beta}}\sum_{i=1}^n\big[D_i\log \mu_i(\boldsymbol{\beta})+(1-D_i)\log\{1-\mu_i(\boldsymbol{\beta})\}\big] \, .
\end{equation*}
As before, let $q_n$ be the size of the subsample from $\mathcal{N}$, $\pi_i$ the sampling probability of individual $i$, $\sum_{i\in\mathcal{N}} \pi_i=1$, and $\mathcal{Q}$ is the index set containing of all the observed cases (i.e., $D_i=1$) and the subsampled non-cases ($D_i=0$). Finally, for $i=1,\ldots,n$, set
\begin{equation*}
	w_i= \begin{cases}
		(\pi_i q_n)^{-1}, & \text{if } D_i=0\\
		1, & \text{if } D_i=1
	\end{cases}
\end{equation*}
as the sampling weights. Then, the estimator $\widetilde{\boldsymbol{\beta}}$ that is based on $\mathcal{Q}$ is obtained by maximizing the pseudo log-likelihood function
\begin{equation}\label{log_like_logistic}
	l^*(\boldsymbol{\beta})=\sum_{i\in\mathcal{Q}} w_i\big[D_i\log \mu_i(\boldsymbol{\beta})+(1-D_i)\log\{1-\mu_i(\boldsymbol{\beta})\}\big] \, .
\end{equation}

The following theorem provides the asymptotic distribution of a general subsampling-based estimator $\widetilde{\boldsymbol{\beta}}$, for any vector of  sampling probabilities, given standard assumptions (see the SM, Section S2). Based on the asymptotic distribution, the optimal sampling probabilities will be derived. Since the optimal sampling probabilities will be shown to involve the true unknown $\boldsymbol{\beta}^o$, we describe the two-step algorithm for logistic regression, which uses approximation of the optimal probabilities. 

\begin{theorem}\label{asymptotic_dist_logistic_rare}
	If Assumptions A.1-A.4 hold, then as $q_n, n\rightarrow\infty$,
	\begin{equation*}
		\sqrt{n}\mathbb{H}^R(\boldsymbol{\pi},\boldsymbol{\beta}^o)^{-1/2}(\widetilde{\boldsymbol{\beta}}-\boldsymbol{\beta}^o)\xrightarrow{D} N(0,\mathbf{I})
	\end{equation*} 
	where
	\begin{equation*}
		\mathbb{H}^R(\boldsymbol{\pi},\boldsymbol{\beta})=\mathbf{M}_X^{-1}(\boldsymbol{\beta})+\frac{n}{q_n}\mathbf{M}_X^{-1}(\boldsymbol{\beta})\mathbf{K}^R(\boldsymbol{\pi},\boldsymbol{\beta})\mathbf{M}_X^{-1}(\boldsymbol{\beta}) \, ,
	\end{equation*}
		\begin{equation*}
		\mathbf{M}_X(\boldsymbol{\beta})=n^{-1}\sum_{i=1}^n \mu_i(\boldsymbol{\beta})\{1-\mu_i(\boldsymbol{\beta})\}\mathbf{X}_i\mathbf{X}_i^T \, ,
	\end{equation*}	
	and
	\begin{equation*}
		\mathbf{K}^R(\boldsymbol{\pi},\boldsymbol{\beta})=\frac{1}{n^2}\bigg\{\sum_{i\in\mathcal{N}}\frac{\mu^2_i(\boldsymbol{\beta})\mathbf{X}_i\mathbf{X}_i^T}{\pi_i}-\sum_{i,j\in\mathcal{N}}  \mu_i(\boldsymbol{\beta})\mu_j(\boldsymbol{\beta})\mathbf{X}_i\mathbf{X}_j^T\bigg\} \, .
	\end{equation*}
\end{theorem}
We now turn to derive optimal sampling probabilities while considering  the A-optimal and L-optimal criteria, as before.

\begin{theorem}\label{proof_optimal_A_logistic_rare}
	The respective A-optimal and L-optimal sampling probability vectors, denoted by $\boldsymbol{\pi}^{R,A}$ and $\boldsymbol{\pi}^{R,L}$, which minimize the trace of $\mathbb{H}^R(\boldsymbol{\pi},\boldsymbol{\beta}^o)$ and $\mathbf{K}^R(\boldsymbol{\pi},\boldsymbol{\beta})$, respectively, are given by
	\begin{equation}\label{A_opt_rare_logistic}
		\pi_m^{R,A}=\frac{\mu_m(\boldsymbol{\beta}^o)\|\mathbf{M}_X^{-1}(\boldsymbol{\beta}^o)\mathbf{X}_m\|_2}{\sum_{j\in\mathcal{N}}\mu_j(\boldsymbol{\beta}^o)\|\mathbf{M}_X^{-1}(\boldsymbol{\beta}^o)\mathbf{X}_j\|_2} \quad \text{ for all } \quad m\in\mathcal{N} 
	\end{equation}
and
	\begin{equation}\label{L_opt_rare_logistic}
		\pi_m^{R,L}=\frac{\mu_m(\boldsymbol{\beta}^o)\|\mathbf{X}_m\|_2}{\sum_{j\in\mathcal{N}}\mu_j(\boldsymbol{\beta}^o)\|\mathbf{X}_j\|_2} \quad \text{ for all } \quad m\in\mathcal{N}.
	\end{equation}
\end{theorem}
Notably, the optimal probabilities expressed in \eqref{A_opt_rare_logistic} and \eqref{L_opt_rare_logistic} bear a resemblance to those derived for the subsampling approach in \cite{wang2018optimal} applied to a balanced design. The discrepancy between these optimal probabilities and their counterparts in \cite{wang2018optimal} stems from the fact that, here, the summation in the denominators is restricted to the set $\mathcal{N}$ rather than the entire sample. Since $\boldsymbol{\pi}^{R,A}$ and $\boldsymbol{\pi}^{R,L}$ involve the unknown $\boldsymbol{\beta}^o$, we suggest the following two-step algorithm, in the spirit of the previous section and \cite{wang2018optimal}:

\begin{algorithm2e}
	\caption{Logistic Regression with Rare Events - Two Step Optimal Subsampling}\label{alg:algorithm2}
	\-\hspace{5mm}
	{\bf Step 1:}
	Sample $q_0$ observations uniformly from $\mathcal{N}$ and combine them with all the observed events to create $\mathcal{Q}{pilot}$. Perform a weighted logistic regression on $\mathcal{Q}{pilot}$, based on Eq. (\ref{log_like_logistic}), and obtain $\widetilde{\boldsymbol{\beta}}_U$. Utilize this estimator to derive approximated optimal sampling probabilities by substituting $\boldsymbol{\beta}^o$ with $\widetilde{\boldsymbol{\beta}}_U$ in Eq. \eqref{A_opt_rare_logistic} or \eqref{L_opt_rare_logistic}.
	
	\-\hspace{5mm}
	{\bf Step 2:}
	Sample $q_n$ observations from $\mathcal{N}$ using the sampling probabilities computed in Step 1. Combine these observations with the observed events to create $\mathcal{Q}$ and conduct a weighted logistic regression on $\mathcal{Q}$, based on Eq. (\ref{log_like_logistic}),  to obtain the two-step estimator $\widetilde{\boldsymbol{\beta}}_{TS}$.
\end{algorithm2e}

Consistency and asymptotic normality of $\widetilde{\boldsymbol{\beta}}_{TS}$ can be shown by following the main steps of \cite{wang2018optimal} and \cite{keret2023analyzing}, as detailed in the SM, Section S2. As for the Cox regression, the value of $q_0$ is recommended to be $c_0  (n - n_0)$ with small values of $c_0$. Once $\widetilde{\boldsymbol{\beta}}_{TS}$ is calculated, inference can be executed by using the subsample counterparts of $\mathbb{H}^R(\boldsymbol{\pi},\widetilde{\boldsymbol{\beta}}_{TS})$ and $\mathbf{K}^R(\boldsymbol{\pi},\widetilde{\boldsymbol{\beta}}_{TS})$, namely
\begin{equation}\label{var_est_two_step_logistic_rare}
	\widetilde{\mathbb{H}}^R(\boldsymbol{\pi},\widetilde{\boldsymbol{\beta}}_{TS})=\widetilde{\mathbf{M}}_X^{-1}(\widetilde{\boldsymbol{\beta}}_{TS})+\frac{n}{q_n}\widetilde{\mathbf{M}}_X^{-1}(\widetilde{\boldsymbol{\beta}}_{TS})\widetilde{\mathbf{K}}^R(\boldsymbol{\pi},\widetilde{\boldsymbol{\beta}}_{TS})\widetilde{\mathbf{M}}_X^{-1}(\widetilde{\boldsymbol{\beta}}_{TS})
\end{equation}
and
\begin{equation*}
	\widetilde{\mathbf{K}}(\boldsymbol{\pi},\widetilde{\boldsymbol{\beta}}_{TS})=\frac{1}{n^2}\Bigg\{\frac{1}{q_n}\sum_{i\in\mathcal{Q}\setminus\mathcal{E}}\frac{\mu_i(\widetilde{\boldsymbol{\beta}}_{TS})^2\mathbf{X}_i(\mathbf{X}_i)^T}{\pi_i^2}-\frac{1}{q_n^2}\sum_{i\in\mathcal{Q}\setminus\mathcal{E}} \frac{\mu_i(\widetilde{\boldsymbol{\beta}}_{TS})\mathbf{X}_i}{\pi_i}\bigg(\sum_{i\in\mathcal{Q}\setminus\mathcal{E}}\frac{\mu_i(\widetilde{\boldsymbol{\beta}}_{TS})\mathbf{X}_i}{\pi_i}\bigg)^T\Bigg\} \, ,
\end{equation*}
where $\mathcal{E}=\{i \, : \, D_i = 1\}$ and 
\begin{equation*}
	\widetilde{\mathbf{M}}_X(\widetilde{\boldsymbol{\beta}}_{TS})=\frac{1}{n}\sum_{i\in\mathcal{Q}} w_i \mu_i(\widetilde{\boldsymbol{\beta}}_{TS})\left\{1-\mu_i(\widetilde{\boldsymbol{\beta}}_{TS})\right\}\mathbf{X}_i(\mathbf{X}_i)^T \, .
\end{equation*}

\subsection{Choosing Subsample Size by Relative Efficiency or Hypothesis Testing}\label{RE_logistic_rare}

In the spirit of Section \ref{RE_Cox}, we can estimate the RE of the two-step estimator relative to the estimator based on the entire dataset, in order to assess the required subsample size of Step 2.   To this end, define
\begin{equation}\label{H_midstep}
	\check{\mathbb{H}}^R(\boldsymbol{\pi}^{opt},\widetilde{\boldsymbol{\beta}}_U)=\check{\mathbf{M}}_X^{-1}(\check{\boldsymbol{\beta}}_U)+\frac{n}{q_0}\check{\mathbf{M}}_X^{-1}(\check{\boldsymbol{\beta}}_U)\check{\mathbf{K}}^R(\boldsymbol{\pi}^{opt},\check{\boldsymbol{\beta}}_U)\check{\mathbf{M}}_X^{-1}(\check{\boldsymbol{\beta}}_U)
\end{equation}
where
\begin{equation}\label{M_X_midstep}
	\check{\mathbf{M}}_X(\widetilde{\boldsymbol{\beta}}_U)=\frac{1}{n}\sum_{i\in\mathcal{Q}_{1.5}}\check{w}_i \mu_i(\widetilde{\boldsymbol{\beta}}_U)\left\{1-\mu_i(\widetilde{\boldsymbol{\beta}}_U)\right\}\mathbf{X}_i \mathbf{X}_i^T \, ,
\end{equation}
\begin{equation*}
	\check{w}_i= \begin{cases}
		(\pi_i^{opt} q_0)^{-1}, & \text{if } D_i=0\\
		1, & \text{if } D_i=1
	\end{cases}
\end{equation*}
and
\begin{equation*}
	\check{\mathbf{K}}^R(\boldsymbol{\pi},\widetilde{\boldsymbol{\beta}}_U)=\frac{1}{n^2}\left\{\frac{1}{q_0}\sum_{i\in\mathcal{Q}_{1.5}\setminus\mathcal{E}}\frac{\mu^2_i(\widetilde{\boldsymbol{\beta}}_U)\mathbf{X}_i(\mathbf{X}_i)^T}{\pi_i^2}-\frac{1}{q_0^2}\sum_{i\in\mathcal{Q}_{1.5}\setminus\mathcal{E}} \frac{\mu_i(\widetilde{\boldsymbol{\beta}}_U)\mathbf{X}_i}{\pi_i}\left(\sum_{i\in\mathcal{Q}_{1.5}\setminus\mathcal{E}}\frac{\mu_i(\widetilde{\boldsymbol{\beta}}_U)\mathbf{X}_i}{\pi_i}
	\right)^T\right\} \, .
\end{equation*}
Finally, we define the RE estimators as
\begin{equation}\label{est_RR_logistic}
	\widehat{RE}(q_n)=\frac{\|n^{-1}\check{\mathbf{M}}_X^{-1}(\widetilde{\boldsymbol{\beta}}_U)+q_n^{-1}\check{\mathbf{M}}_X^{-1}(\widetilde{\boldsymbol{\beta}}_U)\check{\mathbb{H}}^R(\widetilde{\boldsymbol{\pi}}^{opt},\widetilde{\boldsymbol{\beta}}_U)\check{\mathbf{M}}_X^{-1}(\widetilde{\boldsymbol{\beta}}_U)\|_F}{\|n^{-1}\check{\mathbf{M}}_X^{-1}(\widetilde{\boldsymbol{\beta}}_U)\|_F}
\end{equation}
and
\begin{equation}\label{est_RR_logistic_single} \widehat{RE}_p(q_n)=\frac{\big[n^{-1}\check{\mathbf{M}}_X^{-1}(\widetilde{\boldsymbol{\beta}}_U)+q_n^{-1}\check{\mathbf{M}}_X^{-1}(\widetilde{\boldsymbol{\beta}}_U)\check{\mathbb{H}}^R(\widetilde{\boldsymbol{\pi}}^{opt},\widetilde{\boldsymbol{\beta}}_U)\widetilde{\mathbf{M}}_X^{-1}(\widetilde{\boldsymbol{\beta}}_U)\big]_{pp}}{\big[n^{-1}\check{\mathbf{M}}_X^{-1}(\widetilde{\boldsymbol{\beta}}_U)\big]_{pp}} \, .
\end{equation}
The procedure can be easily incorporated within the two-step Algorithm \ref{alg:algorithm2} by the following additional step:

\textbf{Step 1.5}: Sample $q_0$ observations from $\mathcal{N}$ using the optimal sampling probabilities from Step 1. Combine the sampled observations with $\mathcal{E}$ to create $\mathcal{Q}_{1.5}$. Calculate $\check{\mathbf{M}}_X^{-1}(\widetilde{\boldsymbol{\beta}}_U)$ and $\check{\mathbb{H}}^R(\widetilde{\boldsymbol{\pi}}^{opt},\widetilde{\boldsymbol{\beta}}_U)$. Plot $\widehat{RE}(q_n)$ or $\widehat{RE}_p(q_n)$ as a function of $q_n$ and select the minimum $q_n$ that satisfies the required relative efficiency.

The minimal subsample size for testing $H_0 : \beta^o_p =0$ against a two-sided alternative, given $\beta^o_p = \beta_p^*$, a significance level $\alpha$ and a power $\gamma$, is given by 
\begin{equation} \label{q_size_logistic_rare}
	\widetilde{q}_n=\left \lceil{\frac{\left\{\check{\mathbf{M}}_X^{-1}(\widetilde{\boldsymbol{\beta}}_U)\check{\mathbf{K}}(\boldsymbol{\pi}^{opt},\widetilde{\boldsymbol{\beta}}_U)\check{\mathbf{M}}_X^{-1}(\widetilde{\boldsymbol{\beta}}_U)\right\}_{pp}(Z_{1-\alpha/2}+Z_{\gamma})^2}{{{\beta_p^*}}^2-n^{-1}\check{\mathbf{M}}_X^{-1}(\widetilde{\boldsymbol{\beta}}_U)_{pp}(Z_{1-\alpha/2} + Z_{\gamma})^2}}\right \rceil \, .
\end{equation}
A plot of $\widetilde{q}_n$ as a function of $\gamma$ can be easily generated.  The algorithm for a single covariate hypothesis testing is defined by adding the following mid-step to the  two-step Algorithm \ref{alg:algorithm2}:

\textbf{Step 1.5*}: Sample $q_0$ observations from $\mathcal{N}$ using the optimal sampling probabilities of Step 1. Combine these sampled observations with $\mathcal{E}$ to form $\mathcal{Q}_{1.5}$. Compute $\check{\mathbf{M}}_X^{-1}(\widetilde{\boldsymbol{\beta}}_U)$ and $\check{\mathbb{H}}^R(\widetilde{\boldsymbol{\pi}}^{opt},\widetilde{\boldsymbol{\beta}}_U)$. Plot $\widetilde{q}_n$ against $\gamma$. If $\widetilde{q}_n<0$, achieving the required power is unattainable even with the entire dataset $n$. Otherwise, set $q_n = \widetilde{q}_n$.

\section{Logistic Regression with Nearly Balanced Data}\label{non_rare_logistic}

\subsection{The Two-Step Optimal Subsampling Algorithm \citep{wang2018optimal}}\label{two_step_logistic_non_rare}
While \cite{wang2018optimal} presented an optimal two-step subsampling algorithm for logistic regression in the context of nearly balanced data and laid the theoretical asymptotic foundations for this approach, they did not offer a method for selecting the subsample size. This section aims to remedy this gap. To enhance clarity, we begin by providing a summary of their optimal two-step subsampling algorithm. 

In the rare event setting, sampling is performed exclusively from the majority class, whereas in the nearly balanced binary outcome scenario, sampling is conducted from the entire sample. Hence, now we redefine  $\mathcal{Q}$ as the index set of all the observations included in the subsample with sampling  weights  $w_i=(\pi_i q_n)^{-1}, i=1,\dots,n$. Then, the estimator $\widetilde{\boldsymbol{\beta}}$ that is based on $\mathcal{Q}$ is obtained by maximizing the pseudo log-likelihood function \eqref{log_like_logistic}.

Under some regularity assumptions  \citep{wang2018optimal}, they showed that given $\mathcal{F}_n= \{D_i, \mathbf{X}_i \,, \, 1,\ldots,n \}$, the asymptotic distribution of  $\widetilde{\boldsymbol{\beta}}$  is
\begin{equation}\label{logistic_var_cond}
	\sqrt{n}\mathbb{H}^B(\boldsymbol{\pi},\widehat{\boldsymbol{\beta}}_{MLE})^{-1/2}(\widetilde{\boldsymbol{\beta}}-\widehat{\boldsymbol{\beta}}_{MLE})\xrightarrow{D} N(0,\mathbf{I})
\end{equation}
as $n, q_n \rightarrow\infty$, where
\begin{equation*}
	\mathbb{H}^B(\boldsymbol{\pi},\boldsymbol{\beta})=\mathbf{M}_X^{-1}(\boldsymbol{\beta})\mathbf{K}^B(\boldsymbol{\pi},\boldsymbol{\beta})\mathbf{M}_X^{-1}(\boldsymbol{\beta})
\end{equation*}
and
\begin{equation*}
	\mathbf{K}^B(\boldsymbol{\pi},\boldsymbol{\beta})=\frac{1}{n^2}\sum_{i=1}^n w_i\left\{D_i-\mu_i(\boldsymbol{\beta})\right\}^2\mathbf{X}_i\mathbf{X}_i^T \, .
\end{equation*}
Then, the A-optimal and L-optimal subsampling probabilities are given by
\begin{equation} \label{logistic_A_optimal}
	\pi_i^{B,A}=\frac{|D_i-\mu_i(\widehat{\boldsymbol{\beta}}_{MLE})|\|\mathbf{M}_X^{-1}\mathbf{X}_i\|}{\sum_{j=1}^n|D_j-\mu_j(\widehat{\boldsymbol{\beta}}_{MLE})|\|\mathbf{M}_X^{-1}\mathbf{X}_j\|}
	\quad , \quad
	 i = 1,\dots,n \, 
\end{equation}
and
\begin{equation} \label{logistic_L_optimal}
	\pi_i^{B,L}=\frac{|D_i-\mu_i(\widehat{\boldsymbol{\beta}}_{MLE})|\|\mathbf{X}_i\|}{\sum_{j=1}^n|D_j-\mu_j(\widehat{\boldsymbol{\beta}}_{MLE})|\|\mathbf{X}_j\|}
	\quad , \quad
	 i = 1,\dots,n \, .
\end{equation}
Since we wish to avoid evaluating the full-data estimator $\widehat{\boldsymbol{\beta}}_{MLE}$, the following two-step algorithm is given by  \cite{wang2018optimal}:

\begin{algorithm2e}
	\caption{Logistic Regression with Nearly Balanced Data - Two Step Optimal Subsampling}\label{alg:algorithm3}
	\-\hspace{5mm}
	{\bf Step 1:}
	Sample $q_0$ observations using the following probabilities
	\begin{equation} \label{logistic_unif_prob}
		\pi_i^{prop}= \begin{cases}
			(2n_0)^{-1}  & \text{if } D_i=0\\
			(2n_1)^{-1}  & \text{if } D_i=1
		\end{cases} \quad \quad i = 1,\dots,n \, .
	\end{equation}
	where $n_1=n-n_0$. Conduct a weighted logistic regression with the subsample , based on Eq. (\ref{log_like_logistic}), and get $\widetilde{\boldsymbol{\beta}}_{prop}$. Derive the approximated optimal sampling probabilities by substituting $\widehat{\boldsymbol{\beta}}_{MLE}$ with $\widetilde{\boldsymbol{\beta}}_{prop}$ in  \eqref{logistic_A_optimal} or \eqref{logistic_L_optimal}.
	
	\-\hspace{5mm}
	{\bf Step 2:}
	Sample $q_n$ observations from the entire sample using the probabilities  of Step 1 and get $\mathcal{Q}$. Conduct a weighted logistic regression on $\mathcal{Q}$ , based on Eq. (\ref{log_like_logistic}), and obtain the two-step estimator $\widetilde{\boldsymbol{\beta}}_{TS}$.	
\end{algorithm2e}

Once $\widetilde{\boldsymbol{\beta}}_{TS}$ is calculated, inference can be carried out by using the variance estimator $\widetilde{\mathbb{H}}^B(\widetilde{\boldsymbol{\beta}}_{TS})=\widetilde{\mathbf{M}}_X(\widetilde{\boldsymbol{\beta}}_{TS})^{-1}\widetilde{\mathbf{K}}^B(\widetilde{\boldsymbol{\beta}}_{TS})\widetilde{\mathbf{M}}_X(\widetilde{\boldsymbol{\beta}}_{TS})^{-1}$, where
\begin{equation*}
	\widetilde{\mathbf{K}}^B(\widetilde{\boldsymbol{\beta}})=\frac{1}{n^2}\sum_{i\in\mathcal{Q}} w_i^2\left\{D_i-\mu_i(\widetilde{\boldsymbol{\beta}})\right\}^2\mathbf{X}_i \mathbf{X}_i^T \, .
\end{equation*}

Certainly, we can utilize the concepts discussed earlier  to determine the desired values of $q_n$. However, the asymptotic properties outlined in \cite{wang2018optimal} are confined to the conditional space,  conditioning on the entire observed data, $\mathcal{F}_n$. In contrast, our approach for the optimal $q_n$ necessitates the consideration of the asymptotic distribution under the unconditional space. The subsequent theorem presents this result, and the proof is available in the SM, Section S2.
\begin{theorem}\label{proof_asymptotic_logistic_non_rare}
	Given Assumptions A.1-A.3 (see SM, Section S2) and as $q_n, n \rightarrow\infty$,
	\begin{equation*}\label{logistic_var}
		\sqrt{n}\mathbb{H}^B(\boldsymbol{\pi},\boldsymbol{\beta}^o)^{-1/2}(\widetilde{\boldsymbol{\beta}}_{TS}-\boldsymbol{\beta}^o)\xrightarrow{D} N(0,\mathbf{I}) \, .
	\end{equation*}
\end{theorem}

\subsection{Choosing Subsample Size by Relative Efficiency or Hypothesis Testing}\label{sec_RE_logistic_no_rare}

An estimator of the RE of the two-step estimator  relative to the estimator based on the entire datatset is given by
\begin{equation*}\label{RR_logistic_no_rare}
	RE(q_n)=\frac{\|\mathbb{H}^B(\widetilde{\boldsymbol{\beta}}_{TS})\|_F}{\|n^{-1}\mathbf{M}_X(\widehat{\boldsymbol{\beta}}_{MLE})\|_F} \, ,
\end{equation*}
and the respective estimator that focuses  on   the $p^{th}$ covariate is given by
\begin{equation*}\label{RR_single_logistic_no_rare}
	RE_p(q_n)=\frac{\left[\mathbb{H}^B(\widetilde{\boldsymbol{\beta}}_{TS})\right]_{pp}}{\left[n^{-1}\mathbf{M}_X^{-1}(\widehat{\boldsymbol{\beta}}_{MLE})\right]_{pp}} \, .
\end{equation*}
Once again, we substitute $\widetilde{\boldsymbol{\beta}}_{TS}$ and $\widehat{\boldsymbol{\beta}}_{MLE}$ with the consistent estimator $\widetilde{\boldsymbol{\beta}}_{prop}$ from Step 1. To ensure numerical stability in approximating $\mathbb{H}^B(\widetilde{\boldsymbol{\beta}}_{prop})$ and $\mathbf{M}_X^{-1}(\widetilde{\boldsymbol{\beta}}_{prop})$, we recommend utilizing the estimated optimal probabilities of  Step 1 to sample an additional set of size $q_0$, denoted as $\mathcal{Q}_{1.5}$. Let
\begin{equation*}\label{V_logistic_non_rare_1.5}
	\check{\mathbb{H}}^B(\boldsymbol{\pi},\widetilde{\boldsymbol{\beta}})=\check{\mathbf{M}}_X^{-1}(\widetilde{\boldsymbol{\beta}})\check{\mathbf{K}}^B(\boldsymbol{\pi},\widetilde{\boldsymbol{\beta}})\check{\mathbf{M}}_X^{-1}(\widetilde{\boldsymbol{\beta}})
\end{equation*}
where
\begin{equation*}
	\check{\mathbf{K}}^B(\widetilde{\boldsymbol{\beta}})=\frac{1}{n^2}\sum_{i\in\mathcal{Q}_{1.5}} \check{w}_i^2\left\{D_i-\mu_i(\widetilde{\boldsymbol{\beta}})\right\}^2\mathbf{X}_i \mathbf{X}_i^T \, ,
\end{equation*}
and  $\check{w}_i=(q_0 \pi_i)^{-1}$, $i=1,\dots,n$. Finally,
\begin{equation}\label{RE_est_logistic_no_rare}
	\widehat{RE}(q_n)=\frac{\|q_0 q_n^{-1} \check{\mathbb{H}}^B(\widetilde{\boldsymbol{\beta}}_{prop})\|}{\|n^{-1}\check{\mathbf{M}}_X^{-1}(\widetilde{\boldsymbol{\beta}}_{prop})\|} \, .
\end{equation}
Unlike the RE estimator in the rare event setting, Eq. \eqref{RE_est_logistic_no_rare} approaches zero as $q_n \rightarrow \infty$ while keeping $q_0$ and $n$ fixed. Consequently, only practical sizes for $q_n$ should be taken into account. In other words, values of $q_n$ that are close to $n$ should not be considered in the plot of $\widehat{RE}(q_n)$ as a function of $q_n$. This procedure can be seamlessly integrated into the two-step Algorithm \ref{alg:algorithm3} with minimal additional computation time, as outlined below:

\textbf{Step 1.5}: Draw a sample of $q_0$ observations from the entire dataset using the optimal sampling probabilities obtained in Step 1 to create $\mathcal{Q}_{1.5}$. Compute $\check{\mathbb{H}}^B(\widetilde{\boldsymbol{\beta}}_{prop})$ and $\check{\mathbf{M}}_X^{-1}(\widetilde{\boldsymbol{\beta}}_{prop})$. Generate a plot of $\widehat{RE}(q_n)$ against $q_n$ and select the smallest $q_n$ that meets the desired RE.

Similarly, the minimal subsample size for testing $H_0 : \beta^o_p =0$ against a two-sided alternative, given $\beta^o_p = \beta_p^*$, a significance level $\alpha$ and power $\gamma$, is given by
\begin{equation} \label{logistic_q_size}
	\widetilde{q}_n=\left \lceil{\frac{q_0(Z_{1-\alpha/2}+Z_{\gamma})^2 \big[\check{\mathbb{H}}^B(\widetilde{\boldsymbol{\beta}}_{prop})\big]_{pp}}{{\beta_p^*}^2}}\right \rceil  \, .
\end{equation}
A plot of $\widetilde{q_n}$ as a function of $\gamma$ can be easily generated. The values derived from Eq. \eqref{logistic_q_size} might exceed the sample size $n$. Nevertheless, exceeding the sample size may not yield any additional information beyond what is already captured by the full-data MLE, $\widehat{\boldsymbol{\beta}}_{MLE}$. Despite this, a value surpassing $n$ remains informative, indicating that the desired statistical power cannot be attained. In conclusion, the algorithm for a single covariate hypothesis testing is provided by adding the following mid-step to the two-step Algorithm \ref{alg:algorithm3}:

\textbf{Step 1.5*}: Draw a sample of $q_0$ observations from the entire dataset using the optimal sampling probabilities obtained in Step 1 to create $\mathcal{Q}_{1.5}$. Compute $\check{\mathbb{H}}^B(\widetilde{\boldsymbol{\beta}}_{prop})$ and $\check{\mathbf{M}}_X^{-1}(\widetilde{\boldsymbol{\beta}}_{prop})$. Generate a plot of $\widehat{RE}(q_n)$ against $\gamma$ and select the smallest $q_n$ that meets the desired  power.

\section{Simulation Study}\label{sims}

\subsection{Cox Regression}

\subsubsection{Data Generation}\label{sim_cox_RMSE}
The sampling designs are similar to that of  
\cite{keret2023analyzing}. For each of the settings described below, 500 samples were drawn, $n=15,000$ with $\boldsymbol{\beta}^o=(0.3, -0.5, 0.1, -0.1, 0.1, -0.3)^T$. Censoring times were generated from an exponential distribution with rate 0.2, independently of failure times. The instantaneous baseline hazard rate was set to be $\lambda_0(t)=0.001I(t<6)+c_{\lambda_0}I(t\geq_6)$.
The distributions of the covariates and the parameter \(c_{\lambda_0}\) of each setting, I, II, III, were as follows:
\begin{enumerate}
	\item \textbf{Setting I}: $X_j \sim Unif(0,4)$, $j=1,\ldots,6$ and  $c_{\lambda_0}=0.075$. This is a setting of equal variances and no correlation between the covariates.
	\item \textbf{Setting II}: $X_j \sim Unif(0,\theta_j)$, $(\theta_1,\theta_2,\theta_3,\theta_4,\theta_5,\theta_6)=(1,6,2,2,1,6)$ and $c_{\lambda_0}=0.15$. This is a setting of unequal variances and no correlation between covariates.
	\item \textbf{Setting III}: $X_1, X_2$ and $X_3$ are independently sampled from $Unif(0,4)$, $X_4 = 0.5X_1 + 0.5X_2 + \varepsilon_1$,  $X_5 = X_1 + \varepsilon_2$, $X_6 = X_1 + \varepsilon_3$ and $c_{\lambda_0}=0.05$, where $\varepsilon_1 \sim N(0,0.1)$, $\varepsilon_2 \sim N(0,1)$, $\varepsilon_3 \sim N(1,1.5)$, and the $\varepsilon$'s are independent. The strongest correlation between two covariates is about 0.75.
\end{enumerate}

\subsubsection{Results}
Eq. \eqref{RR_est} and \eqref{q_size} become practically valuable only when the approximations made in Step 1.5 and Step 1.5* of Sections \ref{RE_Cox} and \ref{hypothesis_testing_cox}, respectively, closely align with their true values. In Fig. S1 of the SM, we compare the Frobenius norm of three covariance matrices: (i) The covariance matrix of the two-step estimator, $\widetilde{\boldsymbol{\beta}}_{TS}$. (ii) The approximated covariance matrix utilized in Step 1.5. (iii) The empirical covariance matrix of $\widetilde{\boldsymbol{\beta}}_{TS}$. The results are obtained with $q_n=5n_e$ and $q_0=c_0 n_e$, where $c_0$ ranges from 1 to 5. Clearly, the Frobenius norm of Step 1.5  is remarkably close to the covariance matrix of the two-step estimator, and both are in close agreement with the empirical variance, even for small values of $c_0$ such as $c_0=1$.

\begin{figure}[h]
	\centering
		\includegraphics[width=0.95\textwidth]{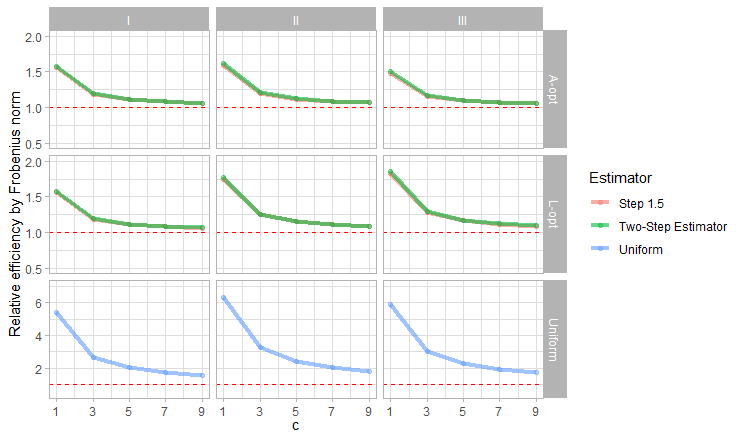}
	\caption{\label{fig:relative_frobenius}
		Simulation results of Cox regression model: RE with and without the approximation of Step 1.5 and the RE of a one-step estimator based on a uniform subsampling from the censored data. The curves of the true and approximated relative efficiency coincide. The dashed red line is $y=1$.}
\end{figure}

A comparison between the RE as defined by Eq. \eqref{RR} and its approximation in Eq. \eqref{RR_est} is summarized in Fig. \ref{fig:relative_frobenius}, where $q_0=2n_e$, $q_n=c n_e$, and $c=1,\ldots,9$. The results indicate that the approximated RE of Step 1.5, (i.e., Eq. \eqref{RR_est}) closely mirrors Eq. \eqref{RR}. The presence of an `elbow' shape around $c=3$ with RE fairly close to $1$ suggests that $q_n=3 n_e$ is sufficiently large under these specific settings. Clearly, the two optimal sampling strategies substantially outperform uniform sampling in terms of RE.

To assess the effectiveness of $\widetilde{q}_n$ as defined in Eq. \eqref{q_size}, we conducted a comparison of the empirical and nominal power of the test for $H_0: \beta_5=0$ against a two-sided alternative  using $\widetilde{q}_n$ based on the proposed three-step estimation algorithm, comprising Steps 1, 1.5*, and 2. The tests were performed with $\alpha=0.05$. Here, we increased the sample size to $n=150,000$, and for Setting I $c_{\lambda_0}=0.005$, while for Settings II and III $c_{\lambda_0}=0.05$. Consequently, the respective event rates were $0.65\%$, $1.3\%$ and $3\%$.

The results are outlined in Table \ref{power_sims_cox} with $q_0=2 n_e$. Clearly, $\widetilde{q}_n$ achieves the intended nominal power. When considering the mean and standard deviation (SD) of $\widetilde{q}_n$, we observe that both optimality criteria exhibit similar performance under Settings I and II, while A surpasses L in Setting III. These results further validate our assertion that in extensive datasets with rare events, only a small fraction of the censored data is practically necessary. For instance, in Setting III, it is demonstrate that subsampling approximately $6000$ censored observations from about 145,550 censored observations is sufficient to achieve a power of $0.95$.

\subsection{Logistic Regression with Rare Events}
\subsubsection{Data Generation}
The sampling designs are similar to that of \cite{wang2018optimal} with some modification to represent settings of rare events.  500 samples were drawn for each setting, each dataset is of size $n=100,000$. The following covariates' distributions were considered:
\begin{enumerate}
	\item \textbf{mzNormal}. $\mathbf{X}$ follows a multivariate normal distribution $N(\mathbf{0},\mathbf{\Sigma})$, where ${\Sigma}_{ij}=0.5^{I(i\neq j)}$. 
	\item \textbf{mixNormal}. $\mathbf{X}$ is a mixture of two multivariate normal distribution, $\mathbf{X} \sim 0.5N(\mathbf{1},\mathbf{\Sigma}) + 0.5N(\mathbf{-1},\mathbf{\Sigma})$ so the distribution of $\mathbf{X}$ is bimodal.
	\item \textbf{T3}. $\mathbf{X}$ follows a multivariate $t$ distribution with 3 degrees of freedom, $\mathbf{X} \sim t_3(\mathbf{0}, \mathbf{\Sigma})/10$. Hence, the distribution of $\mathbf{X}$ has heavy tails.
	\item \textbf{EXP}. Components of $\mathbf{X}$ are independent and each has an exponential distribution with a rate parameter of 2. The distribution of $\mathbf{X}$ is skewed and has a heavier
	tail on the right.
\end{enumerate}

We set $q_0=1,000$ and explored various values for $q_n$, ranging from $1,000$ to $10,000$ in increments of $1,000$. We set $\beta_i=0.5$, $i=1,\ldots,6$, and employed distinct values for the intercept $\beta_0$ to regulate the event rate. Specifically,  $\beta_0=-6$ for mzNormal (yielding an event rate of 2\%), $\beta_0=-5$ for mixNormal (event rate of 2.1\%), $\beta_0=-5$ for T3 (event rate of 1.5\%), and $\beta_0=-11$ for EXP (event rate of 1.3\%).

\subsubsection{Results}
\begin{figure}
	\centering
	\subfloat[RMSE with respect to $\boldsymbol{\beta}^o$]{
	\includegraphics[width=1.0\textwidth]{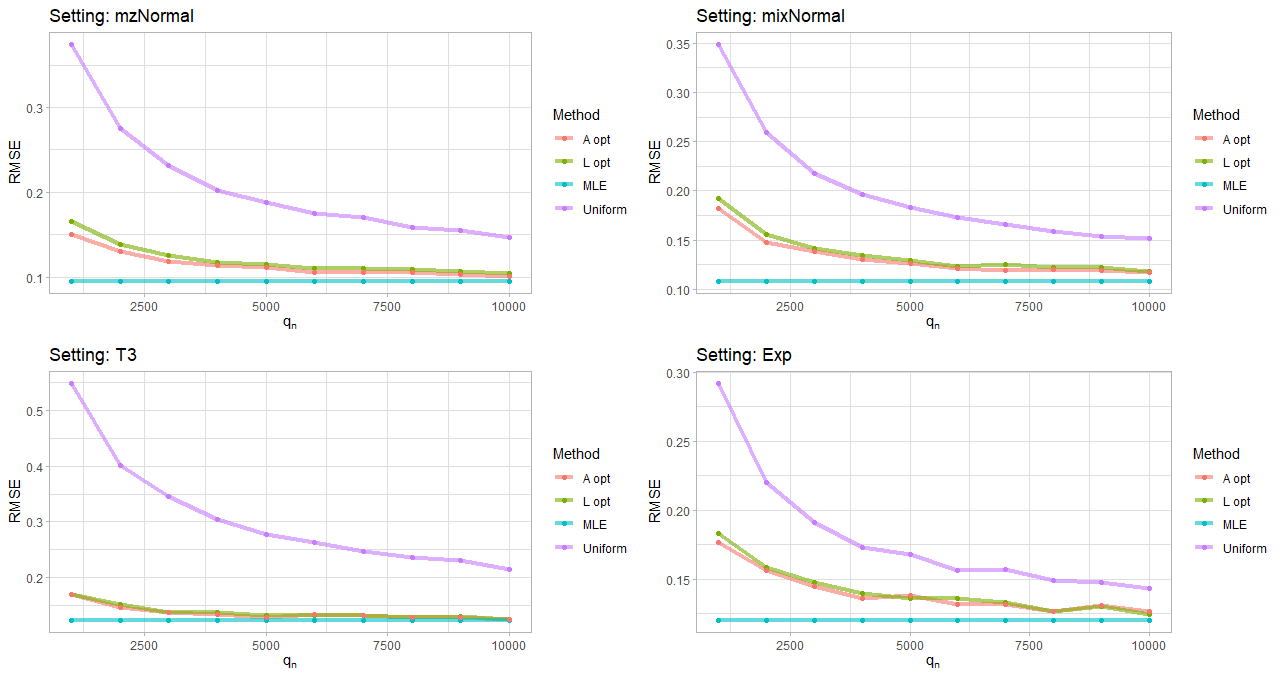}
	}
	
	\subfloat[RMSE with respect to $\widehat{\boldsymbol{\beta}}_{MLE}$]{
	\includegraphics[width=1.0\textwidth]{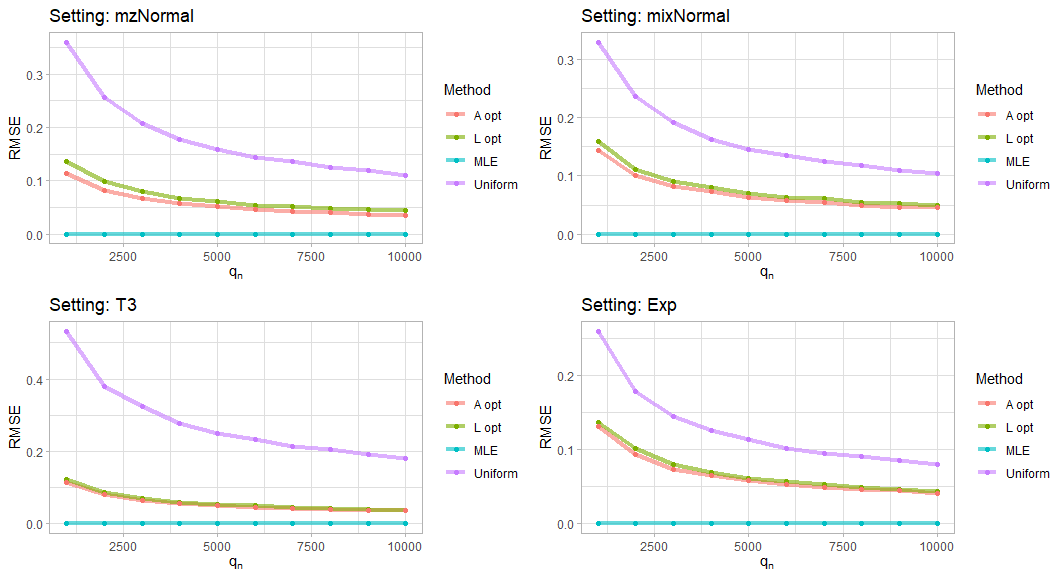}
	}
	\caption{\label{fig:rare_events_rmsel}Simulation results of logistic regression with rare events: RMSE of the two-step estimators with sampling methods A-optimal, L-optimal, full-data MLE, and a one-step estimator with uniform sampling from the non-cases data.}
\end{figure}

The comparison among different estimators involved assessing the empirical root mean squared errors (RMSEs) with respect to $\boldsymbol{\beta}^o$ and $\widehat{\boldsymbol{\beta}}_{PL}$. Namely, $B^{-1}\sum_{j=1}^B\sqrt{\sum_{i=1}^6(\widehat{\beta}_i^{(j)}-\beta_i^o)^2}$ and $B^{-1}\sum_{j=1}^B\sqrt{\sum_{i=1}^6(\widehat{\beta}_i^{(j)}-\widehat{\beta}_{PL,i}^{(j)})^2}$, where $\widehat{\boldsymbol{\beta}}$ represents the relevant estimator, the superscript $(j)$ denotes the $j$'th sample, and $B=500$ signifies the number of repetitions. Fig. \ref{fig:rare_events_rmsel} shows the RMSEs of the two-step estimators of Algorithm \ref{alg:algorithm2} with ${\bf p}^A$ and ${\bf p}^L$, the full-data MLE, and a one-step estimator with uniform subsampling from the non-cases data. Clearly, the optimal subsampling methods outperform uniform subsampling in terms of RMSE, with A-optimal yielding slightly superior results compared to L-optimal, as anticipated. Table \ref{running_logistic_rare} presents a comparison of the running times. Evidently, the optimal subsampling methods are substantially faster than $\widehat{\boldsymbol{\beta}}_{PL}$ while still maintaining low RMSE.

\begin{figure}
	\centering
	\subfloat[Logistic regression - rare event]{
	\includegraphics[width=1.0\textwidth]{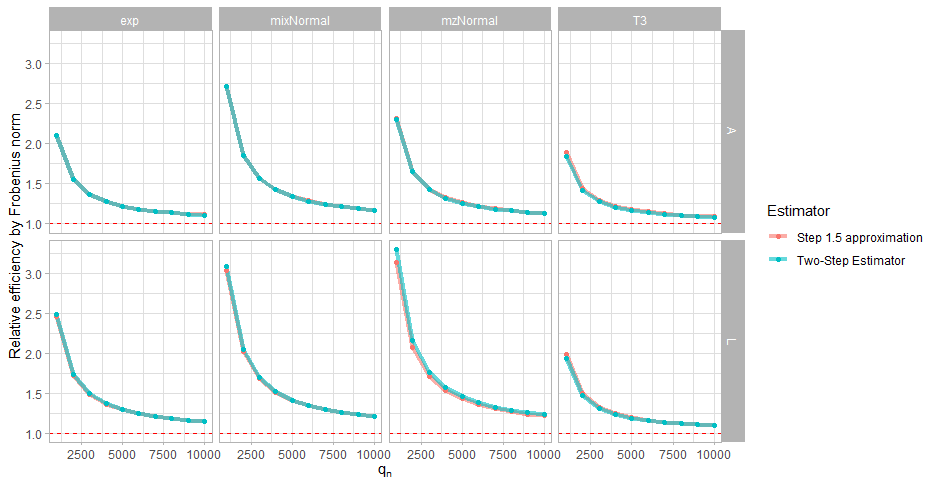}
}

	\subfloat[Logistic regression - nearly balanced]{
	\includegraphics[width=1.0\textwidth]{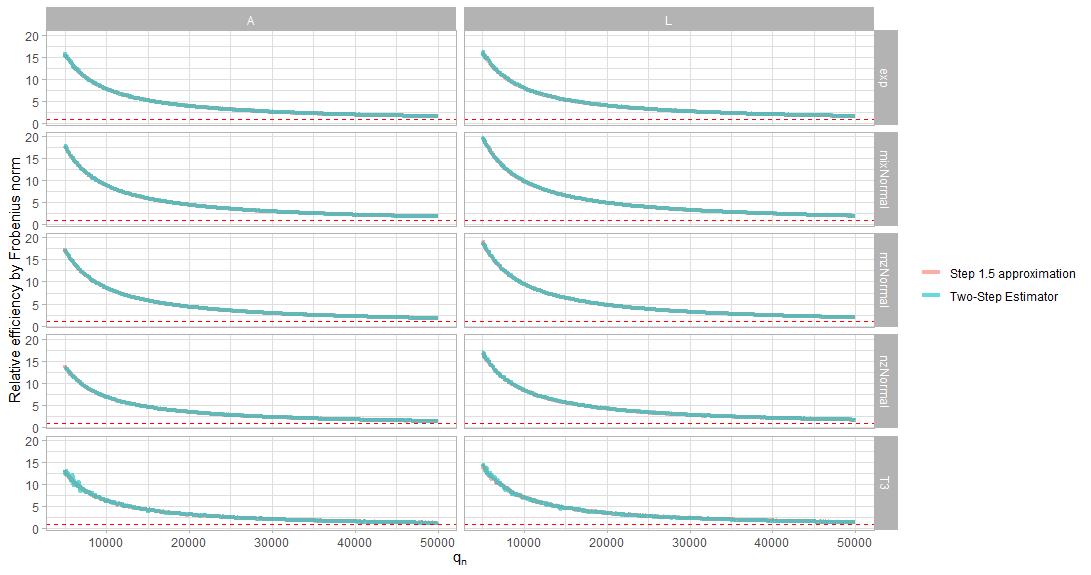}
}	
\caption{\label{fig:RE_results_logistic}
		Simulation results of logistic regression: RE with and without the approximation of Step 1.5. The curves of the true and approximated RE tend to coincide. The dashed red line is $y=1$.}
\end{figure}

Fig. S2 of the SM demonstrate the validity of the variance estimator \eqref{var_est_two_step_logistic_rare}, and the effectiveness of optimal subsampling over uniform subsampling. In Figure \ref{fig:RE_results_logistic}(a) it is demonstrated that Eq. \eqref{est_RR_logistic} provides a good approximation of the RE based on the actual two-step estimator, thereby endorsing the validity of the proposed three-step estimator that includes steps 1, 1.5 and 2.

To evaluate the utility of $\widetilde{q}_n$ derived from Eq. \eqref{q_size_logistic_rare}, a comparison was made between the empirical and nominal power of testing $H_0:\beta_5=0$ against a two-sided alternative and $\alpha=0.05$, using $\widetilde{q}_n$ and the three-step estimation algorithm of Section 3.1 with steps 1, 1.5* and 2. Due to impractical subsample sizes for some higher values of $\gamma$, meaning the required power could not be attained even with the entire sample, the coefficient vector $\boldsymbol{\beta}^o$ was modified:
\begin{enumerate}
	\item \textbf{mzNormal}. $\beta_0=-3.5$ and $\beta_j=0.1$, $i=1,\ldots,6$, with an event rate of 3.2\%.
	\item \textbf{mixNormal}. $\beta_0=-4.5$ and $\beta_j=0.2$, $i=1,\ldots,6$, with an event rate of 1.3\%.
	\item \textbf{T3}. $\beta_0=-3$ and $\beta_j=0.15$, $j=1,\ldots,6$, with an event rate of 5\%.
	\item \textbf{EXP}. $\beta_0=-4$ and $\beta_j=0.15$, $j=1,\ldots,6$, with an event rate of 2.8\%.
\end{enumerate}
The results are summarized in Table \ref{power_sims_rare} employing $q_0=1,000$ with 5,000 repetitions for each $\gamma$ value. Our conclusion is that $\widetilde{q}_n$ yields power close to the nominal level across all scenarios. Regarding the mean and standard deviation of $\widetilde{q}_n$, the A-optimal approach consistently outperforms L-optimal.

\subsection{Logistic Regression with Nearly Balanced Data}

The configurations examined correspond to those outlined in \cite{wang2018optimal} (Section 5.1): mzNormal, nzNormal, mixNormal, T3, and EXP. The setting mzNormal is not balanced, and with event rate of 0.73. For each specified scenario, we generated 500 samples, each consisting of 100,000 observations and $q_0$ was set to 5,000. The results, succinctly illustrated in Fig. \ref{fig:RE_results_logistic}(b), demonstrate a strong agreement between the proposed RE estimator \eqref{RE_est_logistic_no_rare} and the RE based on the actual two-step Algorithm \ref{alg:algorithm3}.

 Table \ref{resluts_log_no_rare} provides a summary of the comparison between empirical and nominal power of testing $H_0:\beta_6=0$ against a two-sided alternative and $\alpha=0.05$, utilizing $\widetilde{q}_n$ and the proposed three-step estimation algorithm outlined in Section 4.2. These results are derived from 5,000 repetitions for each configuration, employing a smaller subsample size for steps 1 and 1.5*, with $q_0$ set to 1,000. Evidently, $\widetilde{q}_n$ provides the desired nominal power, which supports the use of Eq. \eqref{logistic_q_size}. In terms of the mean and standard deviation of $\widetilde{q}_n$,  A-optimal  outperforms L-optimal.

\begin{figure}
	\centering
	\includegraphics[width=1.0\textwidth]{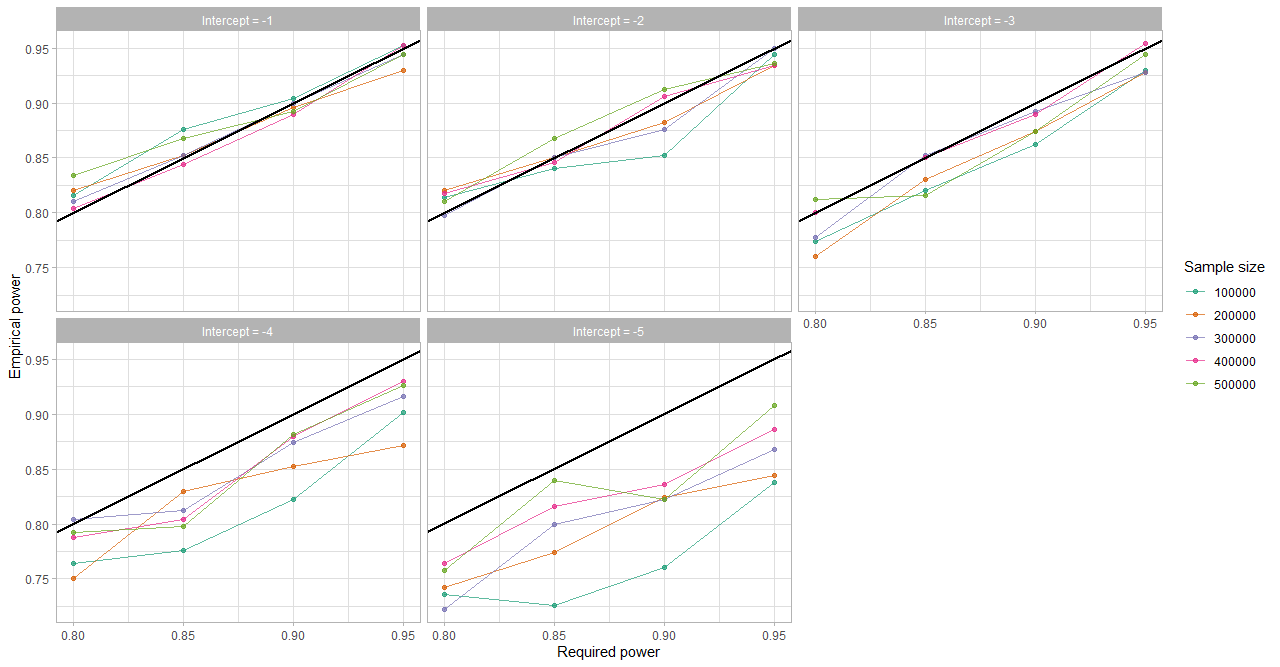}
	\caption{\label{fig:logistic_rare_events}
		Simulation results of logistic regression setting mzNormal with imbalanced data: nominal versus empirical power based on $\widetilde{q}_n$ of Eq. $\eqref{logistic_q_size}$  as a function of $\beta_0^o$ and $n$. The black line is $x=y$.}
\end{figure}

The optimal approach of \cite{wang2018optimal}  involves subsampling from both cases and controls, exhibiting strong performance when dealing with balanced data. The sensitivity of the optimal subsample size, as defined by Eq. \eqref{logistic_q_size}, to imbalanced data is illustrated in Fig. \ref{fig:logistic_rare_events}. The setting mzNormal is explored with varying sample sizes $n=a \times 100,000$, $a=1,\ldots,5$, and diverse values of $\beta_0^o=-5,-4,\ldots,-1$, corresponding to event rates of 0.8\%, 2\%, 5\%, 13\%, and 28\%, respectively. Evidently, Eq. \eqref{logistic_q_size} fails to deliver the required power as the event rate decreases, regardless of the sample size. These findings underscore the imperative need for a distinct consideration of the imbalanced setting, as addressed in this work.

\section{Survival Analysis of UKBiobank  Colorectal Cancer}\label{data_survival}
We conducted an analysis complementing the one presented in \cite{keret2023analyzing} and studied the required subsample size based on RE. The event time is defined as the age at colorectal cancer (CRC) diagnosis, while the censoring time is specified as the age at death before CRC diagnosis or the current age without CRC. The analysis encompasses established environmental CRC risk factors, including body mass index (BMI), smoking status (no/yes), family history of CRC (no/yes), physical activity (no/yes), sex (female/male), alcohol consumption (non or occasional/light frequent drinker/very frequent drinker), education (lower than high school/high school/higher vocational education/college or university graduate/prefer not to answer), NSAIDs drug use (none/Aspirin or Ibuprofen), and post-menopausal hormones (no/yes). Additionally, 139 single-nucleotide polymorphisms (SNPs) associated with CRC through GWAS \citep{jeon2018determining} were included along with six principal components to account for population substructure. The SNPs were standardized to have a mean of zero and unit variance.

Building on the analysis in \cite{keret2023analyzing}, a time-dependent effect $\beta(t)$ is essential for sex, due to violation of the proportional hazard assumption. In total, 180 regression coefficients were considered for the model, with 5,342 observed events  and 479,343 censored observations. However, the introduction of time-dependent coefficients results in the partitioning of each observation into several distinct time-fixed ``dummy-observations", each having an ``entrance" and ``exit" time \citep{Therneau2017-hs}. This creates non-overlapping intervals that reconstruct the original time interval and inflating the dataset to approximately 350 million rows and $n_e=5,342$. Subsampling is then performed from the censored dummy-observations using the reservoir-sampling approach.

\begin{figure}
	\centering
		\subfloat[RE based on Step 1.5]{
		\includegraphics[width=0.6\textwidth]{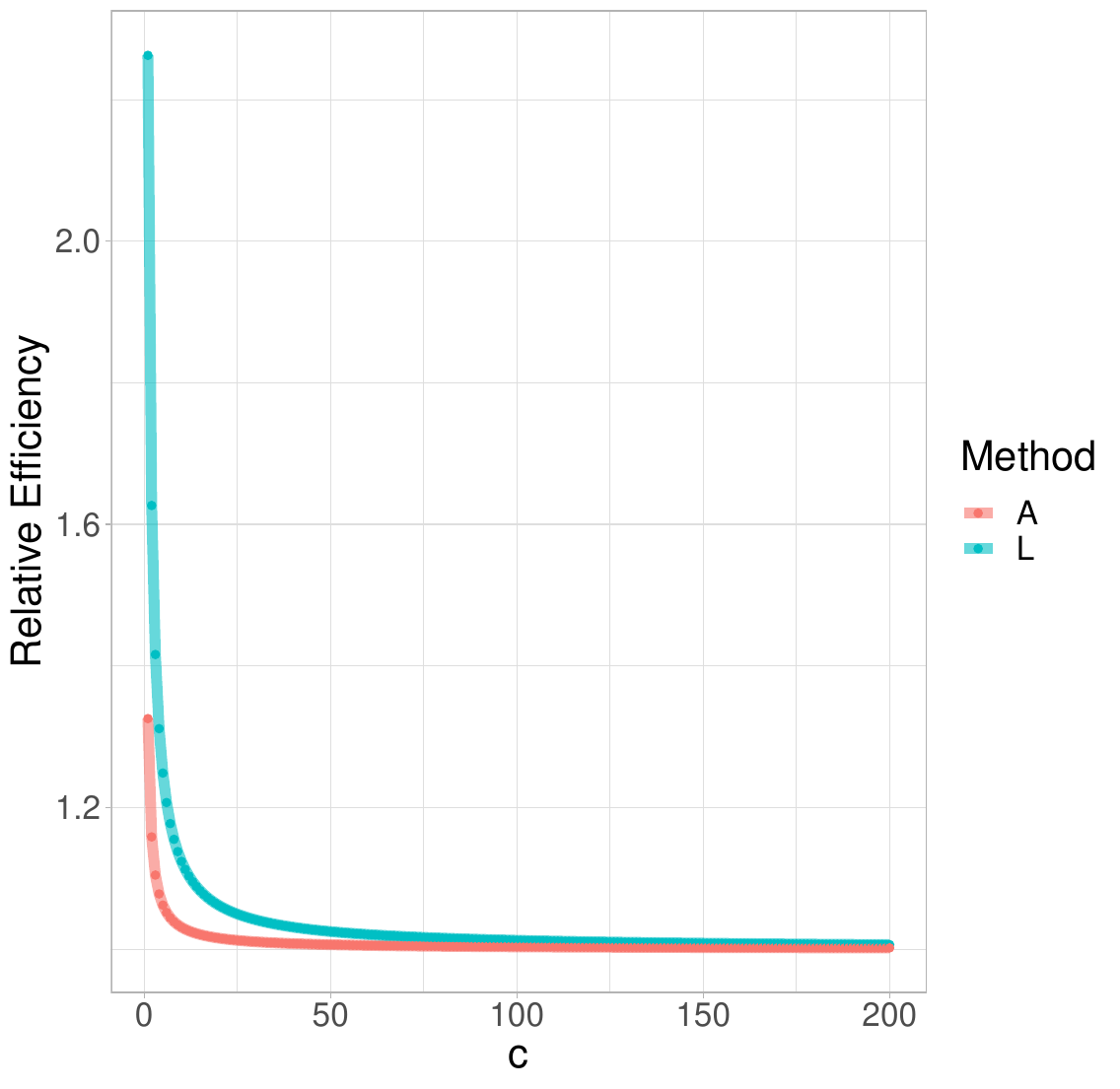}
	}	
	
	\subfloat[Comparison of Estimates]{
		\includegraphics[width=0.5\textwidth]{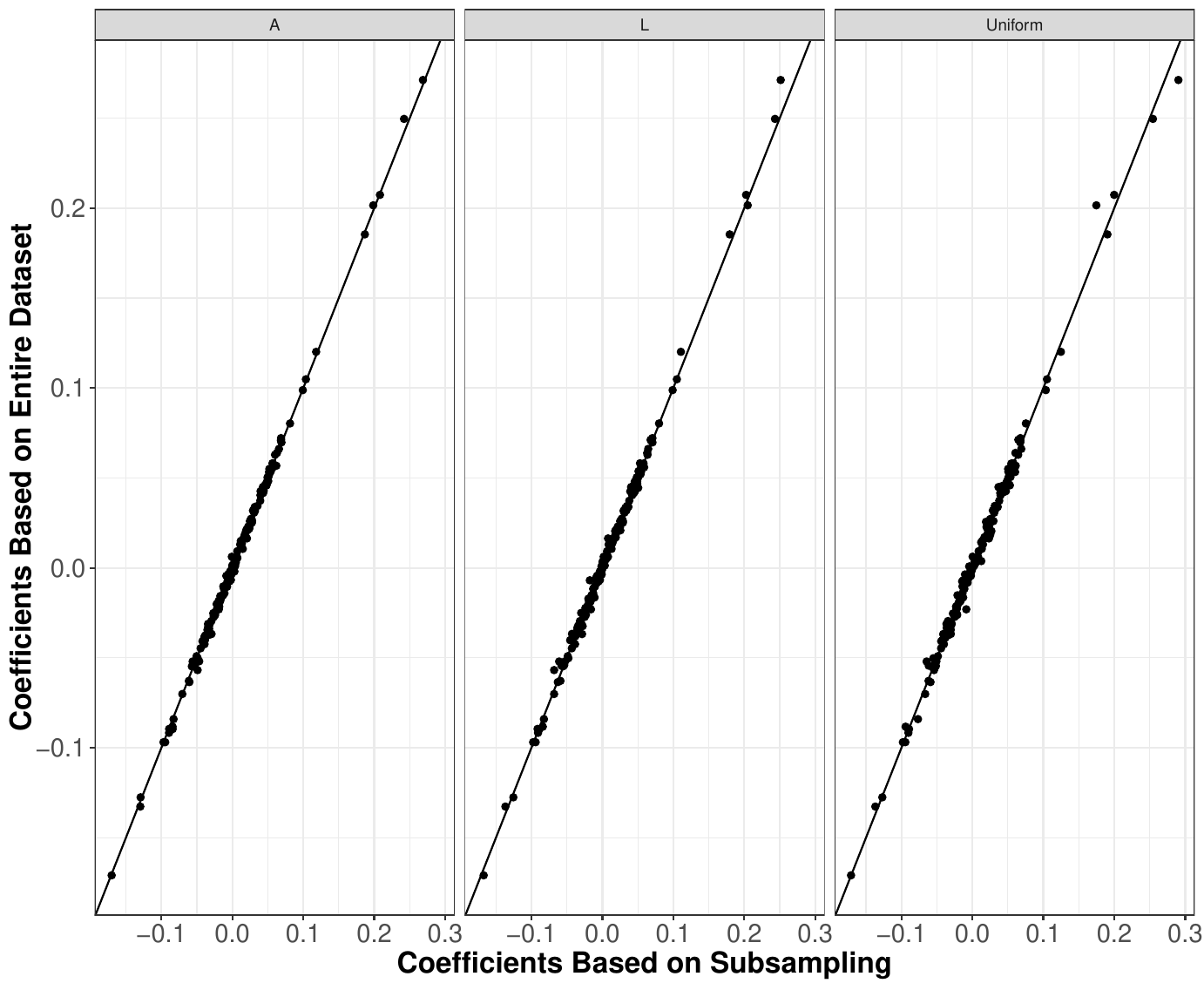}
	}
	\subfloat[Comparison of Standard Errors]{
		\includegraphics[width=0.5\textwidth]{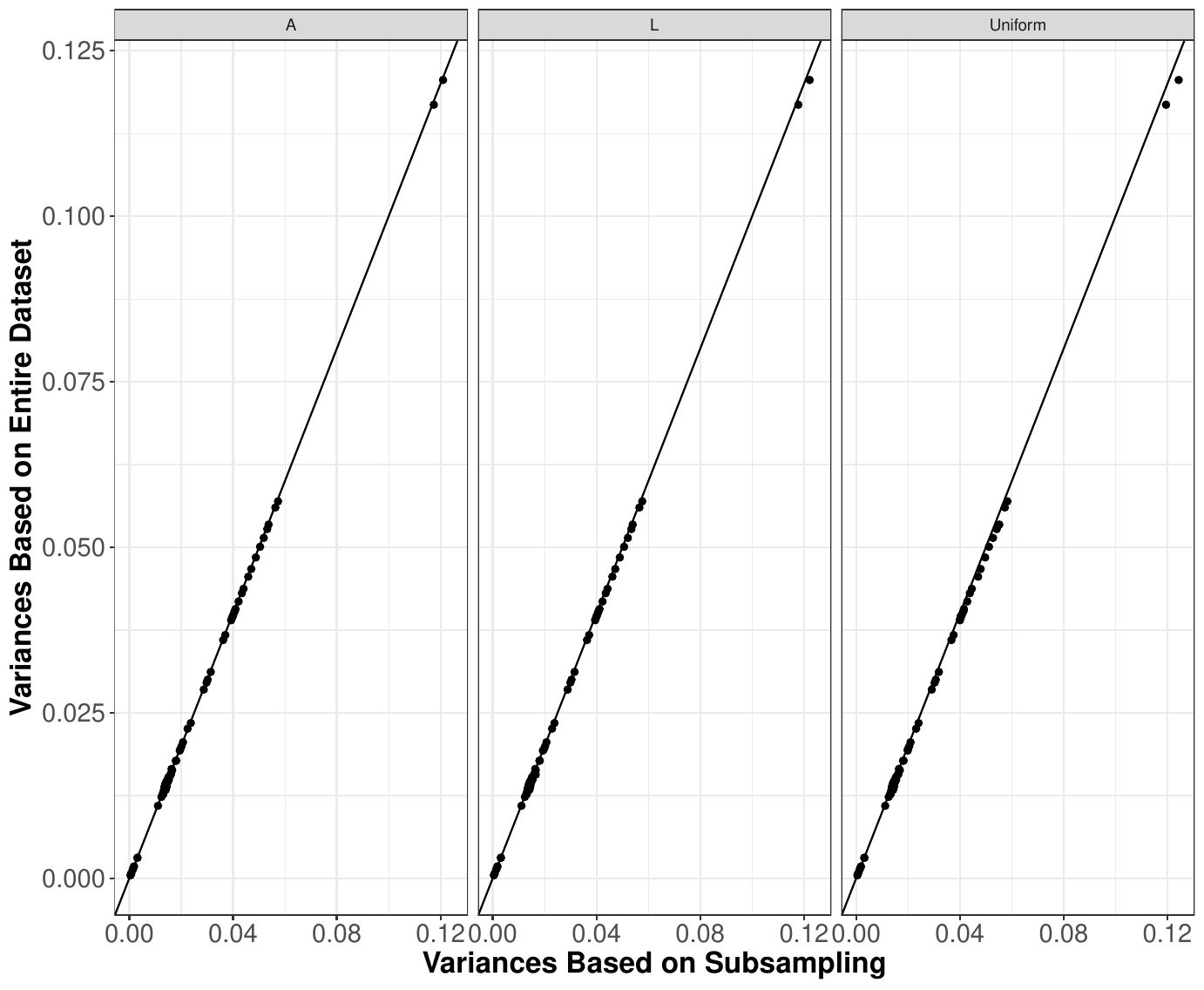}
	}	
	\caption{\label{fig:UKB_relative}
		UKB CRC survival analysis: results of Cox regression analysis.}
\end{figure}

We set $c_0=15$ and investigated the RE based on Step 1.5. The results are summarized in Fig. \ref{fig:UKB_relative} (a) and Table \ref{UKB_results_RR}. Notably, $c=100$ with the L-optimal subsampling approach (i.e., approximately 500K ``dummy-observations" instead of nearly 350 million) proves sufficient. However, in subsequent analyses, we also applied our proposed algorithms for $c=40$ and 160, for comparison purposes. Table \ref{UKB_RMSE_frob} presents the RMSE of the estimators with respect to the full-data PL estimator, the Frobenius norm of the covariance matrices of the estimators, and their running times. Clearly, the optimal methods outperform uniform subsampling regarding both RMSE and Forbenius norm, with the A-optimal method consistently exhibiting somewhat better values than the L-optimal method, as expected.  While the running time required for the full dataset is 14.5 hours, the time required for the L-optimal method with $c=100$ is reduced to 3.287 hours, with minimal loss in terms of efficiency, as demonstrated in Figures \ref{fig:UKB_relative} (b) and (c). 
In summary, this analysis, incorporating Step 1.5, highlights the effectiveness of selecting the optimal 
 $q_n$ according to the RE criterion.

\section{Linked Birth and Infant Death Data - Logistic Regression}\label{data_logistic}
The birth and infant death data sets, sourced from the National Bureau of Economic Research's public-use data archives, combine information from death certificates with corresponding birth certificates for infants under one year old who pass away in the United States, Puerto Rico, The Virgin Islands, and Guam. This linkage aims to leverage the additional information available in birth certificates, such as age, parents' race, birth weight, period of gestation, plurality, prenatal care usage, maternal education, live birth order, marital status, and maternal smoking, to enable more comprehensive analyses of infant mortality patterns.

\begin{figure}
	\centering
	\subfloat[RE based on Step 1.5]{
	\includegraphics[width=0.5\textwidth]{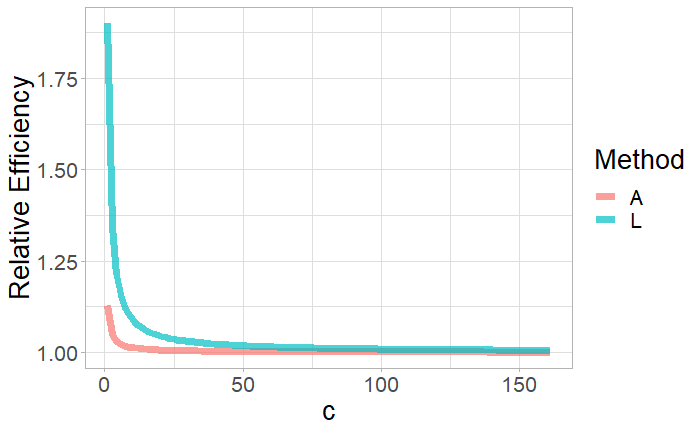}
}	
    \subfloat[Running time as a function of subsample size]{
	\includegraphics[width=0.5\textwidth]{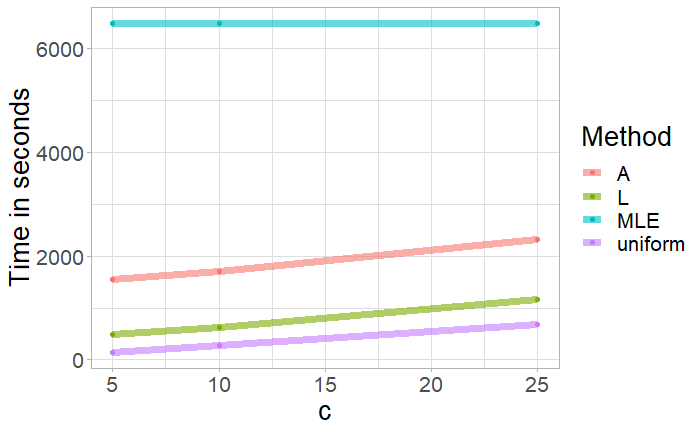}
}
	
\subfloat[RMSE with respect to full-data MLE]{
	\includegraphics[width=0.5\textwidth]{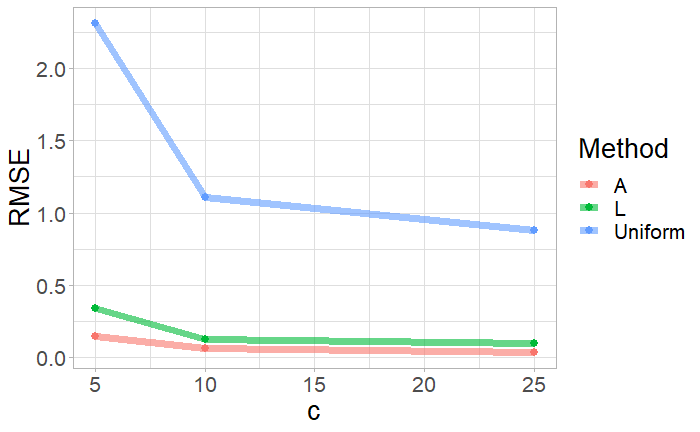}
}	
\subfloat[Comparison of Estimates]{
	\includegraphics[width=0.5\textwidth]{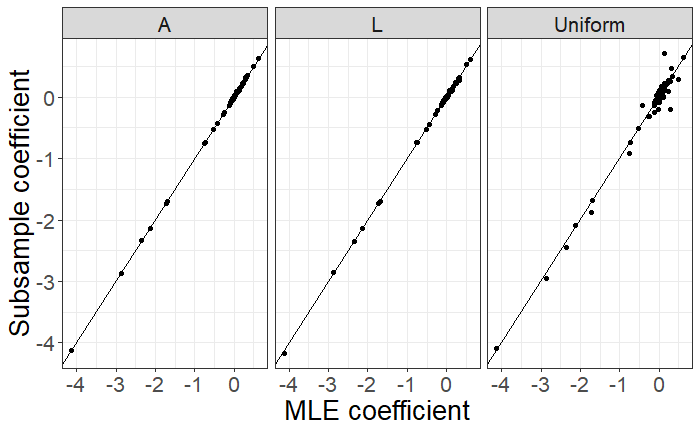}
}

\subfloat[Comparison of Standard Errors]{
    \includegraphics[width=0.55\textwidth]{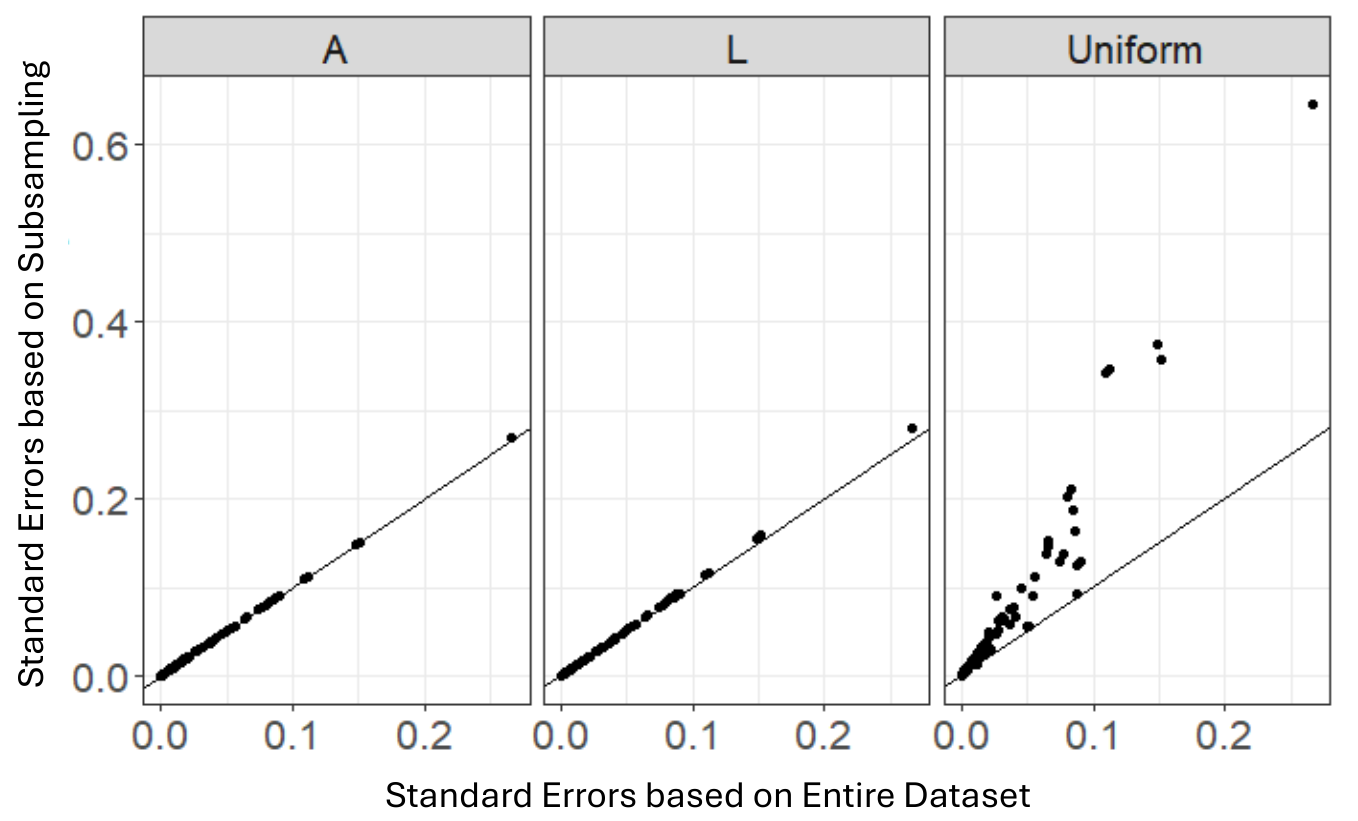}
}
\subfloat[Number of significant coefficients]{
	\includegraphics[width=0.5\textwidth]{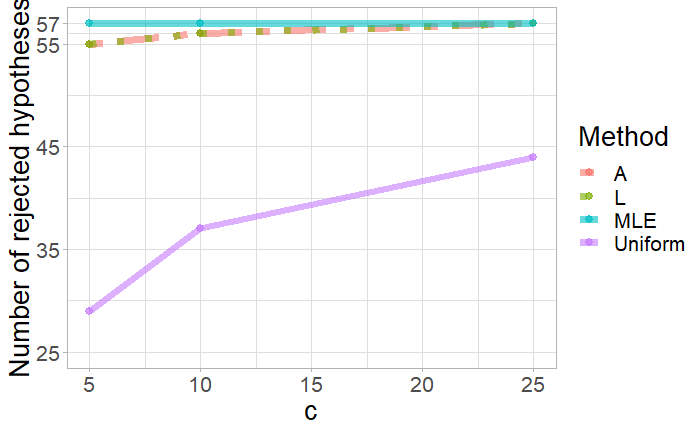}
}
	\caption{\label{fig:infants_results_comb}  Linked birth and infant death cohort data: results of logistic regression analysis}
\end{figure}

The data from years 2007 to 2013 were amalgamated into a single extensive dataset comprising $n=28,586,919$ rows. From the raw data, a set of features was derived, resulting in a covariate matrix with $103$ columns, encompassing 18 interaction terms with sex and 23 interaction terms with birth year. The covariates in the model are summarized in Tables S1--S3 of the SM. The primary outcome of interest is whether an infant passed away before reaching one year of age. Exactly $176,400$ deaths were observed, constituting about $0.6\%$ of event rate, justifying the use of a subsampling algorithm for rare events. The results are summarized in Fig. \ref{fig:infants_results_comb}.

In Fig. \ref{fig:infants_results_comb}(a), the RE based on Step 1.5 is displayed. Notably, the RE exhibits a distinct `elbow' around $q_n=1,500,000$, where the RE is also close to 1. We opted for a slightly higher value, $q_n=1,7640,000$, with $c=10$, indicating that 10 controls were sampled for each event. To offer a more comprehensive assessment of the algorithm's performance, we include results of analyses with $c=5$ and 25. The approximated RE, varying with $c$, is presented in Table \ref{infants_results_RR}. As anticipated, the A-optimal outperforms the L-optimal in terms of RE.

In Fig. \ref{fig:infants_results_comb}(b), the running time of various methods is illustrated as a function of $c$. The effectiveness of optimal subsampling becomes apparent when compared to the full-data MLE. With our chosen $c=10$, the running times for A and L criteria are $1700$ seconds and $628$ seconds, respectively, whereas the full-data MLE estimator takes $6484$ seconds. It is evident that the additional computational time required for optimal subsampling, as opposed to uniform subsampling, is relatively short, especially for the L method. This outcome reinforces the efficacy of the proposed procedure.

In Fig. \ref{fig:infants_results_comb}(c), the RMSE relative to $\widehat{\boldsymbol{\beta}}_{MLE}$ is depicted. The findings validate the judicious selection of $q_n$ since increasing $c$ from $5$ to $10$ substantially reduces the RMSE. However, a further increment to $c=25$ incurs a longer computational time and yields a comparatively modest improvement. Additionally, it is evident that optimal subsampling yields results substantially superior to those obtained through uniform subsampling.

The effectiveness of optimal subsampling methods over uniform subsampling is also evident in Fig.s \ref{fig:infants_results_comb}(d) and \ref{fig:infants_results_comb}(e). In Figure \ref{fig:infants_results_comb}(d), the estimated coefficients of each subsampling method are compared to their  $\widehat{\boldsymbol{\beta}}_{MLE}$ counterparts. Optimal subsampling yields results much closer to the full-data estimator than uniform subsampling. Figure \ref{fig:infants_results_comb}(e) displays the standard errors of  $\widehat{\boldsymbol{\beta}}_{MLE}$ versus the standard errors of their subsampling counterparts. Uniform subsampling produces notably larger standard errors.

We completed the analysis by conducting hypothesis testing, $H_0:\beta_i=0$ versus $H_1: \beta_i\neq0$, $i=1,\dots,103$ with FDR adjustment for multiplicity \citep{benjamini1995controlling}. This process was iterated for $c=5,10$ and $25$. In Figure \ref{fig:infants_results_comb}(f), the total number of rejected hypotheses under each $c$ is presented, contrasting with the number of rejections based on the full-data analysis. Notably, the A-optimal and L-optimal sampling methods outperforms uniform sampling. Even with a relatively small subsample size, the optimal sampling estimator yields results highly similar to those of the full data, surpassing the number of rejections achieved with uniform subsampling. For our chosen $c=10$, both A-optimal and L-optimal methods result in rejecting 56 hypotheses, almost matching the full-data analysis of 57 rejections. In contrast, uniform subsampling at $c=10$ only leads to 37 rejected hypotheses, underscoring the effectiveness of the optimal subsampling.

This dataset possesses a noteworthy characteristic--many of its features consist of rare binary variables. Examples include newborns with congenital anomalies like anencephaly, spina bifida, omphalocele, and Down's syndrome, alongside rare features related to the mother and delivery. Additionally, these features exhibit significant correlations with the outcome of interest, namely, death within the first year of life. The optimal subsampling procedures offer a notable advantage over uniform subsampling by ensuring that observations with rare features associated with the outcome have larger sampling probabilities. Consequently, they are more likely to be included in the subsample, leading to lower variance. In Table \ref{infants_columns}, the 20 rarest features in the data are presented, along with their corresponding proportions in both the full dataset and subsampling procedures for $c=10$. The results affirm that optimal subsamples better capture observations with rare indicators. These insights shed light on the efficiency of our proposed estimators, elucidating their superiority over uniform subsampling.

Regarding the findings derived from the analysis, Tables S4--S6 of the SM present the estimated coefficients for each method, with $c=10$. While the results are organized into three tables for clarity, it is essential to note that the FDR procedure was executed once, encompassing all coefficients collectively.

Among the significant results, both the mother's age and the squared mother's age emerged as noteworthy, corroborating established findings on the impact of maternal age on infant mortality \citep{macdorman1997sudden, standfast1980epidemiology}. This suggests heightened risks associated with motherhood at either a young or advanced age compared to medium age.

Other variables demonstrating significance in our analysis, consistent with prior literature, include lower risk as a function of number of prenatal visits \citep{carter2016number}, gestational weight gain \citep{naeve1979weight, thorsdottir2002weight}, five-minute Apgar score \citep{li2013apgar}, and plurality \citep{ahrens2017plurality}. Conversely, factors known as increasing risk in the literature and affirmed in this study include live birth order \citep{macdorman1997sudden, modin2002birth}, eclampsia \citep{duley2009global}, and certain congenital malformations linked to infant mortality, such as Spina Bifida \citep{pace2019survival}, Omphalocele \citep{marshall2015prevalence}, cleft lip \citep{carlson2013elevated}, and Down's syndrome \citep{sadetzki1999risk}.

Birth year, confined to the years 2007-2013, did not yield a significant effect. Similarly, no distinctions were observed among different months of the year. Treating Sunday as the baseline, negative impacts were noted for all days of the week except Saturday, indicating a significant difference between workdays and weekends. Concerning parental racial attributes, negative significant effects were identified for native-American ancestry in both parents and for African ancestry from the father's side. Additionally, an unknown father's race exhibited statistical significance with a negative effect. In contrast to some prior studies \citep{holmes2020implication, xie2015higher}, our findings indicate a lower risk of infant mortality among Caesarean section. Regarding interaction terms, five sex-interaction terms (weight gain, 5 minutes Apgar score, pre-pregnancy-associated hypertension, induction of labor, and cleft lip) and five birth year-interaction terms (African ancestry for the mother, induction of labor, tocolysis, Anencephaly, and Down's syndrome) were found to be statistically significant.

\section{Discussion}\label{discussion}
This study makes significant enhancements to the efficient two-step algorithms proposed by  \cite{wang2018optimal} and  \cite{keret2023analyzing}. We introduced practical tools for selecting optimal subsample sizes, illustrating their effectiveness through simulations and real-world data. Additionally, we proposed a new subsampling algorithm designed for logistic regression with rare events. This algorithm, which exclusively subsamples among non-cases, demonstrated speed and efficiency compared to full-data maximum-likelihood estimation. Its superiority over uniform subsampling was established in both simulated and real data, as evidenced by lower RMSE and variance. Furthermore, we demonstrated the algorithm's nearly equivalent performance to the full-data estimator in hypothesis testing while significantly reducing computational time.

Similar approaches to those proposed in this study can be extended to other two-step subsampling methods, including algorithms for generalized linear models \citep{ai2018optimal}, quantile regression \citep{ai2021optimal, fan2021optimal}, and quasi-likelihood regression \citep{yu2020optimal}.

Datasets with rare events often pose challenges for classification algorithms primarily oriented toward prediction rather than inference. The subsampling-based algorithm proposed for logistic regression with rare events in this study could serve as a practical tool for sampling probabilities in computationally-intensive methods. Notably, it may be worth exploring its application in methods like random forests \citep{breiman2001random} and gradient boosting \citep{friedman2001greedy}, among others.

\section{Software}
\label{sec5}
R codes for the data analysis and reported simulation results along with a complete documentation  are available at Github site  \url{https://github.com/tal-agassi/optimal-subsampling}.

\section{Supplementary Material}
\label{sec6}

Supplementary material is available online at
\url{http://biostatistics.oxfordjournals.org}.

\section*{Acknowledgments}

The work was supported by the Israel Science Foundation (ISF) grant number 767/21 and by a grant from the Tel Aviv University Center for AI and Data Science (TAD).

{\it Conflict of Interest}: None declared.

\bibliographystyle{biorefs}
\bibliography{mybib}


\newpage

\renewcommand{\baselinestretch}{1} 

\begin{table}
		\centering
			\small
	\begin{tabular}[t]{cccccc}
		\hline
		{Setting} & {Nominal Power} & \multicolumn{2}{c}{Empirical Power} & \multicolumn{2}{c}{Mean (SD) of $\tilde{q}_n$} \\
	    \hline
		 &  & {A}  & {L} & {A}  & {L} \\
		\hline
		I & 0.80 & 0.812  & 0.794 & 857 (45)  & 843 (37) \\
		& 0.83 & 0.818 & 0.814 & 1027 (61)  & 1007 (55)\\
		& 0.85 & 0.852  & 0.842 & 1175 (76)  & 1154 (69) \\
		& 0.87 & 0.882  & 0.882 & 1381 (97)  & 1342 (95) \\
		& 0.90 & 0.896  & 0.874 & 1856 (179)  & 1814 (164) \\
		& 0.91 & 0.920  & 0.890 & 2098 (212)  & 2064 (199) \\
		& 0.93 & 0.936  & 0.934 & 2901 (425)  & 2866 (379) \\
		& 0.95 & 0.952  & 0.960 & 5179 (1550)  & 4922 (1074) \\
		\hline
		II & 0.80 & 0.774  & 0.794 & 1112 (36)  & 1407 (48) \\
		& 0.83 & 0.804  & 0.812 & 1301 (50)  & 1652 (63) \\
		& 0.85 & 0.834  & 0.854 & 1470 (57)  & 1862 (74) \\
		& 0.87 & 0.882  & 0.844 & 1681 (74)  & 2122 (92) \\
		& 0.90 & 0.892  & 0.906 & 2155 (112)  & 2714 (142) \\
		& 0.91 & 0.916  & 0.902 & 2369 (122)  & 2997 (163) \\
		& 0.93 & 0.932  & 0.908 & 3046 (208)  & 3849 (247)\\
		& 0.95 & 0.940  & 0.930 & 4361 (418)  & 5533 (496) \\
		\hline
		III & 0.80 & 0.774  & 0.782 & 1640 (49)  & 2677 (81) \\
		& 0.83 & 0.824  & 0.784 & 1911 (62)  & 3132 (108) \\
		& 0.85 & 0.844  & 0.804 & 2148 (74)  & 3516 (123) \\
		& 0.87 & 0.858  & 0.860 & 2449 (94)  & 3999 (156) \\
		& 0.90 & 0.894  & 0.876 & 3105 (135)  & 5103 (216) \\
		& 0.91 & 0.886  & 0.874 & 3420 (168)  & 5603 (253) \\
		& 0.93 & 0.894  & 0.934 & 4289 (234)  & 7071 (375)\\
		& 0.95 & 0.942  & 0.930 & 6078 (412)  & 9872 (675) \\
		\hline
	\end{tabular}
	\caption{Simulation results of Cox regression model: nominal versus empirical power  based on $\tilde{q}_n$ of Eq. \eqref{q_size} and the proposed three-step estimator of Section \ref{hypothesis_testing_cox}.}
	\label{power_sims_cox}
\end{table}

\begin{table}
	\centering
		\small
	\begin{tabular}[t]{lcccc}
		\hline
	Setting	& MLE & L & A & Uniform\\
		\hline
		mzNormal & 0.815 (0.243) & 0.112 (0.058) & 0.116  (0.055) & 0.047  (0.030)\\
		mixNormal & 0.779 (0.123) & 0.101  (0.045) & 0.111  (0.052) & 0.041  (0.029)\\
		T3 		& 0.740 (0.133)  & 0.102 (0.042) & 0.113 (0.047)  &0.040 (0.029)\\
		EXP	& 1.084  (0.303) & 0.120 (0.036) & 0.130  (0.052) & 0.048  (0.026)\\
		\hline
	\end{tabular}
	\caption{Simulation results of logistic regression with rare events: running times (in hours) with $n=100,000$, $q_0=1,000$, $q_n=10,000$.}
	\label{running_logistic_rare}
\end{table}

\begin{table}
	\centering
		\small
	\begin{tabular}[t]{cccccc}
		\hline
		 {Setting}& {Nominal Power} &  \multicolumn{2}{c}{Empirical Power} & \multicolumn{2}{c}{Mean (SD) of $\tilde{q}_n$} \\
		\hline
		 &  & {A} & {L} & {A} & {L} \\
		\hline
		mzNormal & 0.80 & 0.804 & 0.799 & 1495 (91) & 1648 (107) \\
		& 0.83 & 0.833 & 0.824 & 1715 (113) & 1893 (129) \\
		& 0.85 & 0.843 & 0.849 & 1900 (127) & 2093 (149) \\
		& 0.87 & 0.857 & 0.861 & 2122 (148) & 2345 (175) \\
		& 0.90 & 0.900 & 0.887 & 2591 (196) & 2865 (230) \\
		& 0.93 & 0.921 & 0.931 & 3369 (296) & 3718 (352) \\
		& 0.95 & 0.948 & 0.941 & 4317 (442) & 4758 (524) \\
		\hline
		mixNormal & 0.80 & 0.792 & 0.786 & 821 (58) & 911 (69) \\
		& 0.83 & 0.827 & 0.815 & 958 (74) & 1061 (85) \\
		& 0.85 & 0.844 & 0.846 & 1074 (88) & 1191 (102) \\
		& 0.87 & 0.863 & 0.864 & 1223 (108) & 1353 (128) \\
		& 0.90 & 0.899 & 0.900 & 1542 (156) & 1711 (183) \\
		& 0.93 & 0.930 & 0.935 & 2123 (265) & 2364 (322) \\
		& 0.95 & 0.947 & 0.946 & 2957 (473) & 3289 (585) \\
		\hline
		T3 & 0.80 & 0.797 & 0.798 & 1175 (168) & 1347 (214) \\
		& 0.83 & 0.819 & 0.821 & 1334 (193) & 1536 (247) \\
		& 0.85 & 0.842 & 0.843 & 1467 (221) & 1686 (284) \\
		& 0.87 & 0.853 & 0.850 & 1625 (250) & 1865 (318) \\
		& 0.90 & 0.890 & 0.891 & 1947 (319) & 2243 (404) \\
		& 0.93 & 0.926 & 0.918 & 2449 (421) & 2820 (548) \\
		& 0.95 & 0.947 & 0.951 & 3021 (558) & 3475 (725) \\
		\hline
		EXP & 0.80 & 0.794 & 0.791 & 1263 (225) & 1264 (214) \\
		& 0.83 & 0.828 & 0.831 & 1449 (259) & 1458 (256)\\
		& 0.85 & 0.843 & 0.841 & 1613 (301) & 1624 (296) \\
		& 0.87 & 0.863 & 0.857 & 1812 (346) & 1826 (351) \\
		& 0.90 & 0.892 & 0.891 & 2225 (477) & 2249 (481)\\
		& 0.93 & 0.925 & 0.920 & 2913 (742) & 2924 (727) \\
		& 0.95 & 0.943 & 0.955 & 3814 (1162) & 3826 (1182) \\
		\hline
	\end{tabular}
	\caption{Simulation results of logistic regression with rare events: nominal versus the empirical power based on $\tilde{q}_n$ of Eq. \eqref{q_size_logistic_rare} and the proposed three-step estimator of Section 3.1.}
	\label{power_sims_rare}
\end{table}

\begin{table}
	\centering
		\small
	\begin{tabular}[t]{cccccc}
		\hline
		{Setting} & {Nominal Power} & \multicolumn{2}{c}{Empirical Power} & \multicolumn{2}{c}{Mean (SD) of $\tilde{q}_n$}\\
		\hline
		& & {A} & {L} & {A} & {L} \\
		\hline
		mzNormal & 0.80 & 0.806 & 0.820 & 4126 (230) & 4402 (241) \\
		& 0.83 & 0.854 & 0.818 & 4513 (253) & 4811 (269) \\
		& 0.85 & 0.840 & 0.822 & 4831 (266) & 5111 (285) \\
		& 0.87 & 0.872 & 0.860 & 5114 (271) & 5461 (292) \\
		& 0.89 & 0.880 & 0.872 & 5506 (307) & 5883 (330) \\
		& 0.91 & 0.900 & 0.924 & 5952 (324) & 6339 (352) \\
		& 0.93 & 0.928 & 0.922 & 6502 (353) & 6945 (376)\\
		& 0.95 & 0.958 & 0.940 & 7219 (415) & 7690 (404) \\
		\hline
		nzNormal & 0.80 & 0.784 & 0.752 & 4169 (263) & 4921 (292) \\
		& 0.83 & 0.820 & 0.856 & 4604 (297) & 5392 (353) \\
		& 0.85 & 0.866 & 0.852 & 4855 (319) & 5706 (348) \\
		& 0.87 & 0.874 & 0.880 & 5208 (326) & 6094 (382) \\
		& 0.89 & 0.904 & 0.880 & 5582 (368) & 6591 (410)\\
		& 0.91 & 0.910 & 0.890 & 6049 (398) & 7092 (415) \\
		& 0.93 & 0.920 & 0.920 & 6597 (420) & 7727 (477) \\
		& 0.95 & 0.952 & 0.942 & 7360 (472) & 8587 (535) \\
        \hline	
		mixNormal & 0.80 & 0.842 & 0.812 & 8682 (467) & 9160 (443) \\
		& 0.83 & 0.810 & 0.836 & 9514 (481) & 9952 (534) \\
		& 0.85 & 0.864 & 0.828 & 10120 (548) & 10575 (554) \\
		& 0.87 & 0.856 & 0.864 & 10866 (608) & 11367 (595) \\
		& 0.89 & 0.874 & 0.900 & 11623 (603) & 12181 (608) \\
		& 0.91 & 0.926 & 0.912 & 12565 (624) & 13219 (710) \\
		& 0.93 & 0.930 & 0.932 & 13777 (665) & 14396 (753)\\
		& 0.95 & 0.956 & 0.918 & 15286 (787) & 16076 (857) \\
		\hline
		T3 & 0.80 & 0.816 & 0.780 & 11900 (1534) & 12899 (1668) \\
		& 0.83 & 0.822 & 0.804 & 12966 (1659) & 14200 (1843) \\
		& 0.85 & 0.838 & 0.838 & 13940 (1741) & 15171 (2158) \\
		& 0.87 & 0.854 & 0.850 & 14698 (1917) & 16340 (2112) \\
		& 0.89 & 0.880 & 0.874 & 16012 (1959) & 17557 (2326) \\
		& 0.91 & 0.912 & 0.922 & 17265 (2199) & 19050 (2485) \\
		& 0.93 & 0.928 & 0.916 & 18837 (2371) & 20551 (2805) \\
		& 0.95 & 0.928 & 0.924 & 20634 (2660) & 23076 (3047)\\
		\hline
		exp & 0.80 & 0.804 & 0.808 & 7526 (832) & 7661 (832) \\
		& 0.83 & 0.834 & 0.800 & 8200 (829) & 8396 (918) \\
		& 0.85 & 0.850 & 0.856 & 8652 (940) & 8957 (1010) \\
		& 0.87 & 0.872 & 0.858 & 9281 (997) & 9463 (1089) \\
		& 0.89 & 0.906 & 0.902 & 9938 (1057) & 10190 (1102) \\
		& 0.91 & 0.890 & 0.912 & 10725 (1162) & 11027 (1206) \\
		& 0.93 & 0.932 & 0.910 & 11786 (1217) & 12009 (1230) \\
		& 0.95 & 0.946 & 0.950 & 13195 (1380) & 13385 (1369) \\
		\hline
	\end{tabular}
	\caption{Simulation results of logistic regression with (nearly) balanced data:  nominal versus the empirical power based on $\tilde{q}_n$ of Eq. \eqref{logistic_q_size} and the proposed three-step estimator of Section 4.2.}
	\label{resluts_log_no_rare}
\end{table}

\begin{table}
	\centering
	\small
	\begin{tabular}[t]{ccccc}
		\hline
		$c$ & $q_n$ & A & L & Uniform\\
		\hline
		40  & 213,680  &  1.0076 &  1.0316 & 1.1091\\
		60 & 320,520 & 1.0051 &  1.0206 & 1.0725 \\
		80 & 427,360 & 1.0038 &  1.0154 & 1.0543\\
		100 & 534,200 & 1.0031 & 1.0123 & 1.0434\\
		120 &  641,040 & 1.0025 & 1.0103 & 1.0361\\
		140 &  747,880 & 1.0022 & 1.0088 & 1.0309\\
		160 & 854,720 & 1.0019 & 1.0077 & 1.0271\\
		180 & 961,560 & 1.0017 & 1.0068 & 1.0240\\
		200 & 1,068,400 & 1.0015 & 1.0062 & 1.0216\\
		\hline
	\end{tabular}
	\caption{UKB CRC survival analysis: RE of Step 1.5.}
	\label{UKB_results_RR}
\end{table}

\begin{table}
	\centering
		\small
	\begin{tabular}[t]{cccccccccc}
		\hline
            & \multicolumn{3}{c}{RMSE with respect} & \multicolumn{3}{c}{Frobenius norm} & \multicolumn{3}{c}{Computation Time}\\
            & \multicolumn{3}{c}{ to $\boldsymbol{\widehat\beta}_{PL}$  ($\times 100$)} & \multicolumn{3}{c}{of covariance matrix ($\times 100$)} & \multicolumn{3}{c}{in hours} \\		
		$c$ &  A & L & Uniform  & A & L & Uniform & A & L & Uniform  \\
		\hline
		40 & 6.308 & 7.627 & 11.186 & 2.387 & 2.457 & 2.649 & 5.927 & 2.911 &  0.306 \\
		100 & 3.033  & 4.356 & 5.690 & 2.343 & 2.374 & 2.447 & 5.870 & 3.287 &  0.387  \\
		160 & 2.528 & 3.225 & 7.201 & 2.338 & 2.354 & 2.398 & 5.847 & 3.254 & 0.428 \\
		\hline
	\end{tabular}
	\caption{UKB CRC survival analysis. The Frobenius norm of the covarianve matrix of the entire-data PL estimator, $\hat{\boldsymbol{\beta}}_{PL}$ ($\times 100$), is $2.325$, and its running time is 14.5 hours.}
	\label{UKB_RMSE_frob}
\end{table}		

\begin{table}
	\centering
		\small
	\begin{tabular}[t]{cccc}
		\hline
		$c$ & $q_n$ & A & L \\
		\hline
		5  & 882,000  &  1.023 &  1.180 \\
		10 & 1,764,000 & 1.013 &  1.090 \\
		25 & 4,410,000 & 1.005 &  1.036 \\
		\hline
\end{tabular}
		\caption{Linked birth  and infant death data: RE of Step 1.5.}
		\label{infants_results_RR}
\end{table}		

\begin{table}
	\centering
	\small
	\resizebox{\columnwidth}{!}{%
		\begin{tabular}[t]{lccccc}
			\hline
			\multicolumn{1}{l}{Coefficient} & \multicolumn{1}{c}{Full-sample Proportion} & \multicolumn{2}{c}{Subsample Proportion} & \multicolumn{2}{c}{Ratio}\\
			\hline
			{} & {} & {A} & {L} & {A} & {L}\\
			\hline
			Anencephaly = no & 0.00011 & 0.03529 & 0.00416 & 309.41689 & 36.52090\\
			Spina Bifida = no & 0.00016 & 0.03385 & 0.00227 & 206.53149 & 13.85638\\
			Omphalocele = no & 0.00038 & 0.04523 & 0.00497 & 118.48022 & 13.02339\\
			Downs syndrome = no & 0.00048 & 0.03536 & 0.00710 & 73.33574 & 14.73338\\
			Cleft lip = no & 0.00072 & 0.04208 & 0.00908 & 58.72595 & 12.66362\\
			Residence status = 4 & 0.00190 & 0.00927 & 0.00118 & 4.88057 & 0.62083\\
			Eclampsia = no & 0.00253 & 0.04124 & 0.00754 & 16.28528 & 2.97785\\
			Attendant = other midwife & 0.00638 & 0.01121 & 0.00387 & 1.75569 & 0.60659\\
			Attendant = other & 0.00654 & 0.02102 & 0.01575 & 3.21212 & 2.40667\\
			Forceps delivery = no & 0.00663 & 0.03397 & 0.00433 & 5.12472 & 0.65301\\
			Father's race = american indian & 0.00870 & 0.02241 & 0.01109 & 2.57516 & 1.27423\\
			Mother’s race = american indian & 0.01165 & 0.03025 & 0.01596 & 2.59707 & 1.37015\\
			Birth place = not in hospital & 0.01205 & 0.02719 & 0.01835 & 2.25730 & 1.52346\\
			Tocolysis = no & 0.01211 & 0.05251 & 0.03946 & 4.33686 & 3.25930\\
			Cronic hypertension = no & 0.01325 & 0.04522 & 0.03319 & 3.41423 & 2.50554\\
			Residence status = 3 & 0.02119 & 0.04346 & 0.03647 & 2.05113 & 1.72109\\
			Precipitous labor = no & 0.02471 & 0.04615 & 0.03853 & 1.86780 & 1.55949\\
			Vacuum delivery = no & 0.03009 & 0.03916 & 0.01360 & 1.30158 & 0.45214\\
			Prepregnacny associated hypertension = no & 0.04275 & 0.06545 & 0.05763 & 1.53100 & 1.34794\\
			Meconium = no & 0.04736 & 0.05964 & 0.04199 & 1.25928 & 0.88652\\
			\hline
	\end{tabular}}
	\caption{Linked birth and infant death data: the proportion of the 20 least common binary variables are compared between the entire dataset and the A-optimal and L-optimal subsamples, with $c=10$.}
	\label{infants_columns}
\end{table}

\newpage

\renewcommand\thesection{S\arabic{section}}
\renewcommand{\thetable}{S\arabic{table}}
\renewcommand{\thefigure}{S\arabic{figure}}
\renewcommand{\theequation}{S.\arabic{equation}}

\setcounter{table}{0}
\setcounter{figure}{0}
\setcounter{section}{0}

\clearpage

\section*{Supplementary Material}

\section{Additional Technical Details}

The following functions are required for $\widetilde{\mathbb{V}}_{\widetilde{\boldsymbol{\beta}}}(\textbf{p},\widehat{\boldsymbol{\beta}})$:
\begin{equation*}
	\widetilde{\boldsymbol{\mathcal{I}}}(\boldsymbol{\beta})=
	\frac{1}{n}\frac{\partial^2l^*(\boldsymbol{\beta})}{\partial\boldsymbol{\beta}^T\partial\boldsymbol{\beta}}=
	-\frac{1}{n}\int_0^\tau \left\{\frac{\mathbf{S}_w^{(2)}(\boldsymbol{\beta},t)}{S_w^{(0)}(\boldsymbol{\beta},t)}-\left(\frac{\mathbf{S}_w^{(1)}(\boldsymbol{\beta},t)}{S_w^{(0)}(\boldsymbol{\beta},t)}\right)\left(\frac{\mathbf{S}_w^{(1)}(\boldsymbol{\beta},t)}{S_w^{(0)}(\boldsymbol{\beta},t)}\right)^T\right\}dN.(t)
\end{equation*}
and
\begin{equation*}
	\widetilde{\boldsymbol{\varphi}}(\mathbf{p},\boldsymbol{\beta})=\frac{1}{n^2}\left\{\frac{1}{q}\sum_{i\in\mathcal{Q}\setminus\mathcal{E}}\frac{\widetilde{\mathbf{a}}_i(\boldsymbol{\beta})\widetilde{\mathbf{a}}_i(\boldsymbol{\beta})^T}{p_i^2}-\frac{1}{q^2}\sum_{i\in\mathcal{Q}\setminus\mathcal{E}}\frac{\widetilde{\mathbf{a}}_i(\boldsymbol{\beta})}{p_i}\left(\sum_{i\in\mathcal{Q}\setminus\mathcal{E}}\frac{\widetilde{\mathbf{a}}_i(\boldsymbol{\beta})}{p_i}\right)^T\right\}
\end{equation*}
where
\begin{equation*}
	\widetilde{\mathbf{a}}_i(\boldsymbol{\beta})=\int_{0}^{\tau}\left\{\mathbf{X}_i-\frac{\mathbf{S}_w^{(1)}(\boldsymbol{\beta},t)}{S_w^{(0)}(\boldsymbol{\beta},t)}\right\}\frac{Y_i(t)e^{\boldsymbol{\beta}^T\mathbf{X}_i}}{S_w^{(0)}(\boldsymbol{\beta},t)}dN.(t) \, .
\end{equation*}

The following functions are required for $\widehat{RE}(q_n)$:
$$
\widetilde{\boldsymbol{\mathcal{I}}}^{-1}_{Q_{1.5}}({\boldsymbol{\beta}}) = \frac{1}{n}\frac{\partial^2l^*(\boldsymbol{\beta})}{\partial\boldsymbol{\beta}^T\partial\boldsymbol{\beta}}=
-\frac{1}{n}\int_0^\tau \left\{\frac{\mathbf{S}_{w,Q_{1.5}}^{(2)}(\boldsymbol{\beta},t)}{S_{w,Q_{1.5}}^{(0)}(\boldsymbol{\beta},t)}-\left(\frac{\mathbf{S}_{w,Q_{1.5}}^{(1)}(\boldsymbol{\beta},t)}{S_{w,Q_{1.5}}^{(0)}(\boldsymbol{\beta},t)}\right)\left(\frac{\mathbf{S}_{w,Q_{1.5}}^{(1)}(\boldsymbol{\beta},t)}{S_{w,Q_{1.5}}^{(0)}(\boldsymbol{\beta},t)}\right)^T\right\}dN.(t)
$$ 
and 
$$
\widetilde{\boldsymbol{\varphi}}_{Q_{1.5}}(\mathbf{p},{\boldsymbol{\beta}}) = \frac{1}{n^2}\left\{\frac{1}{q}\sum_{i\in\mathcal{Q}_{1.5}\setminus\mathcal{E}}\frac{\widetilde{\mathbf{a}}_i(\boldsymbol{\beta})\widetilde{\mathbf{a}}_i(\boldsymbol{\beta})^T}{p_i^2}-\frac{1}{q^2}\sum_{i\in\mathcal{Q}_{1.5}\setminus\mathcal{E}}\frac{\widetilde{\mathbf{a}}_i(\boldsymbol{\beta})}{p_i}\left(\sum_{i\in\mathcal{Q}_{1.5}\setminus\mathcal{E}}\frac{\widetilde{\mathbf{a}}_i(\boldsymbol{\beta})}{p_i}\right)^T\right\}
$$
where
\begin{equation*}
	\widetilde{\mathbf{a}}_i(\boldsymbol{\beta})=\int_{0}^{\tau}\left\{\mathbf{X}_i-\frac{\mathbf{S}_{w,Q_{1.5}}^{(1)}(\boldsymbol{\beta},t)}{S_{w,Q_{1.5}}^{(0)}(\boldsymbol{\beta},t)}\right\}\frac{Y_i(t)e^{\boldsymbol{\beta}^T\mathbf{X}_i}}{S_{w,Q_{1.5}}^{(0)}(\boldsymbol{\beta},t)}dN.(t) \, ,
\end{equation*}
\begin{equation*} {\mathbf{S}}_{{w,Q_{1.5}}}^{(k)}(\boldsymbol{\beta},t)=\sum_{i\in\mathcal{Q}_{1.5}}{w}_i e^{\boldsymbol{\beta}^T\mathbf{X}_i}Y_i(t)\mathbf{X}_i^{\otimes k} \quad \quad  k=0,1,2 \, ,
\end{equation*}
and
\begin{equation*}
	w_i= \begin{cases}
		(p_i q_0)^{-1} & \text{if } \Delta_i=0, p_i>0\\
		1 & \text{if } \Delta_i=1
	\end{cases} \quad \quad i=1,\dots,n \, .
\end{equation*}

\section{Logistic Regression - Assumptions and Proofs}
The $\mathbf{X}_i$'s are assumed to be independent and identically distributed and the following additional assumptions are required for the asymptotic results:
\begin{description}
	\item[A.1] As $n\rightarrow\infty$,  $n^{-1}\sum_{i=1}^n\|\mathbf{X}_i\|^3=O_P(1)$ and $\mathbf{M}_X(\boldsymbol{\beta}^o)$ goes in probability to a positive-definite matrix $\boldsymbol{\Sigma}(\boldsymbol{\beta}^o)$, where
	\begin{equation*}
		\mathbf{M}_X(\boldsymbol{\beta})=n^{-1}\sum_{i=1}^n p_i(\boldsymbol{\beta})\big(1-p_i(\boldsymbol{\beta})\big)\mathbf{X}_i\mathbf{X}_i^T.
	\end{equation*}
	
	\item[A.2] $n^{-2}\sum_{i=1}^n \pi_i^{-1}\|\mathbf{X}_i\|^k=O_P(1)$ for $k=2,4$.
	\item[A.3] There exists some $\delta>0$ such that $n^{-2+\delta}\sum_{i=1}^n\pi_i^{-1-\delta}\|\mathbf{X}_i\|^{2+\delta}=O_P(1)$.
	\item[A.4]  $q_n/n$ and $(n-n_0)/n$ converge to small positive constants as $q_n,n\rightarrow\infty$.
\end{description}

The first three assumptions are essentially general moment conditions  \citep{wang2018optimal}. In Assumption A.4 it is assumed that the sampled event rate goes to a positive constant as $n$ goes to infinity.

\subsection{Proof of Theorem 3.1}

This proof follows derivation similar to that of \cite{keret2023analyzing}.  \cite{wang2018optimal} have already shown that $\widetilde{\boldsymbol{\beta}}$ is consistent to $\widehat{\boldsymbol{\beta}}_{MLE}$ in the conditional space, given $\mathcal{F}_n$. We begin by expanding this result and show that $\widetilde{\boldsymbol{\beta}}$ is consistent to $\boldsymbol{\beta}^o$ in the unconditional space. Based on Theorem 1 of  \cite{wang2018optimal}, for any $\epsilon > 0$,
\begin{equation*}
	\lim_{q_n,n \rightarrow\infty} \Pr( {\|\widetilde{\boldsymbol{\beta}}-\widehat{\boldsymbol{\beta}}_{MLE}\|}_2>\epsilon|\mathcal{F}_n)=0 \, .
\end{equation*}
In the unconditional probability space, $\Pr( {\|\widetilde{\boldsymbol{\beta}}-\widehat{\boldsymbol{\beta}}_{MLE}\|}_2>\epsilon|\mathcal{F}_n)$ itself is a random variable. Hence,  denote it by $\zeta_{n,q_n}$ and it follows that
\begin{equation*}
	\Pr(\lim_{q_n,n \rightarrow \infty} \zeta_{n,q_n}=0)=1,
\end{equation*}
in the sense that $\zeta_{n,q_n}\xrightarrow{a.s.}0$ as $q_n,n \rightarrow\infty$. Then, for any $\epsilon>0$,
\begin{equation}\label{beta_tilde_consist}
	\lim_{q_n,n \rightarrow\infty} \Pr( {\|\widetilde{\boldsymbol{\beta}}-\widehat{\boldsymbol{\beta}}_{MLE}\|}_2>\epsilon)=\lim_{q_n,n \rightarrow\infty} E(\zeta_{n,q_n})=E(\lim_{q_n,n \rightarrow\infty} \zeta_{n,q_n})=0
\end{equation}
where the interchange of expectation and limit is allowed due to the dominated convergence theorem, since $\zeta_{n,q_n}$ is trivially bounded by $1$. Next, we write
\begin{equation*}
	\begin{aligned}
		\Pr({\|\widetilde{\boldsymbol{\beta}}-\boldsymbol{\beta}^o\|}_2>\epsilon) & =   \Pr({\|\widetilde{\boldsymbol{\beta}}+\widehat{\boldsymbol{\beta}}_{MLE}-\widehat{\boldsymbol{\beta}}_{MLE}-\boldsymbol{\beta}^o\|}_2>\epsilon) \\
		& \leq \Pr({\|\widetilde{\boldsymbol{\beta}}-\widehat{\boldsymbol{\beta}}_{MLE}\|}_2 + {\|\widehat{\boldsymbol{\beta}}_{MLE}-\boldsymbol{\beta}^o\|}_2>\epsilon) \\
		& \leq \Pr\big(\{{\|\widetilde{\boldsymbol{\beta}}-\widehat{\boldsymbol{\beta}}_{MLE}\|}_2>\epsilon/2\}\cup \{{\|\widehat{\boldsymbol{\beta}}_{MLE}-\boldsymbol{\beta}^o\|}_2>\epsilon/2\}\big) \\
		& \leq \Pr({\|\widetilde{\boldsymbol{\beta}}-\widehat{\boldsymbol{\beta}}_{MLE}\|}_2>\epsilon/2) + \Pr({\|\widehat{\boldsymbol{\beta}}_{MLE}-\boldsymbol{\beta}^o\|}_2>\epsilon/2) \, .
	\end{aligned}
\end{equation*}
Taking limits on both sides, yields
\begin{equation*}
	\begin{aligned}
		& \lim_{q_n,n \rightarrow\infty} \Pr({\|\widetilde{\boldsymbol{\beta}}-\boldsymbol{\beta}^o\|}_2>\epsilon) \\
		& \leq \lim_{q_n,n \rightarrow\infty}\Pr({\|\widetilde{\boldsymbol{\beta}}-\widehat{\boldsymbol{\beta}}_{MLE}\|}_2>\epsilon/2) + \lim_{q_n,n \rightarrow\infty}\Pr({\|\widehat{\boldsymbol{\beta}}_{MLE}-\boldsymbol{\beta}^o\|}_2>\epsilon/2) = 0
	\end{aligned}
\end{equation*}
where the first addend is $0$ due to Equation \eqref{beta_tilde_consist} and the second addend is $0$ based on the well-known properties of  logistic regression MLE. Then, we conclude that
\begin{equation*}
	\lim_{q_n,n \rightarrow\infty} \Pr({\|\widetilde{\boldsymbol{\beta}}-\boldsymbol{\beta}^o\|}_2>\epsilon)=0.
\end{equation*}

Similarly to Eq. (S.12) in  \cite{wang2018optimal}, a Taylor expansion for the subsample-based pseudo-score function evaluated at $\widetilde{\boldsymbol{\beta}}$ around $\boldsymbol{\beta}^o$ instead of $\widehat{\boldsymbol{\beta}}_{MLE}$ gives
\begin{equation}\label{taylor_logistic}
	\widetilde{\boldsymbol{\beta}}-\boldsymbol{\beta}^o=-\widetilde{\mathbf{M}}_X^{-1}(\boldsymbol{\beta}^o)\bigg\{\frac{1}{n}\frac{\partial l^*(\boldsymbol{\beta}^o)}{\partial(\boldsymbol{\beta}^T)}+o_P(\|\widetilde{\boldsymbol{\beta}}-\boldsymbol{\beta}^o\|)\bigg\}
\end{equation}
where the consistency of $\widetilde{\mathbf{M}}_\mathbf{X}(\widetilde{\boldsymbol{\beta}})$ to $\mathbf{M}_\mathbf{X}(\boldsymbol{\beta}^o)$ is derived in a similar manner to the proof of the consistency of $\widetilde{\boldsymbol{\beta}}$ to $\boldsymbol{\beta}^o$ and based on Eq. (S.1) in  Wang et al. \cite{wang2018optimal}.

Denote by $R_i$ the number of times observation $i$ appears in the subsample. Then,
\begin{eqnarray}\label{l*pirook}
	\frac{\partial l^*(\boldsymbol{\beta})}{\partial \boldsymbol{\beta}^T}&=&\sum_{i\in\mathcal{Q}} w_i^*\big(D_i^*-\mu_i^*(\boldsymbol{\beta})\big)\mathbf{X}_i \nonumber  \\ 
	&=&\sum_{i\in\mathcal{E}} w_i\big(1-\mu_i(\boldsymbol{\beta})\big)\mathbf{X}_i-\sum_{i\in\{\mathcal{Q}\setminus\mathcal{E}\}} w_i \mu_i(\boldsymbol{\beta})\mathbf{X}_i \nonumber \\
	&=&\sum_{i\in\mathcal{E}} \big(1-\mu_i(\boldsymbol{\beta})\big)\mathbf{X}_i-\sum_{i\in\{\mathcal{Q}\setminus\mathcal{E}\}} w_i \mu_i(\boldsymbol{\beta})\mathbf{X}_i  \nonumber\\
	&=&\sum_{i\in\mathcal{E}} \big(1-\mu_i(\boldsymbol{\beta})\big)\mathbf{X}_i-\sum_{i\in\{\mathcal{Q}\setminus\mathcal{E}\}} w_i \mu_i(\boldsymbol{\beta})\mathbf{X}_i -\sum_{i\in\mathcal{N}}\mu_i(\boldsymbol{\beta})\mathbf{X}_i+\sum_{i\in\mathcal{N}}\mu_i(\boldsymbol{\beta})\mathbf{X}_i \nonumber\\
	&=&\sum_{i=1}^n\big(D_i-\mu_i(\boldsymbol{\beta})\big)\mathbf{X}_i-\sum_{i\in\{\mathcal{Q}\setminus\mathcal{E}\}} w_i \mu_i(\boldsymbol{\beta})\mathbf{X}_i+\sum_{i\in\mathcal{N}}\mu_i(\boldsymbol{\beta})\mathbf{X}_i \nonumber\\
	&=&\sum_{i=1}^n\big(D_i-\mu_i(\boldsymbol{\beta})\big)\mathbf{X}_i-\sum_{i\in\mathcal{N}} R_i w_i \mu_i(\boldsymbol{\beta})\mathbf{X}_i+\sum_{i\in\mathcal{N}}\mu_i(\boldsymbol{\beta})\mathbf{X}_i \nonumber\\
	&=&\sum_{i=1}^n\big(D_i-\mu_i(\boldsymbol{\beta})\big)\mathbf{X}_i+\sum_{i\in\mathcal{N}}(1-w_i R_i)\mu_i(\boldsymbol{\beta})\mathbf{X}_i \nonumber\\
	&=&
	\frac{\partial l(\boldsymbol{\beta})}{\partial\boldsymbol{\beta}^T}+\sum_{i=1}^n(1-w_i R_i)\mu_i(\boldsymbol{\beta})\mathbf{X}_i.
\end{eqnarray}
Based on Eq.s \eqref{taylor_logistic} and \eqref{l*pirook}, we conclude that
\begin{equation}\label{taylor_explicit}
	\sqrt{n}(\widetilde{\boldsymbol{\beta}}-\boldsymbol{\beta}^o)=-\widetilde{\mathbf{M}}_X^{-1}(\boldsymbol{\beta}^o)\frac{1}{\sqrt{n}}\frac{\partial l(\boldsymbol{\beta}^o)}{\partial\boldsymbol{\beta}^T}-\widetilde{\mathbf{M}}_X^{-1}(\boldsymbol{\beta}^o)\frac{1}{\sqrt{n}}\sum_{i=1}^n(1-w_i R_i)\mu_i(\boldsymbol{\beta})\mathbf{X}_i+o_P(\sqrt{n}{\|\widetilde{\boldsymbol{\beta}}-\boldsymbol{\beta}^o\|}_2).
\end{equation}

Now it will be shown that  $n^{-1/2}\partial l(\boldsymbol{\beta}^o)/\partial\boldsymbol{\beta}^T$ and $n^{-1/2}\sum_{i=1}^n(1-w_i R_i)\mu_i(\boldsymbol{\beta})\mathbf{X}_i$ are asymptotically independent and each one of them is asymptotically normal.  From the asymptotic theory of standard logistic regression,
\begin{equation*}
	-\mathbf{M}_X^{-1/2}(\boldsymbol{\beta}^o)\frac{1}{\sqrt{n}}\frac{\partial l(\boldsymbol{\beta}^o)}{\partial\boldsymbol{\beta}^T}\xrightarrow[]{D}N(0,\mathbf{I}) \, ,
\end{equation*}
and
\begin{equation}\label{first_addended}
	\frac{1}{\sqrt{n}}\frac{\partial l(\boldsymbol{\beta}^o)}{\partial\boldsymbol{\beta}^T}\xrightarrow[]{D}N\big(0,\boldsymbol{\Sigma}(\boldsymbol{\beta}^o)\big) \, .
\end{equation}
Also,  $\sqrt{q_n}/n\sum_{i\in\mathcal{N}}w_i R_i \mu_i(\boldsymbol{\beta})\mathbf{X}_i$ can be alternatively expressed as a sum of independent identically distributed observations in the conditional space, namely
\begin{eqnarray*}
	\frac{\sqrt{q_n}}{n}\sum_{i\in\mathcal{N}}w_i R_i \mu_i(\boldsymbol{\beta})\mathbf{X}_i&=&\frac{\sqrt{q_n}}{n}\sum_{i=1}^{q_n} w_i^*\mu_i^*(\boldsymbol{\beta})\mathbf{X}_i^* \\
	&=&\frac{\sqrt{q_n}}{n}\sum_{i=1}^{q_n}\frac{\mu_i^*(\boldsymbol{\beta})\mathbf{X}_i^*}{\pi_i^* q_n}\\
	&=&\frac{1}{\sqrt{q_n}}\sum_{i=1}^{q_n}\frac{\mu_i^*(\boldsymbol{\beta})\mathbf{X}_i^*}{n\pi_i^*} \\
	&\equiv&\frac{1}{\sqrt{q_n}}\sum_{i=1}^{q_n} \boldsymbol{\omega}_i(\mathbf{\boldsymbol{\pi}},\boldsymbol{\beta}^o).
\end{eqnarray*}
Since the distribution of $\boldsymbol{\omega}_i(\boldsymbol{\pi},\boldsymbol{\beta^o})$ changes as a function of $n$ and $q_n$, the Lindeberg-Feller condition \cite[proposition 2.27]{van2000asymptotic} should be established as it covers the setting of triangular arrays.
First, let us denote $\mathbf{K}^R(\boldsymbol{\pi},\boldsymbol{\beta})\equiv Var(\boldsymbol{\omega}_i(\mathbf{\boldsymbol{\pi}},\boldsymbol{\beta}^o)|\mathcal{F}_n)$. It follows that
\begin{equation*}
	\begin{aligned}
		\mathbf{K}^R(\boldsymbol{\pi},\boldsymbol{\beta})&=E\big(\boldsymbol{\omega}(\mathbf{\boldsymbol{\pi}},\boldsymbol{\beta}^o)\boldsymbol{\omega}^T(\mathbf{\boldsymbol{\pi}},\boldsymbol{\beta}^o)|\mathcal{F}_n\big)-E(\boldsymbol{\omega}(\mathbf{\boldsymbol{\pi}},\boldsymbol{\beta}^o)|\mathcal{F}_n)E(\boldsymbol{\omega}(\mathbf{\boldsymbol{\pi}},\boldsymbol{\beta}^o)|\mathcal{F}_n)^T\\
		&=\frac{1}{n^2} \bigg\{\sum_{i\in\mathcal{N}}\frac{\mu^2_i(\boldsymbol{\beta})\mathbf{X}_i\mathbf{X}_i^T}{\pi_i}-\sum_{i,j\in\mathcal{N}}\mu_i(\boldsymbol{\beta})\mu_j(\boldsymbol{\beta})\mathbf{X}_i\mathbf{X}_j\bigg\}=O_{|\mathcal{F}_n}(1)
	\end{aligned}
\end{equation*}
where the last equation is due to Assumptions A.1 and A.2.

Now, for every $\varepsilon>0$ and some $\delta>0$,
\begin{equation*}
	\begin{aligned}
		\sum_{i=1}^{q_n} &E\{\|q_n^{-1/2}\boldsymbol{\omega}_i(\boldsymbol{\pi},\boldsymbol{\beta}^o)\|_2^2I(q_n^{-1/2}\boldsymbol{\omega}_i(\boldsymbol{\pi},\boldsymbol{\beta}^o)>\varepsilon)|\mathcal{F}_n\}.\\
		&\leq\frac{1}{q_n^{1+\delta/2}\varepsilon^\delta}\sum_{i=1}^q E\Bigg\{\bigg\|\frac{\mu_i^*(\boldsymbol{\beta}^o)\mathbf{X}_i^*}{n\pi_i^*}\bigg\|_2^{2+\delta}\Bigg|\mathcal{F}_n\Bigg\}\\
		&=\frac{1}{q_n^{\delta/2}\epsilon^\delta n^{2+\delta}}\sum_{i\in\mathcal{N}}\frac{\{\mu_i(\boldsymbol{\beta})\}^2\|\mathbf{X}_i\|_2^{2+\delta}}{\pi_i^{\delta+1}}\\
		&\leq\frac{1}{q_n^{\delta/2}\epsilon^\delta n^{2+\delta}}\sum_{i\in\mathcal{N}}\frac{\|\mathbf{X}_i\|_2^{2+\delta}}{\pi_i^{\delta+1}}=o_{P|\mathcal{F}_n}(1)
	\end{aligned}
\end{equation*}
where the first inequality is due to Van der Vaart \cite[p. 21]{van2000asymptotic} and the last equality is due to Assumption A.3. Since $E(1-w_i R_i|\mathcal{F}_n)=0$, it holds that $q_n^{1/2}n^{-1}K(\boldsymbol{\pi},\boldsymbol{\beta})^{-1/2}\sum_{i\in\mathcal{N}}(1-w_i R_i)\mu_i(\boldsymbol{\beta}^o)\mathbf{X}_i$ converges conditionally on $\mathcal{F}_n$ to a standard multivariate distribution. Put differently, for any $\mathbf{u}\in\mathbb{R}^r$,
\begin{equation}\label{cond_converge}
	\Pr\bigg\{\frac{\sqrt{q_n}}{n}\mathbf{K}^R(\boldsymbol{\pi},\boldsymbol{\beta})^{-1/2}\sum_{i\in\mathcal{N}}(1-w_i R_i)\mu_i(\boldsymbol{\beta}^o)\mathbf{X}_i\leq\mathbf{u}|\mathcal{F}_n\bigg\}\rightarrow\Phi(\mathbf{u}) \, .
\end{equation}
where $\Phi$ is the cumulative distribution function of the standard multivariate normal distribution. Since the conditional probability is a random variable in the unconditional space, then due to Eq. \eqref{cond_converge} it converges almost surely to $\Phi(\mathbf{u})$. Being additionally bounded, then due to the dominated convergence theorem, it follows that for any $\mathbf{u}\in\mathbb{R}^r$,
\begin{equation}\label{kappa_is_normal}
	\Pr\bigg\{\frac{\sqrt{q_n}}{n}\mathbf{K}^R(\boldsymbol{\pi},\boldsymbol{\beta})^{-1/2}\sum_{i\in\mathcal{N}}(1-w_i R_i)\mu_i(\boldsymbol{\beta}^o)\mathbf{X}_i\leq\mathbf{u}\bigg\}\rightarrow\Phi(\mathbf{u}) \, .
\end{equation}

Suppose that $\mathbf{K}^R(\boldsymbol{\pi},\boldsymbol{\beta}^o)\xrightarrow{P}\boldsymbol{\Psi}(\boldsymbol{\pi},\boldsymbol{\beta}^o)$ where ${\boldsymbol{\Psi}}(\boldsymbol{\pi},\boldsymbol{\beta}^o)$ is a positive-definite matrix. Denote $\theta$ as the limit of $q_n/n$, which we assumed its existence earlier. Then, from Eq. \eqref{kappa_is_normal} 
\begin{equation}\label{second_addended}
	\frac{1}{\sqrt{n}}\sum_{i\in\mathcal{N}}(1-w_i R_i)\mu_i(\boldsymbol{\beta}^o)\xrightarrow{D}N(0,\theta\boldsymbol{\Psi}(\boldsymbol{\pi},\boldsymbol{\beta}^o)).
\end{equation}
In the following, it will be shown that the two addends are asymptotically independent. Write
\begin{eqnarray}\label{asymptot_indep}
	&\lim_{n,q_n\rightarrow\infty}&\Pr\bigg(\frac{1}{n}\frac{\partial l(\boldsymbol{\beta}^o)}{\partial\boldsymbol{\beta}^T}\leq\mathbf{u} \, ,\, \frac{1}{\sqrt{n}}\sum_{i\in\mathcal{N}}(1-w_iR_i)p_i(\boldsymbol{\beta}^o)\mathbf{X}_i\leq\mathbf{v}\bigg) \nonumber\\
	&=&\lim_{n,q_n\rightarrow\infty}E\bigg(I\bigg\{\frac{1}{n}\frac{\partial l(\boldsymbol{\beta}^o)}{\partial\boldsymbol{\beta}^T}\leq\mathbf{u}\bigg\}\Pr\bigg\{\frac{1}{\sqrt{n}}\sum_{i\in\mathcal{N}}(1-w_iR_i)p_i(\boldsymbol{\beta}^o)\mathbf{X}_i\leq\mathbf{v}|\mathcal{F}_n\bigg\}\bigg)\nonumber\\
	&=&E\bigg(\lim_{n,q_n\rightarrow\infty}I\bigg\{\frac{1}{n}\frac{\partial l(\boldsymbol{\beta}^o)}{\partial\boldsymbol{\beta}^T}\leq\mathbf{u}\bigg\}\lim_{n,q\rightarrow\infty}\Pr\bigg\{\frac{1}{\sqrt{n}}\sum_{i\in\mathcal{N}}(1-w_iR_i)p_i(\boldsymbol{\beta}^o)\mathbf{X}_i\leq\mathbf{v}|\mathcal{F}_n\bigg\}\bigg)\nonumber\\
	&=&E\bigg(\lim_{n,q_n\rightarrow\infty}I\bigg\{\frac{1}{n}\frac{\partial l(\boldsymbol{\beta}^o)}{\partial\boldsymbol{\beta}^T}\leq\mathbf{u}\bigg\}\Phi(\theta^{-1/2}\boldsymbol{\Psi}(\boldsymbol{\pi},\boldsymbol{\beta}^o)^{-1/2}\mathbf{v})\bigg)\nonumber\\
	&=&\lim_{n,q_n\rightarrow\infty}E\bigg(I\bigg\{\frac{1}{n}\frac{\partial l(\boldsymbol{\beta}^o)}{\partial\boldsymbol{\beta}^T}\leq\mathbf{u}\bigg\}\Phi(\theta^{-1/2}\boldsymbol{\Psi}(\boldsymbol{\pi},\boldsymbol{\beta}^o)^{-1/2}\mathbf{v})\bigg)\nonumber\\
	&=&\lim_{n,q_n\rightarrow\infty}\Pr\bigg(\frac{1}{n}\frac{\partial l(\boldsymbol{\beta}^o)}{\partial\boldsymbol{\beta}^T}\leq\mathbf{u}\bigg)\Phi(\theta^{-1/2}\boldsymbol{\Psi}(\boldsymbol{\pi},\boldsymbol{\beta}^o)^{-1/2}\mathbf{v})\nonumber\\
	&=&\Phi\left(\boldsymbol{\Sigma}(\boldsymbol{\beta}^o)^{-1/2}\mathbf{u}\right)\Phi\left(\theta^{-1/2}\boldsymbol{\Psi}(\boldsymbol{\pi},\boldsymbol{\beta}^o)^{-1/2}\mathbf{v}\right)
\end{eqnarray}
and we have used the dominated convergence theorem.

Since $\widetilde{\mathbf{M}}_\mathbf{X}(\boldsymbol{\beta}^o)$ is a consistent estimator of $\mathbf{M}_\mathbf{X}(\boldsymbol{\beta}^o)$, its consistency to $\boldsymbol{\Sigma}(\boldsymbol{\beta}^o)$ could be easily shown. Then, from Slutsky's theorem and Eq.s \eqref{first_addended}, \eqref{second_addended} and \eqref{asymptot_indep} it follows that Eq. \eqref{taylor_explicit} converges in distribution to a multivariate normal distribution with zero mean and a covariance matrix asymptotically equivalent to $\mathbb{H}^R(\boldsymbol{\pi},\boldsymbol{\beta}^o)$. The two variance components correspond to two orthogonal sources of variance,  the variance of the original full-data MLE, and the additional variance generated by the subsampling procedure.

\subsection{Proof of Theorem 3.2}
A-optimal criterion is equivalent to minimizing the asymptotic MSE of $\widetilde{\boldsymbol{\beta}}_{TS}$, which is the trace of $\mathbb{H}^R(\boldsymbol{\pi},\boldsymbol{\beta}^o)$. However,
\begin{equation*}
	Tr\big(\mathbb{H}^R(\boldsymbol{\pi},\boldsymbol{\beta}^o)\big)
	=Tr\bigg(\frac{n}{q_n}\mathbf{M}_X^{-1}(\boldsymbol{\beta}^o)\mathbb{K}^R(\boldsymbol{\pi},\boldsymbol{\beta}^o)\mathbf{M}_X^{-1}(\boldsymbol{\beta}^o)\bigg)+d
\end{equation*}
where $d$ is a constant that does not involve $\boldsymbol{\pi}$, and
\begin{equation*}
	\begin{aligned}
		&Tr\bigg(\frac{n}{q_n}\mathbf{M}_X^{-1}(\boldsymbol{\beta}^o)\mathbb{K}^R(\boldsymbol{\pi},\boldsymbol{\beta}^o)\mathbf{M}_X^{-1}(\boldsymbol{\beta}^o)\bigg)\\
		&\quad = \quad Tr\Bigg(\frac{1}{nq_n}\mathbf{M}_X^{-1}(\boldsymbol{\beta}^o)\bigg\{\sum_{i\in\mathcal{N}}\frac{\mu^2_i(\boldsymbol{\beta}^o)}{\pi_i}\mathbf{X}_i\mathbf{X}_i^T-
		\sum_{i,j\in\mathcal{N}}\mu_i(\boldsymbol{\beta}^o)\mu_j(\boldsymbol{\beta}^o)\mathbf{X}_i\mathbf{X}_j^T\bigg\}\mathbf{M}_X^{-1}(\boldsymbol{\beta}^o)\Bigg).
	\end{aligned}
\end{equation*}
By removing the part that does not involve $\boldsymbol{\pi}$ and the factor $(nq_n)^{-1}$ which does not alter the optimization process, we are left with
\begin{eqnarray*}
	Tr\bigg(\sum_{i\in\mathcal{N}}\frac{\mu^2_i(\boldsymbol{\beta}^o)}{\pi_i}\mathbf{M}_X^{-1}(\boldsymbol{\beta}^o)\mathbf{X}_i\mathbf{X}_i^T\mathbf{M}_X^{-1}(\boldsymbol{\beta}^o)\bigg)&=&\sum_{i\in\mathcal{N}}\frac{\mu^2_i(\boldsymbol{\beta}^o)}{\pi_i}Tr\big(\mathbf{X}_i^T\mathbf{M}_X^{-2}(\boldsymbol{\beta}^o)\mathbf{X}_i\big) \\
	&=&\sum_{i\in\mathcal{N}}\frac{\mu^2_i(\boldsymbol{\beta}^o)}{\pi_i}\|\mathbf{M}_X^{-1}\mathbf{X}_i\|_2^2 \, .
\end{eqnarray*}
Define the following Lagrangian function, with multiplier $\alpha$,
\begin{equation*}
	g(\boldsymbol{\pi})=\sum_{i\in\mathcal{N}}\frac{\mu^2_i(\boldsymbol{\beta}^o)}{\pi_i}\|\mathbf{M}_x^{-1}\mathbf{X}_i\|_2^2
	+\alpha\left(1-\sum_{i\in\mathcal{N}}\pi_i\right) \, .
\end{equation*}
Differentiating $g(\boldsymbol{\pi})$ with respect to $\pi_i$ for any $i\in\mathcal{N}$ and setting the derivative to 0, gives
\begin{equation*}
	\frac{\partial g(\boldsymbol{\pi})}{\partial \pi_i}=-\frac{\mu^2_i(\boldsymbol{\beta}^o)\|\mathbf{M}_x^{-1}\mathbf{X}_i\|_2^2}{\pi_i^2}-\alpha \equiv 0,
\end{equation*}
and
\begin{equation*}
	\pi_i=\frac{\mu_i(\boldsymbol{\beta}^o)\|\mathbf{M}_x^{-1}\mathbf{X}_i\|_2}{\sqrt{-\alpha}} \, .
\end{equation*}
Since $\sum_{i\in\mathcal{N}}\pi_i=1$, 
\begin{equation*}
	\sqrt{-\alpha}=\sum_{i\in\mathcal{N}}\mu_i(\boldsymbol{\beta}^o)\|\mathbf{M}_x^{-1}\mathbf{X}_i\|_2,
\end{equation*}
which yields Eq. (3.9) in the main text. The proof of Eq. (3.10) of the main text follows similarly.

\subsection{Proof of Theorem 4.1}

Following the main steps of the proof of Theorem 2 in \cite{wang2018optimal}, it is straightforward  to show that given $\mathcal{F}_n$,
\begin{equation*}
	\frac{1}{n}\mathbf{K}^R(\boldsymbol{\pi},\boldsymbol{\beta}^o)^{1/2}\frac{\partial l^*(\boldsymbol{\beta}^o)}{\partial \boldsymbol{\beta}^o}=\frac{1}{\sqrt{q_n}}\{Var(\boldsymbol{\eta}_i|\mathcal{F}_n)\}^{-1/2}\sum_{i=1}^{q_n} \boldsymbol{\eta}_i \xrightarrow[]{D}N(0,\mathbf{I})
\end{equation*}
where
\begin{equation*}
	\boldsymbol{\eta_i}\equiv \frac{\{D_i^*-\mu_i^*(\boldsymbol{\beta}^o)\}\mathbf{X}_i^*}{n\pi_i^*} \quad , \quad i=1,\dots,q_n
\end{equation*}
are independent and identically distributed with mean $\mathbf{0}$ and variance $q_n\mathbf{K}^B(\boldsymbol{\pi},\boldsymbol{\beta}^o)$. In other words, for all $\mathbf{u}\in\mathbb{R}^r$,
\begin{equation}\label{cond_prob}
	\Pr\Big\{n^{-1}\mathbf{K}^R(\boldsymbol{\pi},\boldsymbol{\beta}^o)^{1/2}\frac{\partial l^*(\boldsymbol{\beta}^o)}{\partial \boldsymbol{\beta}^o}\leq \mathbf{u}|\mathcal{F}_n\Big\}\xrightarrow[]{P}\Phi(\mathbf{u}) \, .
\end{equation}
The conditional probability in Eq. \eqref{cond_prob} is a bounded random variable, thus convergence in probability to a constant implies convergence in the mean. Therefore,
\begin{equation*}
	\Pr\Big\{n^{-1}\mathbf{K}^R(\boldsymbol{\pi},\boldsymbol{\beta}^o)^{1/2}\frac{\partial l^*(\boldsymbol{\beta}^o)}{\partial \boldsymbol{\beta}^o}\leq \mathbf{u}\Big\}=E\Bigg\{\Pr\Big\{n^{-1}\mathbf{K}^R(\boldsymbol{\pi},\boldsymbol{\beta}^o)^{1/2}\frac{\partial l^*(\boldsymbol{\beta}^o)}{\partial \boldsymbol{\beta}^o}\leq \mathbf{u}\Big\}|\mathcal{F}_n\Bigg\}\xrightarrow[]{}\Phi(\mathbf{u}) \, ,
\end{equation*}
and therefore
\begin{equation*}
	\frac{1}{n}\mathbf{K}^R(\boldsymbol{\pi},\boldsymbol{\beta}^o)^{1/2}\frac{\partial l^*(\boldsymbol{\beta}^o)}{\partial \boldsymbol{\beta}^o} \xrightarrow[]{D}N(0,\mathbf{I})
\end{equation*}
in the unconditional space. The rest of the proof follows directly from \cite{wang2018optimal}.

\section{Additional Simulation Results}
In Fig. S1, we compare the Frobenius norm of three covariance matrices: (i) The covariance matrix of the two-step estimator, $\widetilde{\boldsymbol{\beta}}_{TS}$. (ii) The approximated covariance matrix utilized in Step 1.5. (iii) The empirical covariance matrix of $\widetilde{\boldsymbol{\beta}}_{TS}$. Fig. S2  demonstrates the validity of the variance estimator (3.11), and the effectiveness of optimal subsampling over uniform subsampling.

\begin{figure}[h]
	\centering
	\includegraphics[scale=0.65]{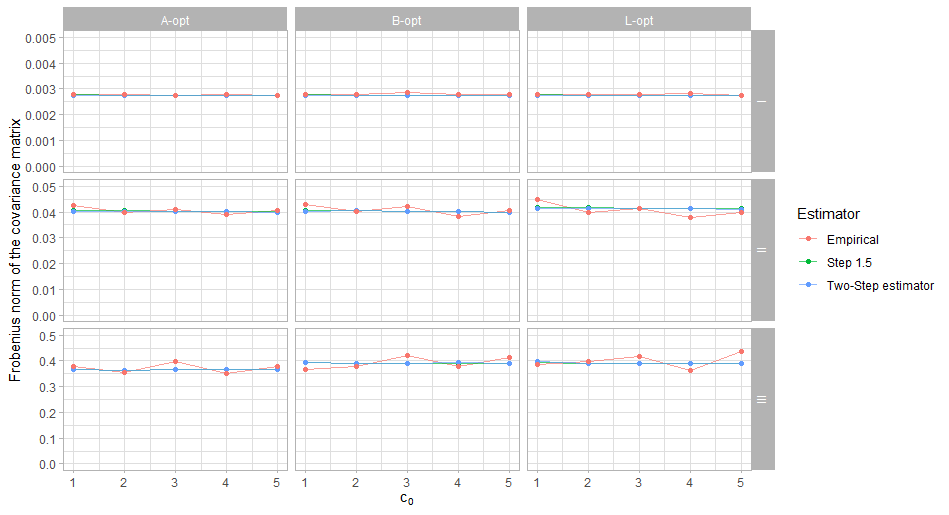}
	\caption{\label{fig:Var_frobenius}
		Simulation results of Cox regression model: Frobenius norm of (i) the estimated covariance matrix of the two-stage estimator; (ii) the empirical covariance matrices of the two-stage estimator; and (iii) the covariance matrix approximation used in Step 1.5, for various values of $c_0$ where $q_0 = c_0\times n_e$.
	}
\end{figure}

\begin{figure}
	\centering
	\includegraphics[width=1.0\textwidth]{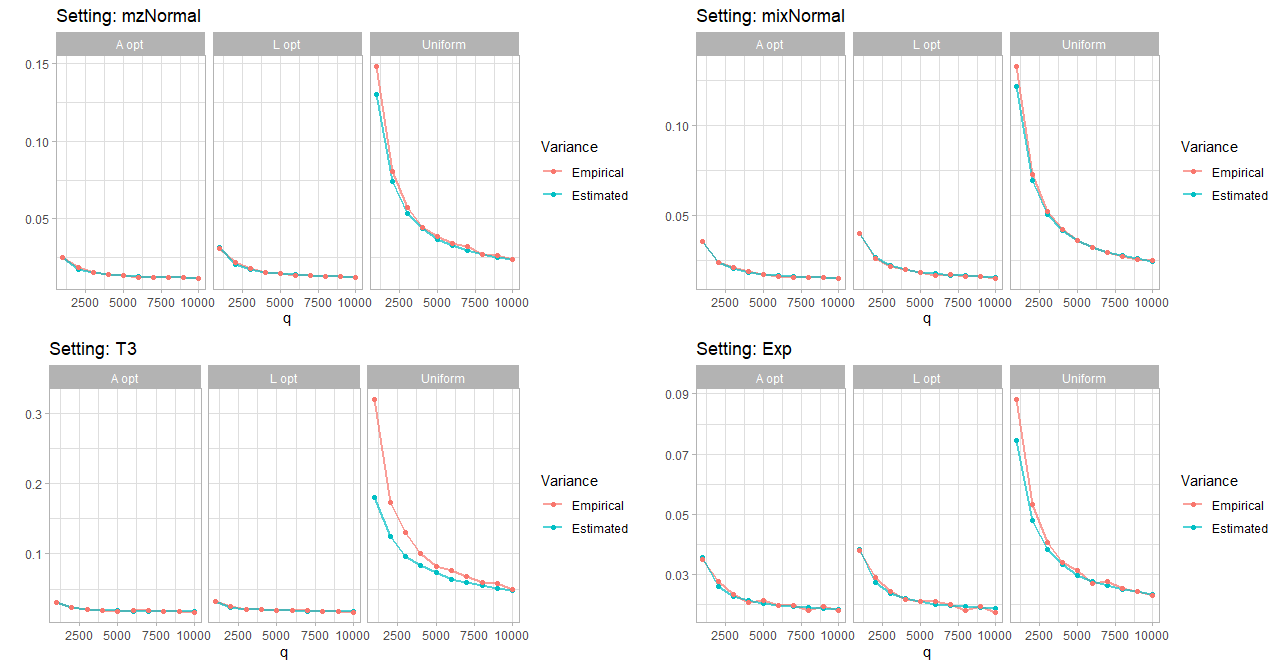}
	\caption{\label{fig:rare_events_msel}Simulation results for the two-step algorithm for logistic regression with rare events: trace of the variance estimator of the two-step estimators versus the trace of the empirical covariance, with subsampling methods A, L and Uniform along with the gold standard estimator based on the full data, denoted by MLE.}
\end{figure}

\renewcommand{\baselinestretch}{1} 

\newpage

\section{Linked Birth and Infant Death Data - Additional Results}
The covariates in the model are summarized in Tables S1--S3.  Tables S4--S6 present the estimated coefficients for each method, with $c=10$. While the results are organized into three tables for clarity, it is essential to note that the FDR procedure was executed once, encompassing all coefficients collectively.

\begin{table}
	\centering
	\footnotesize
	\begin{tabular}{|l|c|c|}
		\hline
		~ & \textbf{Non-events} & \textbf{Events} \\ \hline
		~ & (N=28410519) & (N=176400) \\ \hline
		\textbf{Mother's age (limited 12-50)} & ~ & ~ \\ \hline
		Mean (SD) & 27.71 (6.09) & 26.9 (6.5) \\ \hline
		Median [Min, Max] & 28 [12, 15] & 26 [12, 50] \\ \hline
		\textbf{Live Birth Order} & ~ & ~ \\ \hline
		Mean (SD) & 2.08 (1.24) & 2.19 (1.4) \\ \hline
		Median [Min, Max] & 2 [1, 8] & 2 [1, 8] \\ \hline
		\textbf{Number of Prenatal Visits} & ~ & ~ \\ \hline
		Mean (SD) & 11.26 (3.94) & 8.21 (5.11) \\ \hline
		Median [Min, Max] & 12 [0, 49] & 8 [0,49] \\ \hline
		\textbf{Weight Gain (limited to 99 pounds)} & ~ & ~ \\ \hline
		Mean (SD) & 30.51 (14.32) & 22.95 (15.04) \\ \hline
		Median [Min, Max] & 30 [0, 98] & 22 [0, 98] \\ \hline
		\textbf{Five Minute APGAR Score} & ~ & ~ \\ \hline
		Mean (SD) & 8.84 [0.71] & 5.19 (3.44) \\ \hline
		Median [Min, Max] & 9 [0,10] & 6 [0, 10] \\ \hline
		\textbf{Plurality (limited to 5)} & ~ & ~ \\ \hline
		Mean (SD) & 1.04 (0.19) & 1.65 (0.42) \\ \hline
		Median [Min, Max] & 1 [1, 5] & 1 [1, 5] \\ \hline
		\textbf{Gestation weeks} & ~ & ~ \\ \hline
		Mean (SD) & 38.65 (2.37) & 30.21 (7.68) \\ \hline
		Median [Min, Max] & 39 [17, 47] & 30 [17,47] \\ \hline
		\textbf{Years after 2007} & ~ & ~ \\ \hline
		Mean (SD) & 2.93 (2.01) & 2.85 (2.01) \\ \hline
		Median [Min, Max] & 3 [0, 6] & 3 [0, 6] \\ \hline
		\textbf{Birth month} & ~ & ~ \\ \hline
		January & 8.19\% & 8.36\% \\ \hline
		February & 7.63\% & 7.72\% \\ \hline
		March & 8.31\% & 8.35\% \\ \hline
		April & 7.98\% & 8.19\% \\ \hline
		May & 8.33\% & 8.56\% \\ \hline
		June & 8.32\% & 8.32\% \\ \hline
		July & 8.80\% & 8.66\% \\ \hline
		August & 8.93\% & 8.79\% \\ \hline
		September & 8.66\% & 8.42\% \\ \hline
		October & 8.50\% & 8.52\% \\ \hline
		November & 8.02\% & 7.97\% \\ \hline
		December & 8.33\% & 8.14\% \\ \hline
		\textbf{Birth weekday} & ~ & ~ \\ \hline
		Sunday & 9.31\% & 11.36\% \\ \hline
		Monday & 15.18\% & 14.73\% \\ \hline
		Tuesday & 16.59\% & 15.51\% \\ \hline
		Wednesday & 16.27\% & 15.38\% \\ \hline
		Thursday & 16.21\% & 15.57\% \\ \hline
		Friday & 15.84\% & 15.25\% \\ \hline
		Saturday & 10.60\% & 12.20\% \\ \hline
		\textbf{Birth place} & ~ & ~ \\ \hline
		In hospital & 98.80\% & 98.53\% \\ \hline
		Not in hospital & 1.20\% & 1.47\% \\ \hline
		\textbf{Residence status} & ~ & ~ \\ \hline
		Resident & 72.97\% & 65.82\% \\ \hline
		Interstate nonresident (type 1) & 24.73\% & 30.26\% \\ \hline
		Interstate nonresident (type 2) & 2.11\% & 3.83\% \\ \hline
		Foreign resident & 0.19\% & 0.09\% \\ \hline
	\end{tabular}
	\caption{Linked birth and infant death cohort data: variables summary.}
	\label{infants_summary_1}
\end{table}

\begin{table}[!ht]
	\centering
	\footnotesize
	\begin{tabular}{|l|c|c|}
		\hline
		~ & \textbf{Non-events} & \textbf{Events} \\ \hline
		~ & (N=28410519) & (N=176400) \\ \hline
		\textbf{Mother's race} & ~ & ~ \\ \hline
		White & 76.72\% & 64.51\% \\ \hline
		Black & 15.81\% & 29.58\% \\ \hline
		American Indian / Alaskan Native & 1.16\% & 1.55\% \\ \hline
		Asian / Pacific Islander & 6.31\% & 4.36\% \\ \hline
		\textbf{Mother's marital status} & ~ & ~ \\ \hline
		Married & 59.52\% & 45.56\% \\ \hline
		Not Married & 40.48\% & 54.44\% \\ \hline
		\textbf{Father's race} & ~ & ~ \\ \hline
		White & 63.37\% & 47.25\% \\ \hline
		Black & 11.58\% & 17.65\% \\ \hline
		American Indian / Alaskan Native & 0.87\% & 1.01\% \\ \hline
		Asian / Pacific Islander & 5.25\% & 3.24\% \\ \hline
		Unknown & 18.93\% & 30.85\% \\ \hline
		\textbf{Diabetes} & ~ & ~ \\ \hline
		Yes & 5.14\% & 4.65\% \\ \hline
		No & 94.86\% & 95.35\% \\ \hline
		\textbf{Chronic Hypertension} & ~ & ~ \\ \hline
		Yes & 1.32\% & 2.65\% \\ \hline
		No & 98.68\% & 97.35\% \\ \hline
		\textbf{Prepregnacny Associated Hypertension} & ~ & ~ \\ \hline
		Yes & 4.27\% & 4.82\% \\ \hline
		No & 95.73\% & 95.18\% \\ \hline
		\textbf{Eclampsia} & ~ & ~ \\ \hline
		Yes & 0.25\% & 0.58\% \\ \hline
		No & 99.75\% & 99.42\% \\ \hline
		\textbf{Induction of Labor} & ~ & ~ \\ \hline
		Yes & 23.01\% & 13.01\% \\ \hline
		No & 76.99\% & 86.99\% \\ \hline
		\textbf{Tocolysis} & ~ & ~ \\ \hline
		Yes & 1.19\% & 4.30\% \\ \hline
		No & 98.81\% & 95.70\% \\ \hline
		\textbf{Meconium} & ~ & ~ \\ \hline
		Yes & 4.74\% & 3.71\% \\ \hline
		No & 95.26\% & 96.29\% \\ \hline
		\textbf{Precipitous Labor} & ~ & ~ \\ \hline
		Yes & 2.46\% & 4.46\% \\ \hline
		No & 97.54\% & 95.54\% \\ \hline
		\textbf{Breech} & ~ & ~ \\ \hline
		Yes & 5.36\% & 20.60\% \\ \hline
		No & 94.64\% & 79.40\% \\ \hline
		\textbf{Forceps delivery} & ~ & ~ \\ \hline
		Yes & 0.66\% & 0.38\% \\ \hline
		No & 99.34\% & 99.62\% \\ \hline
		\textbf{Vacuum delivery} & ~ & ~ \\ \hline
		Yes & 3.02\% & 1.08\% \\ \hline
		No & 96.98\% & 98.92\% \\ \hline
		\textbf{Delivery method} & ~ & ~ \\ \hline
		Vaginal & 67.54\% & 60.98\% \\ \hline
		C-Section & 32.46\% & 39.02\% \\ \hline
	\end{tabular}
	\label{infants_summary_2}
	\caption{Linked birth and infant death cohort data: variables summary - continued.}
\end{table}

\begin{table}[!ht]
	\centering
	\footnotesize
	\begin{tabular}{|l|c|c|}
		\hline
		~ & \textbf{Non-events} & \textbf{Events} \\ \hline
		~ & (N=28410519) & (N=176400) \\ \hline
		\textbf{Attendant} & ~ & ~ \\ \hline
		Doctor of Medicine (MD) & 85.40\% & 90.00\% \\ \hline
		Doctor of Osteopathy (DO) & 5.56\% & 4.96\% \\ \hline
		Certified Nurse Midwife (CNM) & 7.75\% & 3.12\% \\ \hline
		Other Midwife & 0.64\% & 0.27\% \\ \hline
		Other & 0.65\% & 1.65\% \\ \hline
		\textbf{Sex} & ~ & ~ \\ \hline
		Female & 48.86\% & 44.17\% \\ \hline
		Male & 51.14\% & 55.83\% \\ \hline
		\textbf{Birth Weight} & ~ & ~ \\ \hline
		227- 1499 grams & 1.15\% & 53.28\% \\ \hline
		1500 – 2499 grams & 6.62\% & 14.70\% \\ \hline
		2500 - 8165 grams & 92.23\% & 32.02\% \\ \hline
		\textbf{Anencephalus} & ~ & ~ \\ \hline
		Yes & 0.01\% & 1.01\% \\ \hline
		No & 99.99\% & 98.99\% \\ \hline
		\textbf{Spina Bifida} & ~ & ~ \\ \hline
		Yes & 0.01\% & 0.25\% \\ \hline
		No & 99.99\% & 99.75\% \\ \hline
		\textbf{Omphalocele} & ~ & ~ \\ \hline
		Yes & 0.03\% & 0.68\% \\ \hline
		No & 99.97\% & 99.32\% \\ \hline
		\textbf{Cleft Lip} & ~ & ~ \\ \hline
		Yes & 0.07\% & 1.08\% \\ \hline
		No & 99.93\% & 98.92\% \\ \hline
		\textbf{Downs Syndrome} & ~ & ~ \\ \hline
		Yes & 0.05\% & 0.55\% \\ \hline
		No & 99.95\% & 99.45\% \\ \hline
	\end{tabular}
	\caption{Linked birth and infant death cohort data: variables summary - continued.}
	\label{infants_summary_3}
\end{table}

\begin{table}
	\centering
	\footnotesize
	\resizebox{\columnwidth}{!}{%
		\begin{tabular}[t]{lrrrrrrrrrrrr}
			\hline
			&  \multicolumn{4}{c}{Estimate} & \multicolumn{4}{c}{Standard Deviation} & \multicolumn{3}{c}{Adjusted P-value}\\
			\hline
			{} & {MLE} & {A} & {L} & {uniform} & {MLE} & {A} & {L} & {uniform} & {MLE} & {A} & {L} &  {uniform}\\
			\hline
			Intercept & 18.5035 & 18.4739 & 18.4791 & 19.0026 & 0.2762 & 0.2782 & 0.2892 & 0.6643 & 0.0000 & 0.0000 & 0.0000 & 0.0000\\
			Mother Age & -0.0938 & -0.0947 & -0.0919 & -0.0944 & 0.0036 & 0.0038 & 0.0038 & 0.0069 & 0.0000 & 0.0000 & 0.0000 & 0.0000\\
			Live birth order & 0.1109 & 0.1120 & 0.1111 & 0.1080 & 0.0023 & 0.0024 & 0.0024 & 0.0046 & 0.0000 & 0.0000 & 0.0000 & 0.0000\\
			Number of prenatal visits & -0.0134 & -0.0132 & -0.0134 & -0.0151 & 0.0007 & 0.0007 & 0.0007 & 0.0013 & 0.0000 & 0.0000 & 0.0000 & 0.0000\\
			Weight gain & -0.0029 & -0.0029 & -0.0030 & -0.0030 & 0.0003 & 0.0003 & 0.0003 & 0.0006 & 0.0000 & 0.0000 & 0.0000 & 0.0000\\
			Five minute APGAR score & -0.5183 & -0.5182 & -0.5180 & -0.5140 & 0.0019 & 0.0019 & 0.0019 & 0.0044 & 0.0000 & 0.0000 & 0.0000 & 0.0000\\
			Plurality & -0.0872 & -0.0846 & -0.0811 & -0.0553 & 0.0122 & 0.0128 & 0.0127 & 0.0250 & 0.0000 & 0.0000 & 0.0000 & 0.0691\\
			Gestation weeks & -0.1307 & -0.1308 & -0.1300 & -0.1282 & 0.0012 & 0.0013 & 0.0013 & 0.0025 & 0.0000 & 0.0000 & 0.0000 & 0.0000\\
			Year & 0.0126 & 0.0260 & 0.0215 & -0.0901 & 0.0657 & 0.0662 & 0.0690 & 0.1523 & 0.8913 & 0.7523 & 0.8268 & 0.7115\\
			Squared mother age & 0.0012 & 0.0012 & 0.0012 & 0.0013 & 0.0001 & 0.0001 & 0.0001 & 0.0001 & 0.0000 & 0.0000 & 0.0000 & 0.0000\\
			Birth place = not in hospital & 0.6151 & 0.6252 & 0.6098 & 0.6420 & 0.0317 & 0.0325 & 0.0331 & 0.0643 & 0.0000 & 0.0000 & 0.0000 & 0.0000\\
			Diabetes = no & -0.0148 & 0.0001 & -0.0172 & -0.0110 & 0.0285 & 0.0293 & 0.0294 & 0.0511 & 0.6896 & 0.9975 & 0.6611 & 0.9378\\
			Chronic hypertension = no & 0.1177 & 0.1152 & 0.0952 & 0.0178 & 0.0402 & 0.0409 & 0.0415 & 0.0768 & 0.0072 & 0.0101 & 0.0418 & 0.9339\\
			Prepregnacny hypertension = no & 0.3343 & 0.3450 & 0.3208 & 0.3400 & 0.0271 & 0.0280 & 0.0282 & 0.0474 & 0.0000 & 0.0000 & 0.0000 & 0.0000\\
			Eclampsia = no & 0.4996 & 0.4983 & 0.5299 & 0.2907 & 0.0770 & 0.0774 & 0.0796 & 0.1371 & 0.0000 & 0.0000 & 0.0000 & 0.0823\\
			Induction of labor = no & 0.0054 & 0.0029 & 0.0020 & -0.0083 & 0.0176 & 0.0184 & 0.0183 & 0.0246 & 0.8081 & 0.9082 & 0.9134 & 0.8705\\
			Tocolysis = no & 0.0992 & 0.1073 & 0.1005 & -0.0120 & 0.0313 & 0.0321 & 0.0327 & 0.0640 & 0.0034 & 0.0019 & 0.0047 & 0.9418\\
			Meconium = no & -0.1216 & -0.1277 & -0.1117 & -0.0889 & 0.0262 & 0.0272 & 0.0275 & 0.0475 & 0.0000 & 0.0000 & 0.0001 & 0.1242\\
			Precipitous labor = no & 0.0361 & 0.0351 & 0.0248 & -0.0213 & 0.0286 & 0.0294 & 0.0298 & 0.0622 & 0.2721 & 0.3129 & 0.5080 & 0.8705\\
			Breech = no & -0.1003 & -0.0980 & -0.1055 & -0.1011 & 0.0147 & 0.0155 & 0.0153 & 0.0330 & 0.0000 & 0.0000 & 0.0000 & 0.0066\\
			Forceps delivery = no & 0.1253 & 0.1161 & 0.1195 & 0.1464 & 0.0898 & 0.0902 & 0.0937 & 0.1282 & 0.2293 & 0.2778 & 0.2840 & 0.4185\\
			Vacuum delivery = no & 0.2982 & 0.3145 & 0.3024 & 0.2598 & 0.0508 & 0.0514 & 0.0529 & 0.0568 & 0.0000 & 0.0000 & 0.0000 & 0.0000\\
			Delivery method = C-Section & -0.0401 & -0.0432 & -0.0377 & -0.0368 & 0.0098 & 0.0103 & 0.0102 & 0.0186 & 0.0001 & 0.0001 & 0.0005 & 0.1008\\
			Sex = male & -0.7590 & -0.7601 & -0.7388 & -0.9213 & 0.2665 & 0.2686 & 0.2796 & 0.6448 & 0.0090 & 0.0099 & 0.0168 & 0.2792\\
			Anencephaly = no & -4.1255 & -4.1271 & -4.1709 & -4.0831 & 0.1118 & 0.1120 & 0.1172 & 0.3471 & 0.0000 & 0.0000 & 0.0000 & 0.0000\\
			Spina Bifida = no & -2.1350 & -2.1323 & -2.1315 & -2.0956 & 0.1488 & 0.1487 & 0.1559 & 0.3737 & 0.0000 & 0.0000 & 0.0000 & 0.0000\\
			Omphalocele = no & -1.7259 & -1.7286 & -1.7383 & -1.8822 & 0.0829 & 0.0831 & 0.0876 & 0.2115 & 0.0000 & 0.0000 & 0.0000 & 0.0000\\
			Cleft lip = no & -2.8745 & -2.8681 & -2.8451 & -2.9524 & 0.0656 & 0.0660 & 0.0685 & 0.1456 & 0.0000 & 0.0000 & 0.0000 & 0.0000\\
			Downs syndrome = no & -2.3438 & -2.3402 & -2.3462 & -2.4419 & 0.0863 & 0.0868 & 0.0886 & 0.1639 & 0.0000 & 0.0000 & 0.0000 & 0.0000\\
			\multicolumn{13}{l}{\textbf{Birth month vs. January}}\\
			\hspace{1em}Birth month = February & 0.0103 & 0.0119 & 0.0123 & 0.0592 & 0.0145 & 0.0153 & 0.0152 & 0.0268 & 0.5611 & 0.5133 & 0.5178 & 0.0691\\
			\hspace{1em}Birth month = March & -0.0305 & -0.0312 & -0.0278 & 0.0051 & 0.0143 & 0.0151 & 0.0149 & 0.0265 & 0.0571 & 0.0703 & 0.1064 & 0.9418\\
			\hspace{1em}Birth month = April & -0.0294 & -0.0275 & -0.0268 & -0.0121 & 0.0144 & 0.0152 & 0.0150 & 0.0269 & 0.0711 & 0.1193 & 0.1184 & 0.8082\\
			\hspace{1em}Birth month = May & -0.0434 & -0.0441 & -0.0402 & -0.0014 & 0.0143 & 0.0150 & 0.0149 & 0.0272 & 0.0051 & 0.0073 & 0.0145 & 0.9683\\
			\hspace{1em}Birth month = June & -0.0344 & -0.0296 & -0.0294 & 0.0006 & 0.0143 & 0.0151 & 0.0149 & 0.0268 & 0.0299 & 0.0883 & 0.0870 & 0.9829\\
			\hspace{1em}Birth month = July & -0.0301 & -0.0259 & -0.0271 & 0.0163 & 0.0141 & 0.0149 & 0.0147 & 0.0265 & 0.0571 & 0.1339 & 0.1064 & 0.7080\\
			\hspace{1em}Birth month = August & -0.0213 & -0.0185 & -0.0241 & 0.0174 & 0.0141 & 0.0148 & 0.0147 & 0.0264 & 0.2002 & 0.2945 & 0.1514 & 0.6974\\
			\hspace{1em}Birth month = September & -0.0180 & -0.0165 & -0.0150 & 0.0418 & 0.0142 & 0.0150 & 0.0148 & 0.0263 & 0.2721 & 0.3511 & 0.4116 & 0.2134\\
			\hspace{1em}Birth month = October & -0.0287 & -0.0275 & -0.0273 & 0.0314 & 0.0142 & 0.0150 & 0.0148 & 0.0262 & 0.0731 & 0.1136 & 0.1064 & 0.3872\\
			\hspace{1em}Birth month = November & -0.0311 & -0.0414 & -0.0302 & 0.0029 & 0.0144 & 0.0152 & 0.0150 & 0.0271 & 0.0561 & 0.0129 & 0.0808 & 0.9524\\
			\hspace{1em}Birth month = December & -0.0409 & -0.0377 & -0.0422 & -0.0248 & 0.0143 & 0.0151 & 0.0149 & 0.0274 & 0.0090 & 0.0240 & 0.0100 & 0.5444\\
			\multicolumn{13}{l}{\textbf{Birth weekday vs. Sunday}}\\
			\hspace{1em}Birth weekday = Monday & 0.0907 & 0.0910 & 0.0978 & 0.1239 & 0.0118 & 0.0125 & 0.0123 & 0.0228 & 0.0000 & 0.0000 & 0.0000 & 0.0000\\
			\hspace{1em}Birth weekday = Tuesday & 0.0981 & 0.1037 & 0.1045 & 0.1035 & 0.0116 & 0.0123 & 0.0122 & 0.0227 & 0.0000 & 0.0000 & 0.0000 & 0.0000\\
			\hspace{1em}Birth weekday = Wednesday & 0.0941 & 0.0971 & 0.0999 & 0.0781 & 0.0117 & 0.0123 & 0.0122 & 0.0226 & 0.0000 & 0.0000 & 0.0000 & 0.0018\\
			\hspace{1em}Birth weekday = Thursday & 0.0902 & 0.0903 & 0.0974 & 0.1145 & 0.0117 & 0.0123 & 0.0122 & 0.0224 & 0.0000 & 0.0000 & 0.0000 & 0.0000\\
			\hspace{1em}Birth weekday = Friday & 0.0755 & 0.0726 & 0.0783 & 0.0711 & 0.0117 & 0.0124 & 0.0122 & 0.0229 & 0.0000 & 0.0000 & 0.0000 & 0.0060\\
			\hspace{1em}Birth weekday = Saturday & 0.0208 & 0.0199 & 0.0205 & 0.0302 & 0.0124 & 0.0131 & 0.0130 & 0.0245 & 0.1504 & 0.2010 & 0.1687 & 0.3731\\
			\multicolumn{13}{l}{\textbf{Resdience status vs. 1}}\\
			\hspace{1em}Residence status = 2 & 0.1156 & 0.1123 & 0.1118 & 0.1159 & 0.0066 & 0.0069 & 0.0068 & 0.0124 & 0.0000 & 0.0000 & 0.0000 & 0.0000\\
			\hspace{1em}Residence status = 3 & 0.2355 & 0.2306 & 0.2443 & 0.2717 & 0.0162 & 0.0169 & 0.0168 & 0.0338 & 0.0000 & 0.0000 & 0.0000 & 0.0000\\
			\hspace{1em}Residence status = 4 & -0.4333 & -0.4285 & -0.4366 & -0.1390 & 0.0873 & 0.0876 & 0.0897 & 0.0931 & 0.0000 & 0.0000 & 0.0000 & 0.2513\\
			\multicolumn{13}{l}{\textbf{Mother's race vs. white}}\\
			\hspace{1em}Mother’s race = black & -0.0156 & -0.0196 & -0.0136 & -0.0678 & 0.0165 & 0.0174 & 0.0174 & 0.0338 & 0.4331 & 0.3441 & 0.5295 & 0.0980\\
			\hspace{1em}Mother’s race = american indian & 0.2387 & 0.2444 & 0.2206 & 0.0930 & 0.0460 & 0.0467 & 0.0482 & 0.0990 & 0.0000 & 0.0000 & 0.0000 & 0.5240\\
			\hspace{1em}Mother’s race = asian & 0.0121 & 0.0207 & 0.0099 & 0.0364 & 0.0362 & 0.0368 & 0.0375 & 0.0575 & 0.7989 & 0.6485 & 0.8578 & 0.7075\\
			Paternity acknowledged = no & 0.0991 & 0.1022 & 0.0988 & 0.0945 & 0.0074 & 0.0078 & 0.0076 & 0.0136 & 0.0000 & 0.0000 & 0.0000 & 0.0000\\
			\multicolumn{13}{l}{\textbf{Father's race vs. white}}\\
			\hspace{1em}Father's race = black & 0.1357 & 0.1409 & 0.1389 & 0.2239 & 0.0199 & 0.0209 & 0.0209 & 0.0387 & 0.0000 & 0.0000 & 0.0000 & 0.0000\\
			\hspace{1em}Father's race = american indian & 0.2200 & 0.2230 & 0.2323 & 0.2298 & 0.0562 & 0.0569 & 0.0591 & 0.1122 & 0.0002 & 0.0002 & 0.0002 & 0.0895\\
			\hspace{1em}Father's race = asian & -0.0198 & -0.0329 & -0.0227 & -0.0100 & 0.0413 & 0.0421 & 0.0426 & 0.0661 & 0.7066 & 0.5132 & 0.6948 & 0.9481\\
			\hspace{1em}Father's race = unknown & 0.1746 & 0.1724 & 0.1732 & 0.2093 & 0.0141 & 0.0148 & 0.0147 & 0.0265 & 0.0000 & 0.0000 & 0.0000 & 0.0000\\
			\multicolumn{13}{l}{\textbf{Attendant vs. MD}}\\
			\hspace{1em}Attendant = DO & -0.0444 & -0.0513 & -0.0462 & -0.0537 & 0.0134 & 0.0140 & 0.0139 & 0.0239 & 0.0020 & 0.0006 & 0.0021 & 0.0656\\
			\hspace{1em}Attendant = CNM & -0.2782 & -0.2861 & -0.2786 & -0.3119 & 0.0157 & 0.0165 & 0.0164 & 0.0241 & 0.0000 & 0.0000 & 0.0000 & 0.0000\\
			\hspace{1em}Attendant = other midwife & -0.2361 & -0.2453 & -0.2209 & -0.3163 & 0.0542 & 0.0549 & 0.0563 & 0.0901 & 0.0000 & 0.0000 & 0.0002 & 0.0015\\
			\hspace{1em}Attendant = other & 0.1822 & 0.1765 & 0.1797 & 0.2198 & 0.0311 & 0.0318 & 0.0325 & 0.0617 & 0.0000 & 0.0000 & 0.0000 & 0.0013\\
			\multicolumn{13}{l}{\textbf{Birth weight recode vs. 1}}\\
			\hspace{1em}Birth weight recode = 2 & -0.7391 & -0.7430 & -0.7353 & -0.7328 & 0.0116 & 0.0122 & 0.0121 & 0.0220 & 0.0000 & 0.0000 & 0.0000 & 0.0000\\
			\hspace{1em}Birth weight recode = 3 & -1.6916 & -1.6899 & -1.6930 & -1.6808 & 0.0141 & 0.0149 & 0.0147 & 0.0265 & 0.0000 & 0.0000 & 0.0000 & 0.0000\\
			\hline
	\end{tabular}}
	\caption{Linked birth and infant death cohort data: results of logistic regression with $c=10$ for non-interaction terms. P-values are adjusted by FDR. }
	\label{infants_full_coefs}
\end{table}

\begin{table}
	\centering
	\footnotesize
	\resizebox{\columnwidth}{!}{%
		\begin{tabular}[t]{lrrrrrrrrrrrr}
			\hline
			&  \multicolumn{4}{c}{Estimate} & \multicolumn{4}{c}{Standard Deviation} & \multicolumn{4}{c}{Adjusted P-value}\\
			\hline
			{} & {MLE} & {A} & {L} & {uniform} & {MLE} & {A} & {L} & {uniform} & {MLE} & {A} & {L} &  {uniform}\\
			\hline
			Weight gain & -0.0012 & -0.0012 & -0.0010 & -0.0013 & 0.0004 & 0.0004 & 0.0004 & 0.0008 & 0.0127 & 0.0155 & 0.0418 & 0.1586\\
			Apgar & 0.0315 & 0.0312 & 0.0302 & 0.0307 & 0.0025 & 0.0026 & 0.0026 & 0.0059 & 0.0000 & 0.0000 & 0.0000 & 0.0000\\
			Plurality & 0.0180 & 0.0165 & 0.0075 & -0.0103 & 0.0163 & 0.0172 & 0.0169 & 0.0340 & 0.3447 & 0.4207 & 0.7435 & 0.8911\\
			Gestation week & -0.0016 & -0.0014 & -0.0022 & -0.0055 & 0.0012 & 0.0013 & 0.0013 & 0.0025 & 0.2518 & 0.3507 & 0.1257 & 0.0701\\
			Diabetes = no & 0.0388 & 0.0253 & 0.0418 & 0.0964 & 0.0266 & 0.0277 & 0.0275 & 0.0470 & 0.2147 & 0.4457 & 0.1877 & 0.0895\\
			Chronic hypertension = no & -0.0586 & -0.0510 & -0.0428 & 0.0197 & 0.0376 & 0.0386 & 0.0388 & 0.0761 & 0.1846 & 0.2645 & 0.3600 & 0.9196\\
			Prepregnacny hypertension = no & -0.0692 & -0.0634 & -0.0620 & -0.0421 & 0.0260 & 0.0272 & 0.0271 & 0.0478 & 0.0152 & 0.0373 & 0.0418 & 0.5539\\
			Eclampsia = no & -0.0270 & -0.0385 & -0.0374 & -0.0806 & 0.0740 & 0.0745 & 0.0771 & 0.1286 & 0.7827 & 0.6763 & 0.7254 & 0.7075\\
			Induction of labor = no & 0.0409 & 0.0411 & 0.0469 & 0.0427 & 0.0173 & 0.0182 & 0.0180 & 0.0246 & 0.0328 & 0.0440 & 0.0183 & 0.1586\\
			Tocolysis = no & -0.0056 & -0.0120 & -0.0076 & 0.0724 & 0.0312 & 0.0323 & 0.0326 & 0.0676 & 0.8921 & 0.7620 & 0.8655 & 0.4548\\
			Forceps delivery = no & 0.0692 & 0.0866 & 0.1151 & 0.0757 & 0.0875 & 0.0879 & 0.0917 & 0.1252 & 0.5132 & 0.4116 & 0.2903 & 0.7088\\
			Vacuum delivery = no & -0.0215 & -0.0308 & -0.0060 & -0.0080 & 0.0495 & 0.0502 & 0.0518 & 0.0557 & 0.7347 & 0.6169 & 0.9134 & 0.9481\\
			Delivery method = C-Section & -0.0217 & -0.0212 & -0.0242 & -0.0183 & 0.0126 & 0.0133 & 0.0131 & 0.0241 & 0.1380 & 0.1824 & 0.1064 & 0.6312\\
			Anencephaly = no & 0.1447 & 0.1466 & 0.1047 & 0.7030 & 0.1095 & 0.1096 & 0.1141 & 0.3429 & 0.2518 & 0.2616 & 0.4612 & 0.0895\\
			Spina Bifida = no & 0.2920 & 0.2926 & 0.2826 & -0.1952 & 0.1514 & 0.1517 & 0.1594 & 0.3568 & 0.0889 & 0.0946 & 0.1202 & 0.7321\\
			Omphalocele = no & -0.0071 & -0.0054 & -0.0096 & -0.1927 & 0.0799 & 0.0804 & 0.0846 & 0.2031 & 0.9473 & 0.9647 & 0.9134 & 0.5240\\
			Cleft lip = no & 0.3197 & 0.3159 & 0.2768 & 0.4703 & 0.0644 & 0.0649 & 0.0672 & 0.1370 & 0.0000 & 0.0000 & 0.0001 & 0.0019\\
			Downs syndrome = no & 0.0677 & 0.0748 & 0.1029 & 0.1770 & 0.0852 & 0.0857 & 0.0880 & 0.1879 & 0.5132 & 0.4630 & 0.3274 & 0.5240\\
			\hline
	\end{tabular}}
	\caption{Linked birth and infant death cohort data: results of logistic regression with $c=10$ for interaction terms with sex (male). P-values are adjusted by FDR.}
	\label{infants_sex_coefs}
\end{table}

\begin{table}
	\centering
	\footnotesize
	\resizebox{\columnwidth}{!}{%
		\begin{tabular}[t]{lrrrrrrrrrrrr}
			\hline
			&  \multicolumn{4}{c}{Coefficient} & \multicolumn{4}{c}{Coefficient sd} & \multicolumn{3}{c}{P-value}\\
			\hline
			{} & {MLE} & {A} & {L} & {uniform} & {MLE} & {A} & {L} & {uniform} & {MLE} & {A} & {L} &  {uniform}\\
			\hline
			Father's race = black & -0.0079 & -0.0084 & -0.0078 & -0.0265 & 0.0057 & 0.0060 & 0.0059 & 0.0112 & 0.2293 & 0.2400 & 0.2751 & 0.0498\\
			Father's race = american indian & -0.0267 & -0.0250 & -0.0322 & -0.0249 & 0.0161 & 0.0163 & 0.0168 & 0.0305 & 0.1522 & 0.1969 & 0.0986 & 0.5991\\
			Father's race = asian & -0.0128 & -0.0105 & -0.0114 & -0.0257 & 0.0116 & 0.0119 & 0.0120 & 0.0187 & 0.3447 & 0.4625 & 0.4449 & 0.2938\\
			Father's race = unknown & -0.0029 & -0.0028 & -0.0019 & -0.0129 & 0.0039 & 0.0041 & 0.0041 & 0.0073 & 0.5362 & 0.5733 & 0.7345 & 0.1541\\
			Mother’s race = black & -0.0174 & -0.0169 & -0.0169 & -0.0012 & 0.0047 & 0.0050 & 0.0050 & 0.0098 & 0.0006 & 0.0016 & 0.0016 & 0.9489\\
			Mother’s race = american indian & -0.0188 & -0.0196 & -0.0103 & 0.0091 & 0.0132 & 0.0134 & 0.0138 & 0.0270 & 0.2228 & 0.2193 & 0.5481 & 0.8705\\
			Mother’s race = asian & -0.0049 & -0.0032 & -0.0034 & -0.0028 & 0.0102 & 0.0104 & 0.0105 & 0.0163 & 0.7066 & 0.8068 & 0.8237 & 0.9448\\
			Diabetes = no & -0.0058 & -0.0082 & -0.0045 & -0.0162 & 0.0066 & 0.0068 & 0.0068 & 0.0118 & 0.4589 & 0.3129 & 0.6081 & 0.2938\\
			Chronic hypertension = no & -0.0028 & -0.0047 & 0.0019 & 0.0066 & 0.0094 & 0.0096 & 0.0096 & 0.0185 & 0.8081 & 0.6930 & 0.8772 & 0.8705\\
			Prepregnacny hypertension = no & 0.0162 & 0.0116 & 0.0198 & 0.0127 & 0.0064 & 0.0066 & 0.0066 & 0.0115 & 0.0208 & 0.1339 & 0.0062 & 0.4381\\
			Eclampsia = no & -0.0232 & -0.0204 & -0.0319 & 0.0234 & 0.0186 & 0.0187 & 0.0192 & 0.0330 & 0.2758 & 0.3535 & 0.1484 & 0.6627\\
			Induction of labor = no & -0.0157 & -0.0150 & -0.0147 & -0.0125 & 0.0042 & 0.0044 & 0.0043 & 0.0060 & 0.0004 & 0.0015 & 0.0016 & 0.0861\\
			Tocolysis = no & 0.0277 & 0.0239 & 0.0259 & 0.0440 & 0.0077 & 0.0079 & 0.0080 & 0.0160 & 0.0007 & 0.0056 & 0.0028 & 0.0182\\
			Meconium = no & -0.0008 & -0.0004 & -0.0033 & -0.0074 & 0.0072 & 0.0075 & 0.0076 & 0.0129 & 0.9445 & 0.9658 & 0.7450 & 0.7177\\
			Precipitous labor = no & -0.0106 & -0.0109 & -0.0071 & -0.0009 & 0.0077 & 0.0080 & 0.0081 & 0.0165 & 0.2367 & 0.2543 & 0.4840 & 0.9683\\
			Breech = no & -0.0002 & -0.0011 & 0.0010 & -0.0012 & 0.0040 & 0.0043 & 0.0042 & 0.0092 & 0.9685 & 0.8443 & 0.8655 & 0.9481\\
			Forceps delivery = no & -0.0199 & -0.0184 & -0.0264 & -0.0241 & 0.0215 & 0.0216 & 0.0225 & 0.0304 & 0.4390 & 0.4727 & 0.3274 & 0.6090\\
			Vacuum delivery = no & -0.0176 & -0.0187 & -0.0234 & -0.0137 & 0.0120 & 0.0121 & 0.0125 & 0.0135 & 0.2109 & 0.1968 & 0.1064 & 0.4905\\
			Anencephaly = no & -0.1061 & -0.1052 & -0.0938 & -0.2437 & 0.0271 & 0.0272 & 0.0284 & 0.0899 & 0.0002 & 0.0003 & 0.0022 & 0.0194\\
			Spina Bifida = no & 0.0532 & 0.0520 & 0.0503 & 0.1032 & 0.0373 & 0.0374 & 0.0393 & 0.0749 & 0.2228 & 0.2449 & 0.2840 & 0.2938\\
			Omphalocele = no & -0.0008 & -0.0014 & 0.0043 & 0.1153 & 0.0200 & 0.0200 & 0.0212 & 0.0496 & 0.9685 & 0.9647 & 0.8772 & 0.0549\\
			Cleft lip = no & 0.0085 & 0.0071 & 0.0022 & -0.0026 & 0.0159 & 0.0160 & 0.0166 & 0.0349 & 0.6826 & 0.7202 & 0.9134 & 0.9675\\
			Downs syndrome = no & 0.0559 & 0.0533 & 0.0492 & 0.0827 & 0.0211 & 0.0212 & 0.0218 & 0.0428 & 0.0152 & 0.0235 & 0.0443 & 0.1105\\
			\hline
	\end{tabular}}
	\caption{Linked birth and infant death cohort data: results of logistic regression with $c=10$ for interaction terms with birth year. P-values are adjusted by FDR.}
	\label{infants_year_coefs}
\end{table}

\end{document}